# Information Flow and Market Risk: Transformer-Based Evidence on Sentiment and Volatility Dynamics

*Szymon Lis*

*Robert Ślepaczuk*

*Paweł Sakowski*

*University of Warsaw, Faculty of Economic Sciences, Department of Quantitative Finance*

*sm.lis@student.uw.edu.pl*

## Abstract

This paper evaluates whether market-wide ModernFinBERT news variables improve volatility forecasts for an ex-post selected panel of eleven U.S.-listed stocks using source data from 2000–2022. At $h = 1, 5, 22$, the target is the annualized square root of an average of future trailing 252-session Yang–Zhang variance estimates. Sentiment, directionality, posterior entropy, intensity, and news volume are compared with persistence, VIX, and AAII through matched HAR-X models and additional regularized, tree-based, naive, and stacking specifications.

Hyperparameters are selected once on the first 252 complete observations. Five chronological blocks support expanding and rolling comparisons. The primary expanding family contains 792 BH-adjusted tests, with joint-window and dependence sensitivities. Naive has the lowest mean QLIKE at each horizon. HAR-VIX is the most frequent expanding cell winner, but not the model with the lowest mean loss. HAR-Text improves on HAR in two of 33 expanding cells; neither favorable QLIKE rejection survives longer HAC. High-state comparisons are relatively less adverse to text but do not establish generally positive gains.

Stacking contributes 66 tests (33 cells $\times$ two losses) to this fixed family. Because its near-constant forecasts generate very small p-values in many adverse comparisons, those tests can affect the BH rank thresholds applied to the remaining models. They are retained only to preserve the ex-ante model universe; no substantive evidence claim is based on Stacking performance.

ModernFinBERT is a fixed ex-post measurement instrument with an undocumented training-data cutoff. The existing 2,700-text agreement audit is not expert ground truth and is not regenerated here. SHAP explains fitted forecasts, not causal effects or incremental accuracy. Conclusions are conditional on the long-window target, initial-only selection, sparse refits, calibration, and numerical safeguards.

learning

## 1 Introduction

Forecasting financial-market volatility is central to derivatives valuation, portfolio allocation, risk management, margin setting, and systemic-risk monitoring. In contrast to expected returns, volatility displays persistence, clustering, long-memory-like behavior, leverage effects, and pronounced state dependence, which make it comparatively more forecastable and economically relevant (Andersen et al., 2001, 2003; Bollerslev, 1986; Christoffersen, 2012; Engle, 1982; McNeil et al., 2015). The development of realized-volatility measures further strengthened this forecasting literature by providing observable proxies for latent integrated variance (Andersen et al., 2001; Barndorff-Nielsen & Shephard, 2002). Within this class of models, the heterogeneous autoregressive model of realized volatility (HAR-RV) has become a particularly demanding benchmark because it captures persistence across daily, weekly, and monthly horizons in a parsimonious specification (Andersen et al., 2007; Bollerslev et al., 2016; Corsi, 2009).

The strength of HAR-type benchmarks raises a substantive question for information-rich forecasting systems: do additional predictors contain genuinely incremental information, or do they mainly increase model complexity without improving out-of-sample accuracy? This question is especially important for textual data. Financial news is economically plausible as a source of volatility information because news arrivals, uncertainty, disagreement, and heterogeneous interpretation can affect trading activity and risk perceptions (Bekaert & Harvey, 2000; French et al., 1987; O. Kim & Verrecchia, 1994; Ross, 1989). Textual measures may therefore capture dimensions of the information environment that are not fully represented by lagged realized volatility.

The preceding empirical phases sharpen this motivation. Article 2 (Section 5) shows that VIX-based market uncertainty is more consistently related to subsequent trading dynamics than the examined optimism indicators are to abnormal returns. Article 3 (Section 8) finds that text-derived emotions do not provide robust short-horizon directional signals, while market-state and drawdown information is more useful for exposure management. These findings motivate a distinct question for the present study: whether market-wide text adds more reliably to forecasts of risk and volatility than it does to directional return prediction.

The empirical literature provides reasons both for optimism and for caution. Media pessimism, news-based uncertainty, and disagreement have been associated with returns, trading activity, and volatility (Audrino et al., 2020; García, 2013; Tetlock, 2007). At the same time, evidence on machine-learning volatility forecasts is model- and design-dependent. Neural-network and broader machine-learning approaches can improve realized-volatility forecasts in some settings, but their gains are not uniform across models, horizons, and evaluation designs (Bucci, 2020). Separately, nonlinear HAR extensions and explicit treatment of realized-volatility measurement error can improve on the canonical HAR benchmark (Buccheri & Corsi, 2021). These results motivate direct out-of-sample comparison rather than a presumption that a modern language model will improve forecasts relative to a strong persistence benchmark.

This paper therefore asks a deliberately narrow question: *does market-wide financial news processed with a modern transformer provide incremental out-of-sample information for forecasting individual-stock realized volatility?* The emphasis is on the predictive value of the textual signal rather than on the predictive capacity of machine learning in general. To make that distinction explicit, the transformer-based news variables are evaluated alongside two simpler market-wide information channels: VIX-based implied volatility and AAII survey sentiment. VIX summarizes forward-looking option-market expectations (Busch et al., 2011; Christensen & Prabhala, 1998; Whaley, 2000), whereas AAII provides a direct survey-based measure of investor sentiment (M. Baker & Wurgler, 2007; Brown & Cliff, 2004). Comparing these channels helps determine whether any observed gains are specific to transformer-derived text or instead arise from more conventional information sources.

## 1.1 Research hypotheses

The empirical analysis evaluates four hypotheses. The hypotheses distinguish incremental textual information from the predictive value of conventional market-wide indicators, model flexibility, and forecast combination.

**H4.1: Incremental textual forecasting value.** Model specifications that augment volatility-persistence predictors with ModernFinBERT-derived news variables should achieve lower out-of-sample forecast losses than the corresponding available non-text benchmarks.

**H4.2: Comparative information-channel performance.** If transformer-derived news contains economically distinct information, its forecasting contribution should be at least as recurrent as the contribution of survey-based AAII sentiment and VIX-based market-implied volatility across assets and forecast horizons.

**H4.3: Stability across forecasting environments.** Any forecasting gain associated with textual information should remain visible across securities, one-, five-, and twenty-two-day horizons, and rolling and expanding walk-forward estimation schemes rather than being confined to an isolated ticker–horizon–model configuration.

**H4.4: State-dependent textual value.** If textual sentiment, directionality, and posterior uncertainty become particularly informative during changes in market risk, text-augmented specifications should exhibit relatively stronger forecast performance in the ex-post high-volatility regime than in the low-volatility regime.

The hypotheses are assessed using out-of-sample forecast losses and formal forecast comparisons; the interpretive role and limitations of SHAP are defined in Section 3.12.

The textual signal is market-wide rather than firm-specific. Daily New York Times financial news is processed with ModernFinBERT to construct directional sentiment, directionality, posterior entropy, sentiment intensity, and news-volume variables. The same daily information-flow measures are then used to forecast Yang–Zhang realized volatility for a panel of large U.S. equities. This design imposes a demanding test. A broad news signal must add information beyond each stock's own volatility persistence and must remain useful despite being common across firms with different sectoral exposures and information environments.

The forecasting exercise compares the baseline HAR model with matched HAR-X specifications, Elastic Net, Random Forest, XGBoost, a naive benchmark, and forecast-combination models over one-day, five-day, and twenty-two-day horizons. Rolling and expanding walk-forward schemes are considered because the usefulness of historical information may depend on the trade-off between parameter stability and adaptation to structural change (Pesaran & Timmermann, 2007). Forecast performance is evaluated primarily with QLIKE. Squared-error losses enter the secondary DM comparisons and winner counts, while RMSE and MAE are retained as archived diagnostics. Formal forecast-comparison procedures are used to distinguish economically small numerical differences from more persistent forecasting gains (Clark & West, 2007; Diebold & Mariano, 1995; Giacomini & White, 2006; Patton, 2011). All estimated forecasting specifications are evaluated under the same horizon-specific, fully purged walk-forward protocol. Both model selection and final out-of-sample evaluation preserve chronological ordering and impose a gap corresponding to the forecast horizon. Statistical conclusions are additionally based on multiplicity-adjusted forecast-comparison procedures rather than on raw minimum-loss rankings alone.

We also examine whether forecast performance and predictor attribution vary across low- and high-volatility states. Volatility dynamics differ substantially between tranquil and stressed periods, and the relevance of news, sentiment, and nonlinear interactions may be state-dependent (Ang & Bekaert, 2002; Guidolin, 2011; Hamilton, 1989; Marcucci, 2005). The regime analysis is used as an ex-post diagnostic rather than as a real-time forecasting input. It therefore describes where forecast errors and feature attributions differ, without treating full-sample regime labels as implementable trading signals.

This paper contributes to the volatility-forecasting literature in five respects.

First, it evaluates the out-of-sample forecasting performance of market-wide transformer-based news signals relative to a strong HAR benchmark and matched linear HAR-X specifications for the VIX, AAII, text, and full information blocks. These linear comparisons isolate the incremental contribution of each information channel without simultaneously changing the functional form of the benchmark. The text-augmented and full information sets are also evaluated with tree-based models. VIX-only and AAII-only extensions are compared within the matched linear family; no RF-VIX specification is added.

Second, it separates three conceptually different information channels: transformer-derived news variables, AAII survey sentiment, and VIX-based market-implied volatility. This separation prevents forecasting gains generated by survey data, option-market information, or forecast combinations from being attributed to textual sentiment.

Third, it evaluates these information channels within a common multi-stock, multi-horizon walk-forward framework. The comparison covers one-day, five-day, and twenty-two-day Yang–Zhang volatility targets, multiple model classes, and both rolling and expanding estimation schemes. This permits an assessment of whether any incremental value is stable across assets, horizons, and estimation environments rather than confined to an isolated configuration.

Fourth, it complements aggregate forecast losses with ex-post regime-stratified diagnostics. This analysis examines whether the relative performance of HAR, text-based models, survey-based models, VIX specifications, and forecast combinations differs between tranquil and stressed volatility states, while maintaining a clear distinction between descriptive regime analysis and implementable real-time forecasting.

Fifth, it combines predictive evaluation with SHAP-based model interpretation (Lundberg & Lee, 2017). The analysis distinguishes between two questions that are often conflated: whether a variable receives substantial attribution inside a nonlinear model, and whether adding that information produces a robust out-of-sample forecasting gain.

Taken together, the design separates within-model attribution from incremental forecasting performance and makes the dependence on the long-window target and initial parameter-selection rule explicit. The empirical assessment below concerns that fully stated procedure. Neither model sophistication nor feature importance alone establishes the forecasting value of an information channel.

The remainder of the paper is organized as follows. Section 2 describes the equity, VIX, AAII, and New York Times data and the construction of the transformer-based textual variables. Section 3 presents the forecasting targets, feature sets, models, walk-forward evaluation design, forecast-comparison procedures, regime diagnostics, and SHAP analysis. Section 4 reports the full-sample and regime-stratified forecasting evidence and interprets the feature-attribution results. Section 5 concludes.

# 2 Dataset

## 2.1 Data Sources, Dataset Construction, and Corpus Diagnostics

Our empirical design requires a dataset capable of supporting a multi-horizon volatility forecasting task in which stock-specific realized volatility is predicted using three complementary information channels: historical volatility dynamics, market-wide risk expectations, and textual news-based sentiment. Existing volatility datasets typically provide rich market variables but limited textual information, while existing financial text datasets often lack direct alignment with daily asset-level forecasting targets. This motivates the construction of a daily panel dataset that links equity prices, implied volatility, investor sentiment, and large-scale news sentiment over a common sample period.

The final dataset is constructed at the daily frequency for AAPL, ABBV, ABT, AMZN, JNJ, JPM, LLY, MA, UNH, WMT, and XOM. Each equity is evaluated at forecast horizons $h = 1$, $h = 5$, and $h = 22$, yielding a complete empirical panel of $11 \times 3 = 33$ observed ticker–horizon cells. No ticker–horizon cell is imputed.

The equity universe is a purposively selected ex-post panel of 11 currently large U.S.-listed firms. It is not a point-in-time reconstruction of the historical large-cap universe. Accordingly, the empirical conclusions are restricted to the firms included in this panel and should not be generalized to all firms that qualified as large U.S. equities at different points during 2000–2022.

The source-data period spans 1 January 2000 through 31 December 2022. The effective modeling sample begins only after the construction of the 252-session Yang–Zhang initialization and subsequent lag construction. Parameter selection and stacking calibration further precede the common out-of-sample period. Consequently, effective model samples generally begin in 2001 for the long-history securities and later for ABBV and MA. The distinction between the source-data period and the effective estimation period is maintained throughout the analysis.

Daily open, high, low, close, adjusted close, and volume data are obtained from Yahoo Finance. The sample begins in January 2000 and ends in December 2022, subject to security-level data availability.

To capture market-wide uncertainty, we include the CBOE Volatility Index (VIX), which serves as a forward-looking proxy for aggregate option-implied volatility. In addition, we incorporate survey-based investor sentiment from the American Association of Individual Investors (AAII), including bullish, bearish, and neutral sentiment shares, as well as the bull–bear spread. All market-wide variables are transformed into lagged and smoothed predictors over one-day, five-day, and twenty-two-day windows, matching the forecasting horizons considered in the study.

The textual component is based on a large-scale New York Times article corpus obtained from the Kaggle dataset "NYT Articles 2.1M 2000–Present". The raw corpus contains article-level metadata, including publication timestamp, source URL, headline, abstract, snippet, news desk, section, subsection, and lead paragraph.

To ensure data quality and financial relevance, the textual corpus is processed in one ordered sequence before the forecasting features are formed. First, we retain observations with a valid article URL, a parseable publication timestamp, and non-empty textual content in at least one of the headline, abstract, snippet, or lead-paragraph fields; when available, articles are additionally required to contain at least 80 words and to belong to standard editorial formats (articles or blog posts). Second, we apply a finance-relevance screen based jointly on editorial metadata and textual content: an article is retained if it originates from a business-oriented desk or section (e.g., *Business*, *DealBook*, *Finance*, or *Markets*) or if its textual fields contain terms from the predefined financial lexicon covering accounting, macroeconomic, and capital-market terminology. Third, headline, abstract, lead paragraph, and snippet are normalized and concatenated with explicit delimiter tokens; boilerplate and formatting artifacts are removed and the sequence is truncated to the fixed model input length. Fourth, duplicate URLs and exact duplicate texts are checked separately and malformed or empty records are excluded. Fifth, a language diagnostic is performed on a sample of documents to assess non-English contamination. Sixth, each retained article is mapped to an eligible U.S. equity trading session using the timestamp rule defined once in Section 3.4; article-level measures are then aggregated at that assigned-session level. Finally, the daily textual predictors are lagged and transformed into the prespecified rolling features using only information available before the forecast origin.

The textual variables are constructed as market-wide information-flow measures rather than firm-specific news signals. The same daily textual signal is merged into each stock-level forecasting dataset. This design is intentional: the objective is to measure whether broad financial-news sentiment, directionality, posterior uncertainty, and news intensity contain incremental information for individual equity volatility beyond stock-specific volatility persistence and market-wide implied volatility. The resulting document-level corpus is therefore aggregated into market-wide predictors; the forecasting design does not construct firm-specific news features.

The sentiment classifier produces posterior probabilities for three classes: bullish, bearish, and neutral. These probabilities are used to construct five textual information measures. Directional sentiment is defined as the difference between bullish and bearish posterior probabilities. Directionality is defined as the posterior mass assigned to either directional class, irrespective of sign. Posterior uncertainty is measured separately using the normalized Shannon entropy of the three-class probability distribution. Sentiment intensity is the absolute value of directional sentiment, while news volume is the number of valid articles assigned to a trading session.

For article $a$, these quantities are defined as

$$sentiment_a = p_a^+ - p_a^-,$$

$$directionality_a = 1 - p_a^0,$$

$$text_entropy_a = -\frac{p_a^+ \ln p_a^+ + p_a^- \ln p_a^- + p_a^0 \ln p_a^0}{\ln 3},$$

and

$$intensity_a = |sentiment_a|.$$

The normalized entropy lies in the interval $[0, 1]$. Values close to zero indicate a concentrated classifier posterior, whereas values close to one indicate high posterior uncertainty across the three sentiment classes. News volume is the number of valid articles assigned to each trading session. The detailed UTC-to-Eastern-Time conversion and session-assignment rule is specified in Section 3.4; here, the relevant point is that daily textual variables are aggregated by assigned trading session rather than by raw calendar publication date.

Directionality and posterior entropy measure conceptually distinct properties. A strongly bullish or strongly bearish article can have high directionality but low posterior entropy, whereas an article for which the classifier assigns similar probabilities to all three classes has high posterior entropy.

The raw analyzed corpus contains 363,030 documents and 31 metadata fields. As reported in Table 4.1, the corpus contains approximately 18.18 million tokens and a vocabulary of 185,573 unique tokens. The average document length is 50.08 tokens, while the median length is 27 tokens, indicating a strongly right-skewed distribution. The 99th percentile of document length is 272 tokens, while the maximum observed length is 88,822 tokens. Approximately 7,291 documents, or 2.01%, are empty after preprocessing, and 3,622 observations, or 1.00%, lie above the 99th percentile of the length distribution.

Table 4.1: Summary statistics of the textual corpus

| Metric | Value |
|---|---:|
| Number of documents | 363,030 |
| Number of tokens | 18,180,668 |
| Vocabulary size | 185,573 |
| Number of sentiment classes | 3 |
| Label-distribution entropy (bits) | 1.264 |
| Lexical diversity (TTR) | 0.0103 |
| Average document length | 50.08 |
| Median document length | 27 |
| 99th percentile document length | 272 |
| Maximum document length | 88,822 |
| Empty documents | 7,291 |
| Empty documents (%) | 2.01 |
| Outliers above p99 | 3,622 |

The document-length distribution is highly skewed and exhibits a heavy right tail, which is characteristic of heterogeneous news corpora combining short news briefs with substantially longer feature articles. Figure 4.1 shows that most observations are concentrated between 10 and 50 tokens, while extremely long documents constitute a small but non-negligible subset of the sample. To improve robustness and reduce the influence of extreme outliers, subsequent diagnostics are reported both on the full sample and on distributions truncated at the 99th percentile. Figure 4.2 further confirms that the extreme right tail is sparse but very long.

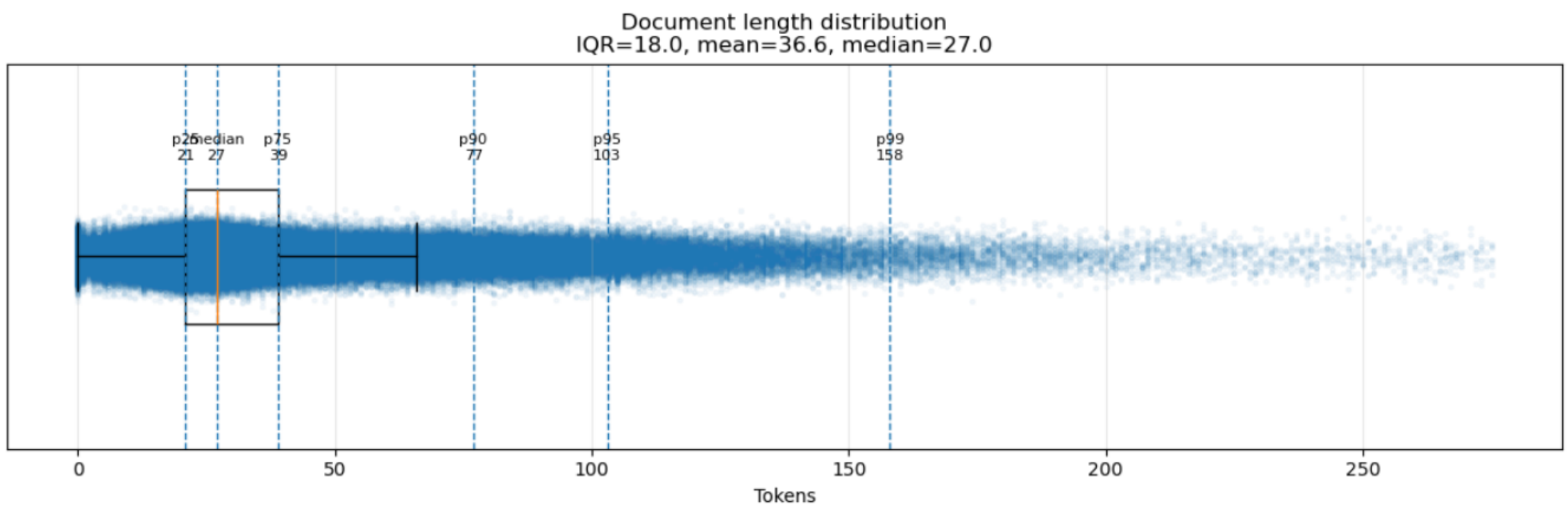

Figure 4.1: Document length distribution clipped at the 99th percentile and boxplot excluding values above the 99th percentile.

The sentiment-label distribution is imbalanced but economically plausible for general financial news. As shown in Table 4.2 and Figure 4.3, neutral articles dominate the corpus, accounting for 234,173 observations, or 64.51%. Bearish articles account for 86,012 observations, or 23.69%, while bullish articles account for 42,845 observations, or 11.80%. The class imbalance ratio is 5.47, and the entropy of the corpus-level label distribution is 1.264 bits. This corpus-level descriptive statistic is distinct from the normalized article-level posterior entropy used as a forecasting predictor.

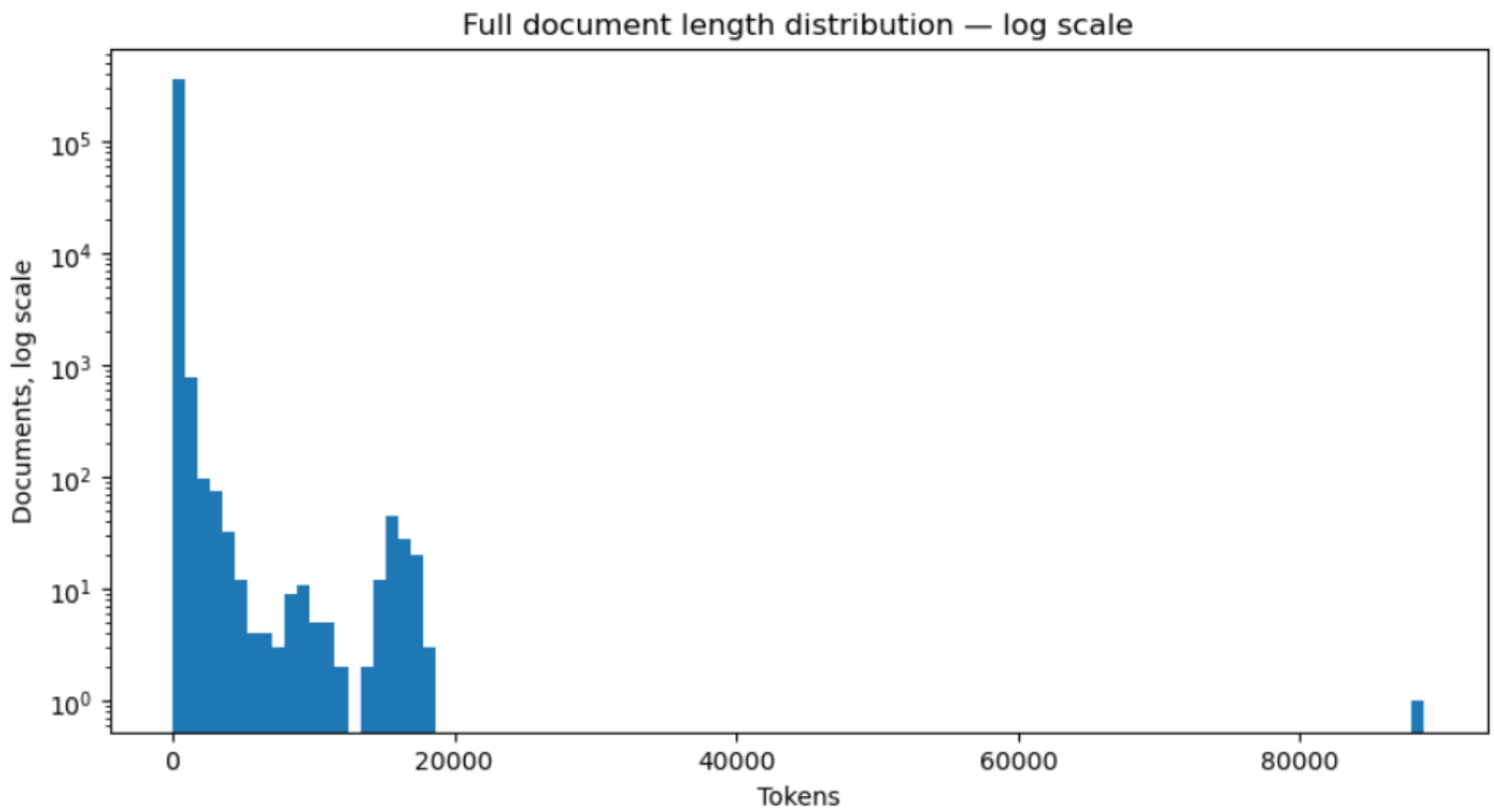

Figure 4.2: Outlier tail above the 99th percentile and full document length distribution on a logarithmic scale.

Table 4.2: Distribution of sentiment labels

| Sentiment class | Documents | Percentage (%) |
|---|---:|---:|
| Neutral | 234,173 | 64.51 |
| Bearish | 86,012 | 23.69 |
| Bullish | 42,845 | 11.80 |
| Total | 363,030 | 100.00 |

Document length also varies across sentiment classes. Table 4.3 reports length statistics by label, while Figure 4.4 visualizes the corresponding distributions truncated at the 99th percentile. Bearish articles have an average length of 42.83 tokens and a median length of 28 tokens. Bullish articles are slightly shorter, with an average length of 37.03 tokens and a median of 26 tokens. Neutral articles are substantially longer on average, with a mean of 55.13 tokens and a considerably heavier right tail. This pattern suggests that neutral articles are more likely to include descriptive or explanatory reporting, whereas strongly directional sentiment tends to appear in shorter and more event-focused news items.

The domain distribution confirms that the dataset is primarily sourced from New York Times domains. Figure 4.5 shows that the dominant source is `www.nytimes.com`, followed by NYT-affiliated subdomains such as `dealbook.nytimes.com`. This source concentration is intentional because the study focuses on professionally curated financial and economic news rather than heterogeneous user-generated content.

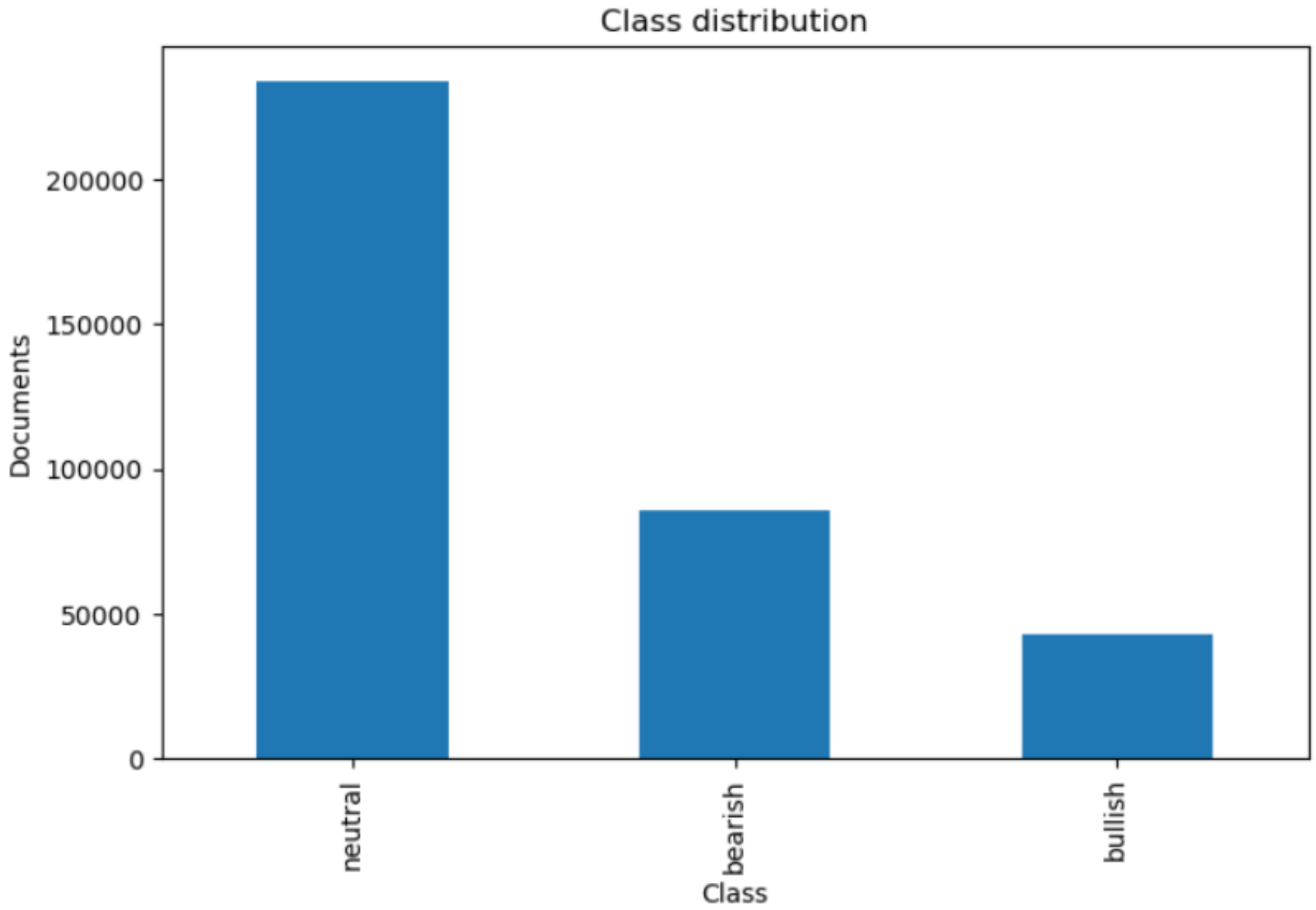

Figure 4.3: Distribution of sentiment labels in the analyzed textual corpus.

Table 4.3: Document length statistics by sentiment class

| Class | Mean | Std | Min | Median | P90 | P95 | P99 | Max |
|---|---|---|---|---|---|---|---|---|
| Bearish | 42.83 | 104.93 | 0 | 28 | 81 | 107 | 182 | 16,034 |
| Bullish | 37.03 | 80.48 | 0 | 26 | 68 | 93 | 141 | 10,037 |
| Neutral | 55.13 | 424.42 | 0 | 26 | 83 | 112 | 366 | 88,822 |

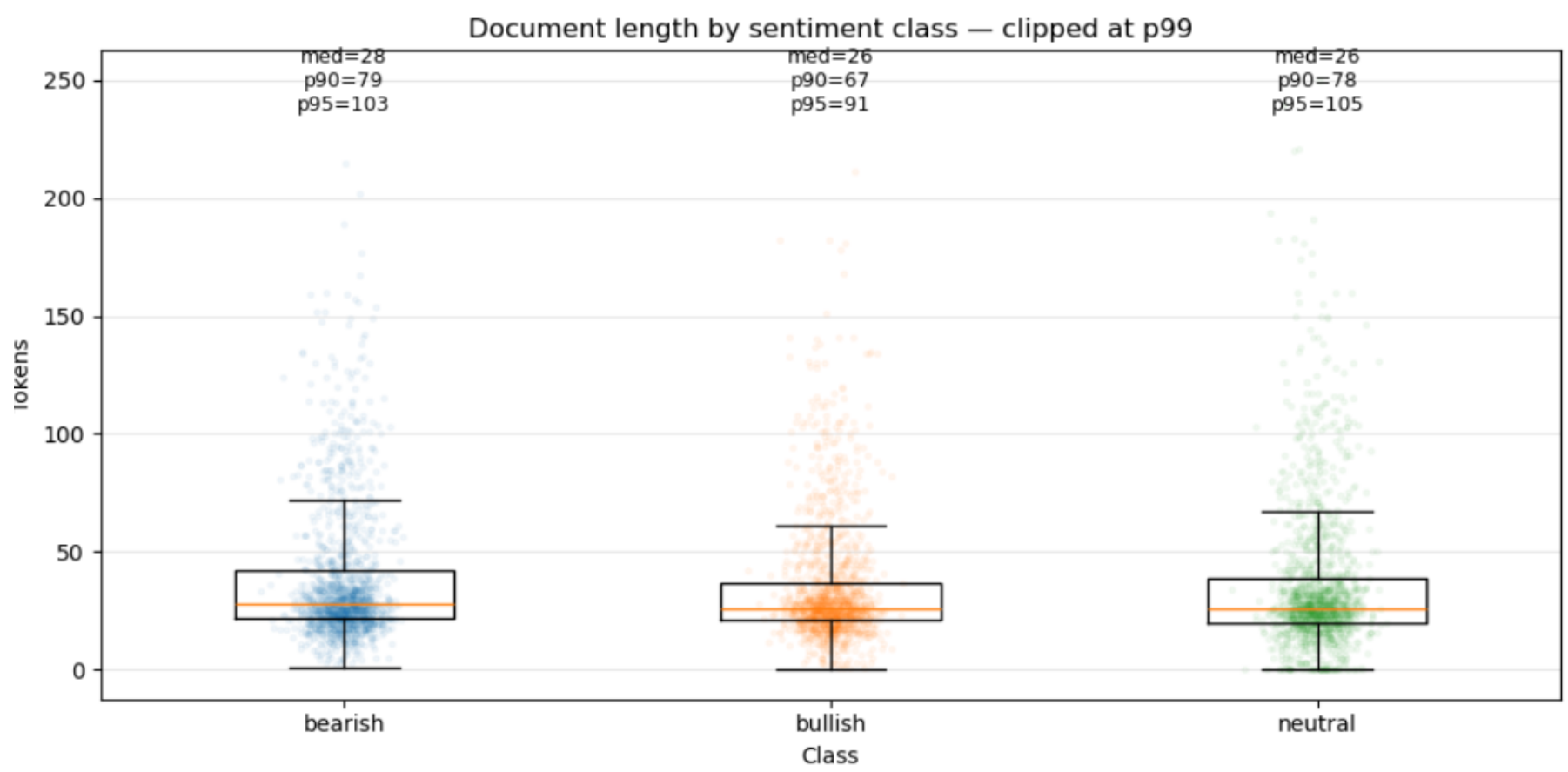

Figure 4.4: Document length by sentiment class, with values clipped at the 99th percentile.

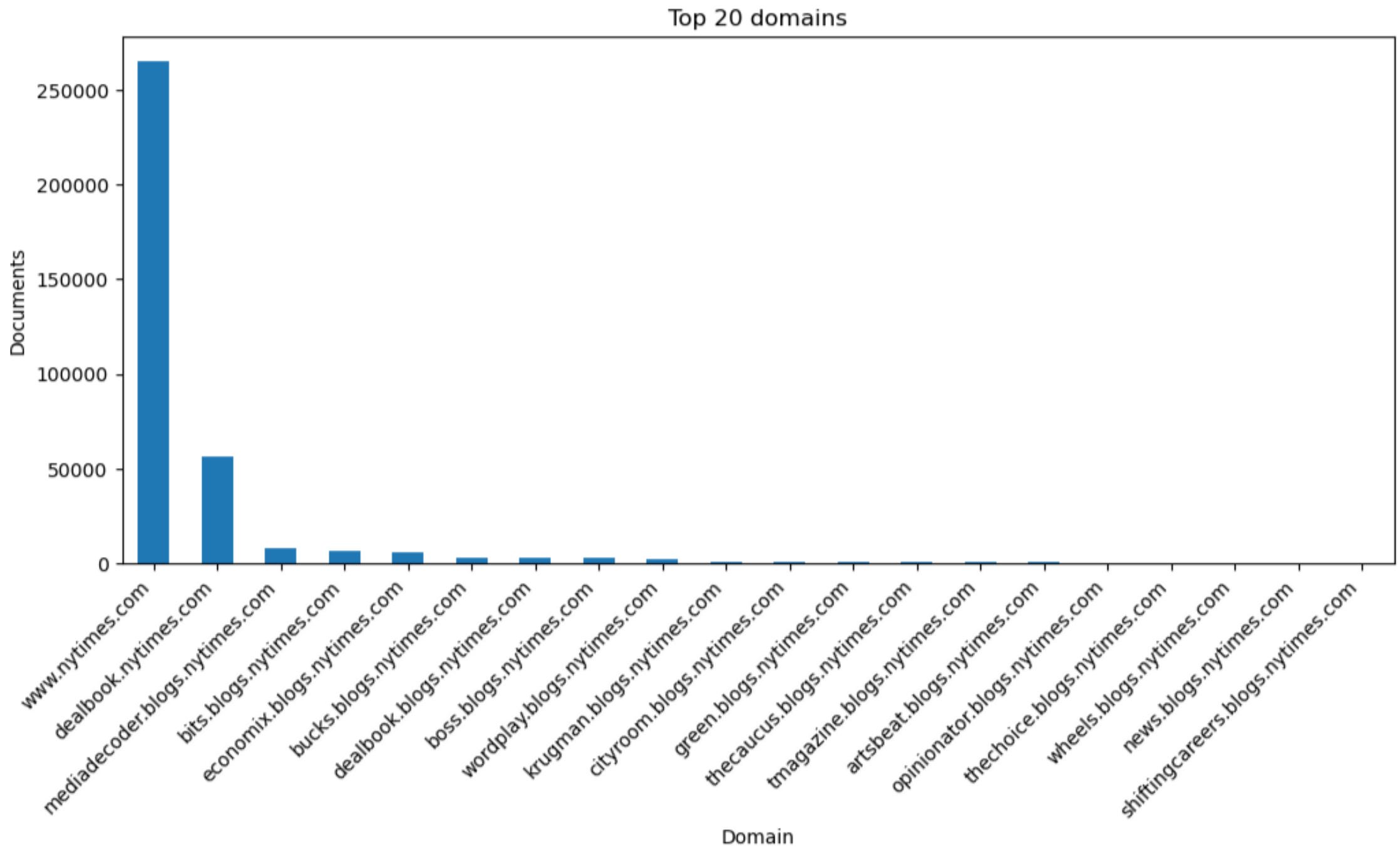

Figure 4.5: Top 20 source domains in the analyzed corpus.

For corpus-diagnostic purposes only, the documents were also divided into fixed 80/10/10 partitions. These partitions are not training, validation, and test sets in the usual supervised-learning sense: they are not used to train, fine-tune, select, or validate ModernFinBERT, and they do not enter the walk-forward volatility forecasts. The associated counts describe ModernFinBERT-predicted classes rather than manually verified labels. Partition-level class frequencies, duplicate diagnostics, vocabulary overlap, and out-of-vocabulary counts are reported in Appendix A.1.

Leakage prevention constitutes a central component. Exact duplicate diagnostics identify 15,333 duplicated texts, corresponding to 4.22% of the corpus, while duplicate URLs are not observed. Near-duplicate analysis using a similarity threshold of 0.90 detects only two highly similar document pairs within a diagnostic sample. Furthermore, overlap diagnostics based on Jaccard similarity indicate only moderate vocabulary overlap between splits. These diagnostics are summarized in Table 4.4.

Several metadata fields contain missing values, although these do not directly compromise the sentiment extraction task because the lead paragraph serves as the analyzed textual variable. Table 4.5 reports that missingness is highest for subsection name, print page, and print section, while the lead paragraph field exhibits only limited missingness.

For the downstream volatility forecasting exercise, random document splits are not used. Instead, all forecasting models are evaluated using walk-forward expanding and rolling temporal windows. This design ensures that each prediction is generated exclusively using information available prior to the forecast origin and prevents look-ahead bias. In addition, all lagged and rolling textual features are constructed strictly from past observations, eliminating temporal leakage between predictors and future volatility targets.

Table 4.4: Dataset overlap and deduplication diagnostics

| Metric | Value |
|---|---|
| Exact duplicate texts | 15,333 |
| Exact duplicate texts (%) | 4.22 |
| Exact duplicate URLs | 0 |
| Near-duplicate pairs (sample) | 2 |
| Near-duplicate threshold | 0.90 |
| 80%–Partition A Jaccard overlap | 0.310 |
| 80%–Partition B Jaccard overlap | 0.297 |
| Partition A–Partition B Jaccard overlap | 0.444 |
| Partition A OOV tokens vs. 80% partition | 11,869 |
| Partition B OOV tokens vs. 80% partition | 8,168 |

Table 4.5: Missing-value statistics for selected metadata fields

| Variable | Missing count | Missing percentage (%) |
|---|---|---|
| Subsection name | 272,701 | 75.12 |
| Print page | 131,743 | 36.29 |
| Print section | 131,509 | 36.23 |
| Snippet | 44,708 | 12.32 |
| News desk | 43,448 | 11.97 |
| Material type | 21,781 | 6.00 |
| Lead paragraph | 8,111 | 2.23 |
| Section name | 2 | 0.00 |

Overall, the resulting dataset is appropriate for the research question for three reasons. First, it aligns daily realized-volatility targets with information available before the forecast origin, enabling a realistic out-of-sample forecasting framework. Second, it combines classical volatility predictors with market-implied uncertainty, survey sentiment, and large-scale textual sentiment, thereby allowing us to compare the out-of-sample performance of models using news-based information with the available non-text benchmarks. Third, the long sample period spans multiple macroeconomic and financial regimes, enabling regime-aware analysis and improving the external validity of the empirical findings.

## 2.2 Blinded agreement audit of ModernFinBERT classifications

A blinded agreement audit was conducted on a simple random sample without replacement comprising $N = 2,700$ records. The purpose of this exercise was not to establish absolute classification accuracy against expert-annotated ground truth. Instead, it assessed the extent to which ModernFinBERT classifications agreed with an independently applied, automated semantic reference procedure that was blinded to the model outputs. Each record received exactly one reference label: bearish, neutral, or bullish.

The validation procedure was blinded to the ModernFinBERT outputs. During label assignment, only the content of the text field was visible. The model-generated bearish, neutral, and bullish probabilities and the model-assigned sentiment label were masked from the reference-label procedure.

The reference labels were generated using ChatGPT 5.6 Sol, accessed in August 2026. Because the interface did not expose a reproducibility seed or decoding parameters, exact deterministic reproduction of individual reference labels is not claimed. The annotation procedure received only the textual content of each sampled record. The ModernFinBERT probabilities, ModernFinBERT class label, publication date, subsequent market outcomes, and all remaining metadata were excluded from the annotation input.

The following instruction was used to assign the blinded validation labels:

> Assign exactly one label to the following financial text: bearish, neutral, or bullish.
>
> Use only the information contained in the text. Do not use probability scores, existing sentiment labels, dates, subsequent market outcomes, or any other metadata.
>
> Apply the following definitions:
>
> **Bearish:** deterioration in financial performance or economic conditions, losses, liquidity problems, declining activity, adverse legal or regulatory decisions, increasing risk, or other predominantly negative economic implications.
>
> **Bullish:** improvement in financial performance or economic conditions, growth, favorable transactions, stronger demand, value creation, or other predominantly positive economic implications.
>
> **Neutral:** descriptive content, materially mixed implications, or insufficiently clear directional economic implications.

> Return exactly one label and a short validation note explaining the dominant economic implication.

The procedure should nevertheless not be interpreted as a substitute for independent expert annotation. Its role is to provide a controlled consistency check and error-profile diagnostic rather than a definitive estimate of classification validity.

ModernFinBERT produced three class probabilities for each observation: $p_{\text{bearish}}$, $p_{\text{neutral}}$, and $p_{\text{bullish}}$. The predicted sentiment class was defined as the class associated with the highest of these three probabilities:

$$\widehat{y}_i = \underset{c\in\{\text{bearish},\text{neutral},\text{bullish}\}}{\arg\max} p_{i,c}, \tag{4.1}$$

where $p_{i,c}$ denotes the probability assigned by ModernFinBERT to class $c$ for observation $i$.

No minimum probability or confidence threshold was imposed. Consequently, every observation received exactly one ModernFinBERT prediction, irrespective of the absolute value of the highest class probability.

A consistency check confirmed that the recorded ModernFinBERT label corresponded to the class with the highest predicted probability for all 2,700 observations. No probability ties were observed.

The ModernFinBERT predictions were compared with the blinded validation labels using overall accuracy, balanced accuracy, class-specific precision, recall, and $F_1$ score, macro-averaged $F_1$, weighted $F_1$, Cohen's $\kappa$, the multiclass Brier score, and multiclass log loss.

Overall accuracy was calculated as the proportion of observations for which the ModernFinBERT prediction matched the validation label. Balanced accuracy was calculated as the unweighted mean of the class-specific recall values. Macro-averaged $F_1$ was calculated as the unweighted mean of the three class-specific $F_1$ scores.

Cohen's $\kappa$ was used to quantify agreement beyond the level expected from the marginal class distributions. The multiclass log loss was calculated as

$$\text{LogLoss} = -\frac{1}{N}\sum_{i=1}^{N}\sum_{c=1}^{C} y_{i,c}\log\left(p_{i,c}\right). \tag{4.2}$$

For the supplementary confidence analysis, confidence was defined as the largest class probability:

$$\text{Confidence}_i = \max\left(p_{i,\text{bearish}}, p_{i,\text{neutral}}, p_{i,\text{bullish}}\right). \tag{4.3}$$

Confidence was used only as a diagnostic variable. It was not used to exclude observations or modify the predicted class.

The largest class in the blinded validation sample was neutral, with 1698 observations, corresponding to 62.9% of the sample. The bearish class contained 578 observations, or 21.4%, while the bullish class contained 424 observations, or 15.7%.

ModernFinBERT predicted 1661 neutral observations, corresponding to 61.5% of the sample, 691 bearish observations, corresponding to 25.6%, and 348 bullish observations, corresponding to 12.9%.

Relative to the blinded validation labels, ModernFinBERT therefore assigned the bearish class more frequently and the bullish class less frequently. Table 4.6 reports the complete class counts and shares.

Table 4.6: Distribution of blinded validation labels and ModernFinBERT predictions.

| | Blinded validation | | ModernFinBERT | |
|---|---|---|---|---|
| Class | $N$ | Share | $N$ | Share |
| Bearish | 578 | 21.4% | 691 | 25.6% |
| Neutral | 1698 | 62.9% | 1661 | 61.5% |
| Bullish | 424 | 15.7% | 348 | 12.9% |
| Total | 2,700 | 100.0% | 2,700 | 100.0% |

ModernFinBERT assigned the same class as the blinded semantic reference procedure in 1558 of the 2,700 observations. The resulting overall agreement rate was 0.577.

Balanced accuracy was 0.484, indicating that classification performance was not uniform across the three classes. Macro F1 was 0.481, whereas weighted F1 was 0.577. The difference between these two measures reflects the class imbalance and the substantially stronger performance of the model for the dominant neutral class.

Cohen's coefficient of agreement was $\kappa = 0.214$, indicating limited agreement beyond the level expected from the marginal class distributions. This low level of agreement constitutes weak corroborative evidence for the semantic validity of the generated labels and should not be interpreted as strong validation of the classifier for the analyzed corpus.

The multiclass Brier score was 0.656, while multiclass log loss was 1.331. These probability-based measures indicate that classification errors were accompanied by imperfect probability estimates. Table 4.7 summarizes the overall agreement and probability-based metrics.

The strongest performance was observed for the neutral class. Precision for neutral predictions was 0.697, recall was 0.682, and F1 was 0.689.

Performance was weaker for the bearish class. Bearish precision was 0.402, recall was 0.481, and F1 was 0.438.

The weakest results were obtained for the bullish class. Bullish precision was 0.351, recall was 0.288, and F1 was 0.316. Thus, ModernFinBERT correctly identified fewer than one third of the observations classified as bullish by the blinded validation procedure. Table 4.8 reports all class-specific metrics.

Table 4.7: Overall ModernFinBERT classification performance relative to the blinded validation labels.

| Metric | Value |
|---|---|
| Number of observations | 2,700 |
| Correct classifications | 1558 |
| Accuracy | 0.577 |
| Balanced accuracy | 0.484 |
| Macro $F_1$ | 0.481 |
| Weighted $F_1$ | 0.577 |
| Cohen's $\kappa$ | 0.214 |
| Multiclass Brier score | 0.656 |
| Multiclass log loss | 1.331 |

Table 4.8: Class-specific ModernFinBERT classification metrics.

| Class | Precision | Recall | $F_1$ | Support |
|---|---|---|---|---|
| Bearish | 0.402 | 0.481 | 0.438 | 578 |
| Neutral | 0.697 | 0.682 | 0.689 | 1698 |
| Bullish | 0.351 | 0.288 | 0.316 | 424 |
| Macro average | 0.484 | 0.484 | 0.481 | 2,700 |
| Weighted average | 0.578 | 0.577 | 0.577 | 2,700 |

The confusion matrix shows that most classification errors occurred between a directional sentiment class and the neutral class.

Among the 578 observations classified as bearish in the blinded validation procedure, ModernFinBERT correctly classified 278 as bearish, classified 265 as neutral, and classified 35 as bullish.

Among the 1698 validation-neutral observations, the model correctly classified 1158 as neutral, while 349 were classified as bearish and 191 as bullish.

Among the 424 observations classified as bullish in the blinded validation procedure, ModernFinBERT correctly classified 122 as bullish, classified 238 as neutral, and classified 64 as bearish. The complete confusion matrix is reported in Table 4.9.

Table 4.9: Confusion matrix for ModernFinBERT relative to the blinded validation labels. Rows represent validation labels and columns represent ModernFinBERT predictions.

| | ModernFinBERT prediction | | |
|---|---|---|---|
| Validation label | Bearish | Neutral | Bullish |
| Bearish | 278 | 265 | 35 |
| Neutral | 349 | 1158 | 191 |
| Bullish | 64 | 238 | 122 |

The most frequent error for both directional classes was classification as neutral. This pattern suggests that ModernFinBERT frequently reduced the strength of directional economic signals by assigning them to the neutral category.

Direct confusion between the two opposing sentiment classes was less frequent. Thirty-five bearish observations were classified as bullish, while 64 bullish observations were classified as bearish. Together, these opposite-direction errors accounted for 99 observations, corresponding to 3.7% of the validation sample.

These results do not establish strong absolute classification validity. Instead, they reveal material disagreement between the transformer-based classifier and the independent semantic reference procedure, particularly for directional classes. The audit therefore performs a diagnostic rather than a confirmatory function: it identifies where the text-derived measurements are most and least reliable and constrains the interpretation of the subsequent forecasting results.

The weak identification of bullish observations is especially important because it implies that the positive-sentiment component may be measured with greater error than the neutral or bearish components. Any weak forecasting performance of directional sentiment variables may therefore reflect both a genuine absence of predictive content and attenuation caused by classification noise.

## 2.3 Performance by Model Confidence

Classification performance improved as the maximum predicted probability increased. Accuracy was 0.449 for observations with confidence below 0.60 and increased to 0.655 for observations with confidence of at least 0.90.

Balanced accuracy increased from 0.412 in the lowest-confidence group to 0.536 in the highest-confidence group. Macro F1 similarly increased from 0.383 to 0.536.

The highest-confidence interval contained 1387 observations, corresponding to 51.4% of the sample. Despite the stronger relative performance in this subgroup, approximately 34.5% of predictions with confidence of at least 0.90 remained inconsistent with the blinded validation labels.

High confidence should therefore not be interpreted as a guarantee of correct classification. Nevertheless, the monotonic relationship between confidence and performance indicates that the maximum class probability contains useful information about prediction reliability. Table 4.10 reports the confidence-bin results.

The results indicate that ModernFinBERT was substantially more effective at identifying neutral content than directional financial sentiment. The difference between overall accuracy and balanced accuracy demonstrates that aggregate performance was strongly influenced by the dominant neutral class.

The relatively low recall values for the bearish and bullish classes suggest that ModernFinBERT frequently classified directional economic information as neutral. This effect was particularly pronounced for the bullish class, for which only 28.8% of the blinded validation observations were correctly identified.

Table 4.10: ModernFinBERT classification performance by maximum predicted class probability.

| Confidence | $N$ | Share | Mean confidence | Accuracy | Balanced accuracy | Macro $F_1$ |
|---|---|---|---|---|---|---|
| $< 0.60$ | 354 | 13.1% | 0.537 | 0.449 | 0.412 | 0.383 |
| 0.60–0.75 | 389 | 14.4% | 0.678 | 0.470 | 0.417 | 0.405 |
| 0.75–0.90 | 570 | 21.1% | 0.835 | 0.540 | 0.445 | 0.452 |
| $\geq 0.90$ | 1387 | 51.4% | 0.967 | 0.655 | 0.536 | 0.536 |

The model also displayed an asymmetry between positive and negative sentiment. Relative to the blinded validation labels, ModernFinBERT generated more bearish predictions and fewer bullish predictions. This pattern may reflect a comparatively conservative tendency toward negative classification or a greater sensitivity to negative financial language.

The low bullish F1 score is especially relevant for empirical applications in which positive and negative financial information is expected to produce distinct effects. In such applications, the use of the unadjusted three-class prediction may lead to an underrepresentation of positive economic signals.

The difference between macro F1 and weighted F1 further demonstrates that overall performance measures may overstate model quality for the minority classes. Accuracy and weighted F1 should therefore be interpreted together with balanced accuracy and class-specific precision, recall, and F1 scores.

The confidence analysis indicates that the maximum predicted probability may be used as a diagnostic measure of prediction reliability. However, because every class was assigned using the maximum-probability rule, observations with similar class probabilities still received a definitive label. Low-confidence predictions may therefore represent ambiguous cases rather than strongly directional sentiment signals.

For empirical applications, ModernFinBERT predictions should be interpreted as model-based sentiment indicators rather than error-free labels. The findings support the use of robustness checks, particularly when bearish and bullish observations are analyzed separately.

Possible robustness procedures include estimating models on higher-confidence subsets, using the continuous class probabilities instead of only the discrete label, combining directional classes in alternative specifications, calibrating the probability outputs on an independent sample, or validating an additional subsample through expert human annotation.

Because the predicted class was selected using the maximum-probability rule, no observations were removed from the primary analysis. The confidence analysis should therefore be treated as supplementary evidence concerning reliability rather than as a sample-selection procedure.

The interpretation of the results is subject to several limitations. First, the validation sample was imbalanced, with neutral observations accounting for almost two thirds of the records. This imbalance affected the aggregate measures and increased the contribution of the neutral class to weighted performance metrics.

Second, the neutral class was internally heterogeneous. It included purely descriptive texts, texts with materially mixed positive and negative implications, and texts for which the economic direction was insufficiently clear. This heterogeneity may have increased disagreement at the boundary between neutral and directional sentiment.

Third, the reference labels were produced using the automated annotation procedure documented in Section 2.2. Because this procedure does not constitute independent expert annotation, the reported metrics measure agreement with that specified reference procedure rather than absolute accuracy relative to an undisputed ground truth.

Fourth, the Brier score and log loss depend not only on classification accuracy but also on the calibration of the ModernFinBERT probability estimates. Further evaluation using an independent sample, multiple human annotators, and formal probability-calibration methods would be necessary to assess absolute classification validity and calibration quality.

In the blinded validation sample of 2,700 observations, ModernFinBERT achieved an accuracy of 0.577, balanced accuracy of 0.484, and macro F1 of 0.481. Performance was strongest for neutral observations and weakest for bullish observations.

The model frequently classified directional financial content as neutral and produced more bearish and fewer bullish predictions than the blinded validation procedure. Higher model confidence was associated with improved classification performance, although substantial disagreement remained even among the highest-confidence predictions.

Overall, the findings indicate that ModernFinBERT provides informative but imperfect financial sentiment classifications. Its outputs appear most reliable for neutral content and should be interpreted cautiously when distinguishing between positive and negative economic implications.

**Data provenance and analytical scope.** Forecasting uses the reconciled market exports and the session-level text aggregates available in the study archive. The ModernFinBERT measurement layer, document-retrieval history, and corpus-diagnostic quantities are treated as fixed inputs and are not independently regenerated. Corpus summaries therefore describe their stated source populations, whereas the common-case modeling and OOS samples are reported separately in Appendix A.1. The computational claims of this chapter are limited to the forecasting and evaluation stages that can be reproduced from these retained inputs.

# 3 Methodology

The analysis implements the target and lagged information sets defined below, with one initial hyperparameter search followed by frozen-setting refits. All reported results use this single YZ-252 level-target protocol.

## 3.1 Empirical evaluation of the hypotheses

H4.1 is evaluated first through the architecture-matched HAR-Text versus HAR comparison. H4.2 compares the recurrence of gains from text, VIX, and AAII under the same linear estimator and evaluation dates. Regularized and tree-based models provide additional functional-form comparisons that enter the fixed 792-test family but do not by themselves define any hypothesis verdict; Elastic Net is linear, not nonlinear. H4.3 concerns heterogeneity across stocks, horizons, folds, and the separately reported rolling-window arm. H4.4 concerns ex-post low- and high-volatility strata and cannot establish a real-time regime-timing result. QLIKE is the primary criterion throughout; the statistical comparison hierarchy is specified in Section 3.10.

## 3.2 Forecasting target

The volatility target is constructed from daily open, high, low, and close prices using the standard multi-period Yang–Zhang volatility estimator. For stock $i$ on trading day $s$, let $O_{i,s}$, $H_{i,s}$, $L_{i,s}$, and $C_{i,s}$ denote the open, high, low, and close prices, respectively. The overnight and open-to-close log-return components are defined as

$$o_{i,s} = \log\left(\frac{O_{i,s}}{C_{i,s-1}}\right), \qquad c_{i,s} = \log\left(\frac{C_{i,s}}{O_{i,s}}\right). \tag{4.4}$$

The corresponding high-to-open and low-to-open log-price ranges are

$$u_{i,s} = \log\left(\frac{H_{i,s}}{O_{i,s}}\right), \qquad d_{i,s} = \log\left(\frac{L_{i,s}}{O_{i,s}}\right). \tag{4.5}$$

At each date $t$, the Yang–Zhang estimator is calculated over a trailing window of $N$ trading sessions, indexed by $s = t - N + 1, \ldots, t$. The window-specific means of the overnight and open-to-close returns are

$$\bar{o}_{i,t}^{(N)} = \frac{1}{N} \sum_{s=t-N+1}^{t} o_{i,s}, \qquad \bar{c}_{i,t}^{(N)} = \frac{1}{N} \sum_{s=t-N+1}^{t} c_{i,s}. \tag{4.6}$$

The overnight variance component is

$$\sigma_{o,i,t}^{2(N)} = \frac{1}{N-1} \sum_{s=t-N+1}^{t} \left(o_{i,s} - \bar{o}_{i,t}^{(N)}\right)^2, \tag{4.7}$$

and the open-to-close variance component is

$$\sigma_{c,i,t}^{2(N)} = \frac{1}{N-1} \sum_{s=t-N+1}^{t} \left(c_{i,s} - \bar{c}_{i,t}^{(N)}\right)^2. \tag{4.8}$$

The Rogers–Satchell component is calculated as

$$\sigma^{2(N)}_{RS,i,t} = \frac{1}{N} \sum_{s=t-N+1}^{t} \left[ u_{i,s} \left( u_{i,s} - c_{i,s} \right) + d_{i,s} \left( d_{i,s} - c_{i,s} \right) \right]. \tag{4.9}$$

The Yang–Zhang variance estimate is then

$$rv^{YZ}_{i,t} = \sigma^{2(N)}_{o,i,t} + k_N \sigma^{2(N)}_{c,i,t} + \left(1 - k_N\right) \sigma^{2(N)}_{RS,i,t}, \tag{4.10}$$

where

$$k_N = \frac{0.34}{1.34 + \frac{N+1}{N-1}}. \tag{4.11}$$

In the empirical implementation, the estimator is calculated using a trailing window of

$$N = 252 \tag{4.12}$$

trading sessions. This choice corresponds to approximately one trading year. The Yang–Zhang measure is therefore a multi-period OHLC estimator rather than a single-day quadratic decomposition. At each date, its overnight, open-to-close, and Rogers–Satchell components are re-estimated using only the observations contained in the corresponding trailing window.

For each forecasting horizon $h \in \{1, 5, 22\}$, the forward-looking volatility target is defined as the annualized square root of the average future Yang–Zhang variance:

$$y^{(h)}_{i,t} = \sqrt{252 \cdot \frac{1}{h} \sum_{j=1}^{h} rv^{YZ}_{i,t+j}}. \tag{4.13}$$

The forecasting targets are the annualized one-day, five-day, and twenty-two-day forward Yang–Zhang volatility measures, respectively. The five- and twenty-two-day targets are averages of overlapping future variance estimates. This overlap induces serial dependence in forecast errors and loss differentials and is taken into account in the purging and statistical inference procedures described below.

A target is retained only when all $h$ future variance estimates are available. No terminal missing value is silently omitted from the mean. The target is a forward average of trailing 252-session estimates, not the variance of only the next $h$ returns. The known historical part of these overlapping estimates makes this a highly persistent, smoothed forecast object. All models fit its annualized-volatility level.

The original indexing convention is retained: the average runs from $t+1$ through $t+h$, whereas predictors end at $t-1$. Thus, at $h = 1$, the target estimator is dated two sessions after the latest volatility input. The nominal horizon is measured from origin $t$, not from the date of that lagged input.

Predictors are lagged before rolling transformations and use observations through $t-1$. Training labels must end before the next evaluation origin. Purging enforces that availability condition but does not remove shared historical observations from the 252-session target windows. This distinction motivates the longer-bandwidth inference sensitivity analysis.

### 3.3 Textual sentiment aggregation

We use ModernFinBERT as an external pre-trained financial sentiment classifier and do not fine-tune it within the volatility forecasting pipeline. The model is based on ModernBERT and is designed for financial sentiment classification across heterogeneous financial text domains, including news, social media, crypto-related text, stock-market commentary, and macroeconomic content. We therefore treat its outputs as externally generated, model-based sentiment measurements rather than as labels estimated within the present forecasting exercise. They are not regarded as historically point-in-time measurements or as semantically validated labels for the New York Times corpus. This limitation is upstream of the downstream forecasting design. Chronological splits, lagged predictors, horizon-specific purging, and training-window-only preprocessing prevent the forecasting models from using future observations contained in the analytical panel. They cannot, however, rule out the possibility that the fixed ModernFinBERT weights encode linguistic or economic information originating after a historical forecast date or from documents overlapping the 2000–2022 evaluation corpus.

Consequently, the text-based results should be interpreted as an ex-post measurement experiment: they assess whether features produced today by a fixed language model are associated with subsequent historical volatility outcomes. They do not establish the performance of a text-classification system that was historically available and trained only on information preceding each forecast origin. Their semantic validity is assessed only through the limited agreement audit described above and should not be regarded as established against expert-annotated ground truth.

**ModernFinBERT model provenance and temporal boundary.** The textual variables were constructed using the publicly available `tabularisai/ModernFinBERT` model distributed through the Hugging Face model repository. ModernFinBERT is a three-class financial sentiment classifier based on the ModernBERT architecture and fine-tuned using a combination of real and synthetic financial data. According to the model card, the training material covers several financial domains, including financial news, social-media content, cryptocurrencies, and macroeconomic text.

The exact temporal coverage of the pre-training and fine-tuning corpora is not disclosed by the model developers. In particular, the model card does not report a verified training-data cutoff date or provide a complete inventory of the documents included in the training corpus. The earliest independently verifiable date associated with the exact model weights used in this study is July 4, 2025.

Consequently, July 4, 2025 should be interpreted as the confirmed model release date rather than as the training-data cutoff. It cannot be established from the available model documentation that the model was trained exclusively on information predating the end of the empirical news sample. The model is therefore used in this study as a fixed ex-post textual measurement instrument rather than as a language model that would have been available to an investor in real time during 2000–2022.

Table 4.11 reports the benchmark performance disclosed in the public ModernFinBERT model card. The reported results indicate that ModernFinBERT performs strongly across standard financial-sentiment benchmarks, with an average ROC-AUC of 0.91 across the three datasets reported by the model provider. These metrics provide external validation for using the model as a sentiment-extraction layer in the subsequent volatility forecasting pipeline.

Table 4.11: External benchmark performance of ModernFinBERT

| Dataset | Accuracy | F1-score | Precision | Recall | ROC-AUC |
|---|---|---|---|---|---|
| FIQA | 0.80 | 0.61 | 0.64 | 0.88 | 0.96 |
| Twitter | 0.71 | 0.70 | 0.68 | 0.81 | 0.94 |
| JeanBaptiste | 0.74 | 0.58 | 0.71 | 0.56 | 0.84 |
| Average | 0.75 | 0.63 | 0.68 | 0.75 | 0.91 |

Notes: The table reports benchmark metrics from the public ModernFinBERT model card. ModernFinBERT is used in this study as a pre-trained financial sentiment classifier. The model is not fine-tuned on the New York Times corpus within the volatility forecasting pipeline. Source: https://huggingface.co/tabularisai/ModernFinBERT.

The study does not observe the complete pre-training corpus or a verified training-data cut-off for the external classifier. It therefore does not claim that every document used to train ModernFinBERT predates every observation in the volatility-forecasting sample. The empirical exercise evaluates the usefulness of the supplied classifier outputs rather than the historical purity of the upstream language-model training process.

The textual information set is obtained from a large-scale New York Times corpus processed with ModernFinBERT. ModernFinBERT produces article-level probabilities for three sentiment classes: bullish, bearish, and neutral. The volatility forecasting treats these probabilities as external, precomputed sentiment signals and does not retrain the sentiment classifier as part of the volatility forecasting exercise.

For article $a$, let $p_a^+$, $p_a^-$, and $p_a^0$ denote the ModernFinBERT posterior probabilities of bullish, bearish, and neutral sentiment, respectively. The article-level directional sentiment, directionality, posterior entropy, and sentiment intensity measures are defined as:

$$sentiment_a = p_a^+ - p_a^-, \tag{4.14}$$

$$directionality_a = 1 - p_a^0, \tag{4.15}$$

$$text_entropy_a = -\frac{p_a^+ \ln p_a^+ + p_a^- \ln p_a^- + p_a^0 \ln p_a^0}{\ln 3}, \tag{4.16}$$

$$intensity_a = |sentiment_a|. \tag{4.17}$$

The directionality measure captures the posterior probability assigned to the two directional classes jointly and should not be interpreted as uncertainty. Posterior uncertainty is instead represented by $text_entropy_a$, which is normalized by $\ln 3$ and therefore ranges from zero to one.

Article-level textual measures are aggregated by assigned trading session rather than by raw publication date. Let $\mathcal{A}_t$ denote the set of articles assigned to trading session $t$ under the timestamp-to-session rule defined in Section 3.4.

$$directionality_t = \frac{1}{N_t} \sum_{a \in \mathcal{A}_t} directionality_a, \tag{4.18}$$

$$text_entropy_t = \frac{1}{N_t} \sum_{a \in \mathcal{A}_t} text_entropy_a. \tag{4.19}$$

$$intensity_t = \frac{1}{N_t} \sum_{a \in \mathcal{A}_t} intensity_a, \tag{4.20}$$

where $\mathcal{A}_t$ denotes the set of articles assigned to trading session $t$ after timestamp conversion and session mapping, and $N_t = |\mathcal{A}_t|$. Daily news volume is defined as

$$news_volume_t = N_t. \tag{4.21}$$

Trading sessions without valid assigned articles are assigned zero values for sentiment, directionality, posterior entropy, intensity, and news volume. Because $news_volume_t = 0$ identifies such sessions, a zero value of $text_entropy_t$ on a no-news day should be interpreted as the absence of textual observations rather than as evidence of classifier certainty. The resulting daily textual variables are merged into the stock-level trading-day dataset.

The NYT-based textual features are interpreted as market-wide information proxies rather than firm-specific news measures. The same daily textual signal is therefore merged into each stock-level dataset. This design captures broad news sentiment, directionality, posterior uncertainty, and attention conditions that may affect equity volatility through market-wide information channels.

### 3.4 Anti-leakage protocol and temporal alignment

The pipeline is designed to prevent look-ahead bias in the construction of forecasting features. Stock-level volatility, trading volume, VIX, AAII variables, and text-based variables are transformed into lagged and rolling predictors using only past information. For a series $z_t$, the available lag operators are

$$z_{t-1}, \qquad z_{t-5}, \qquad z_{t-22}, \tag{4.22}$$

and

$$\overline{z}_{t,5} = \frac{1}{5}\sum_{j=1}^{5} z_{t-j}, \qquad \overline{z}_{t,22} = \frac{1}{22}\sum_{j=1}^{22} z_{t-j}. \tag{4.23}$$

The BASE block uses all five transformations above. The VIX, AAII, and TEXT blocks use only lag one and the two rolling means, as specified in Section 3.5. Thus, the generic operators do not imply that every block contains five features per series. In the implementation, each predictor series is first lagged by one trading day before rolling averages are computed. Hence, the five-day and twenty-two-day rolling features at time $t$ exclude contemporaneous and future values. Daily VIX observations are treated as end-of-day variables and are shifted by one trading day before entering any forecast. Thus, even if the recorded VIX close reflects information through 16:15 Eastern Time, whereas the regular equity session closes at 16:00, the day-$t$ VIX value is first used for a forecast formed on day $t+1$. No same-day equity forecast uses the contemporaneous VIX close.

The session assignment determines the date on which information first enters the daily textual series; it does not imply contemporaneous use in the forecasting equation. As described above, the forecasting predictors are subsequently lagged before model estimation. Thus, the assigned-session textual signal enters a forecast only through the prespecified lagged or trailing features, preventing information published during session $t$ from being used retrospectively to forecast an outcome already realized within that session.

To preserve the chronological availability of textual information without discarding off-session news, each article is mapped explicitly from its publication timestamp to a U.S. equity trading session. Publication timestamps are converted from UTC to `America/New_York` time and matched to the U.S. trading calendar. An article published on a trading day at or before 16:00 ET is assigned to that same trading session. This rule includes overnight and pre-open publications on a valid trading day because the information is available before the close of that session. An article published after 16:00 ET is assigned to the next trading session. Articles published on weekends, exchange holidays, or other non-trading days are also assigned to the next available trading session.

This convention ensures that after-close and non-trading-day information is retained without being attributed to a session that had already ended. Consequently, a Friday post-close article enters the textual signal assigned to the next trading session, normally Monday, whereas a Monday pre-open article is assigned to Monday. The procedure therefore preserves off-session news while maintaining a chronological mapping between publication availability and the subsequent trading opportunity.

The timestamp audit covers 359,908 articles within the 2000–2022 study period. Of these, 233,808 publications occur on a trading day at or before 16:00 ET and are assigned to that session. The remaining 126,100 publications fall into off-session timing categories: 57,057 are published after the market close, 63,455 on weekends, and 5,588 on exchange holidays or other non-trading days. A total of 359,897 articles can be mapped to a trading session within the study period. Eleven publications occurring after the final available trading session of 2022 have no subsequent in-sample session and therefore remain unassigned. Table 4.12 summarizes the timing categories and assignment rules.

Table 4.12: Audit of article-to-trading-session assignment

| Publication timing category | Articles | Assignment rule |
|---|---|---|
| Trading day, at or before 16:00 ET | 233,808 | Same trading session |
| After 16:00 ET | 57,057 | Next trading session |
| Weekend | 63,455 | Next trading session |
| Exchange holiday / non-trading day | 5,588 | Next trading session |
| Study-period articles | 359,908 | – |
| Successfully assigned | 359,897 | – |
| After final 2022 session, no in-sample successor | 11 | Unassigned |

The AAII source is the supplied chronological survey series. Its recorded dates are used for a backward as-of merge; no date is filled from a future release. The derived daily predictors are subsequently shifted by one trading day, like all other blocks. This avoids the day/month reversals in the alternative XLSX export without applying an undocumented date repair. Missing survey entries are not replaced by zero. Historical intraday release times are not independently verified, so point-in-time claims are limited to the supplied date field and the stated lag convention.

The last $h$ eligible origins before a validation or test block are purged before fitting or scaling. This makes the latest training target end before the next evaluation origin. The 252-session estimator nevertheless shares historical OHLC observations across adjacent labels. Such overlap is not in itself future-information leakage, but it can induce long dependence in the loss differential. Both the horizon-specific and longer-bandwidth analyses are therefore retained.

## 3.5 Feature sets

The HAR-style baseline includes both volatility persistence and trading-volume controls. Let $v_{i,t} = \sqrt{252 rv_{i,t}^{YZ}}$ and $w_{i,t} = \log(1 + \text{Volume}_{i,t})$. Define

$$\overline{z}_{t,k} = \frac{1}{k}\sum_{j=1}^{k} z_{t-j}, \qquad B(z)_t = [z_{t-1}, z_{t-5}, z_{t-22}, \overline{z}_{t,5}, \overline{z}_{t,22}].$$

The baseline is $\mathbf{x}_{i,t}^{BASE} = [B(v_i)_t, B(w_i)_t]$, with ten predictors. Thus, HAR here denotes an augmented HAR-style benchmark, not the three-regressor canonical HAR-RV. The volume transformation is the log trading volume used in the source pipeline.

For each additional daily series define $A(z)_t = [z_{t-1}, \overline{z}_{t,5}, \overline{z}_{t,22}]$. The VIX block is $A(\text{VIX})_t$, using the VIX level and containing three predictors. The AAII block applies $A(\cdot)$ to bullish, bearish, neutral, and bull-minus-bear survey shares, yielding twelve predictors. The TEXT block applies the same transformation to directional sentiment, directionality, posterior entropy, intensity, and news volume, yielding fifteen predictors. The full input block used by HAR-Full and XGB-Full concatenates BASE, VIX, AAII, and TEXT and contains forty predictors. Every rolling mean ends at $t-1$; no contemporaneous VIX close or same-session news aggregate enters the forecast at $t$.

AAII shares and their spread are algebraically dependent. This does not alter OLS fitted values when the design is handled by a pseudoinverse, but individual coefficients and feature attributions should not be interpreted as uniquely identified structural effects. The information-channel comparisons concern the entire block. Market-implied VIX, survey-based AAII, and transformer-derived text remain conceptually distinct information sources.

### 3.6 Forecasting models

The following eleven specifications form the single model universe. Numbers in parentheses indicate predictor counts. Machine-readable identifiers are mapped to these same display names in all tables and figures.

| Specification | Estimator | Inputs |
|---|---|---|
| HAR | OLS | BASE (10) |
| HAR-VIX | OLS | BASE + VIX (13) |
| HAR-AAII | OLS | BASE + AAII (22) |
| HAR-Text | OLS | BASE + TEXT (25) |
| HAR-Full | OLS | Full input block (40) |
| Elastic Net-Text | Elastic Net | BASE + TEXT (25) |
| RF-Text | Random Forest | BASE + TEXT (25) |
| XGB-Text | XGBoost | BASE + TEXT (25) |
| XGB-Full | XGBoost | Full input block (40) |
| Naive | Persistence | Last admissible volatility (1) |
| Stacking | Nonnegative Ridge | Five base forecasts (5) |

All models predict the annualized volatility level $y_{i,t}^{(h)}$ defined in Section 3.2. There is no log-target transformation or exponentiation of fitted values. HAR and its four matched extensions estimate

$$y_{i,t}^{(h)} = \alpha_i^{(h)} + \boldsymbol{\beta}_i^{(h)\prime}\mathbf{x}_{i,t} + \varepsilon_{i,t}^{(h)} \tag{4.24}$$

by OLS, with the indicated information block. Predictors are standardized using the relevant training sample for numerical stability; an intercept is included. The matched comparisons hold the estimator and BASE block fixed. The naive benchmark is

$$\widehat{y}_{i,t,\mathrm{Naive}}^{(h)} = v_{i,t-1}. \tag{4.25}$$

Elastic Net minimizes

$$\min_{a,\boldsymbol{\beta}} \frac{1}{2n} \sum_{t\in\mathcal{T}} (y_t - a - \widetilde{\mathbf{x}}_t'\boldsymbol{\beta})^2 + \alpha\rho\|\boldsymbol{\beta}\|_1 + \frac{\alpha(1-\rho)}{2}\|\boldsymbol{\beta}\|_2^2, \tag{4.26}$$

where $\widetilde{\mathbf{x}}$ is standardized on the training subset and the intercept is unpenalized. RF uses bootstrap regression trees; XGBoost uses gradient-boosted trees with squared-error objective. Elastic Net is a linear regularized model and is not classified as a nonlinear specification.

Stacking uses the fixed bases HAR, HAR-VIX, HAR-AAII, HAR-Text, and XGB-Full. This set represents the baseline, each single-channel linear extension, and one full-information nonlinear model; it is not selected from outer-test rankings. Other text specifications are evaluated separately, not declared inferior by their exclusion from the combination. The meta-learner solves

$$(\widehat{\gamma}_0, \widehat{\gamma}) = \arg \min_{\gamma_0, \gamma \geq 0} \left\{ \sum_{t \in \mathcal{C}} \left( y_t - \gamma_0 - \sum_{m=1}^{5} \gamma_m \widehat{y}_{m,t} \right)^2 + \sum_{m=1}^{5} \gamma_m^2 \right\}. \tag{4.27}$$

The Ridge penalty is fixed at $\lambda = 1$, the intercept is unpenalized, and weights do not have to sum to one. Because the base forecasts are left on their raw annualized-volatility scale and the calibration sample is short, this penalty is large relative to the observed cross-origin variation in the base forecasts. The resulting coefficient sums are therefore strongly shrunk toward zero and the specification behaves approximately as an intercept-only, near-constant reference combination rather than as an unconstrained forecast average. This feature is reported explicitly and the poor stacking performance is interpreted as a property of this prespecified calibration rule, not as evidence against forecast combination in general. The calibration set $\mathcal{C}$ contains the last 63 admissible origins in the outer training sample. Base models for these calibration forecasts are fitted before $\mathcal{C}$, with a purge of $h$; their settings have already been fixed by the initial parameter search. Expanding calibration fits use all earlier admissible observations, while rolling calibration fits use the preceding 252. Bases are then refitted on the relevant full outer training window to predict its test block. Neither the bases nor the meta-learner is trained on outer-test outcomes. Calibration predictions, weights, intercepts, and fitted objects are retained for every outer fold.

For the rolling stacking specification, the 252-observation label refers to the final base-model fit. Constructing its 63 calibration forecasts requires an additional earlier 252-observation base fit; the combined calibration history is therefore not restricted to the final 252 origins alone.

## 3.7 Initial hyperparameter selection and subsequent re-estimation

Hyperparameters are selected *once*, separately for each stock, horizon, and model information set, using only the first 252 complete modeling observations. They remain fixed in every later refit and in both window schemes. Subsequent re-estimation changes fitted coefficients or trees, not the hyperparameter search. No re-tuning is performed inside the outer evaluation folds.

Three chronological validation splits are formed inside the initial sample. Each validation block contains 63 observations, and the inner training sizes are $63 - h$, $126 - h$, and $189 - h$. The last $h$ candidate training origins before each validation block are purged. Scaling is fitted separately on the admissible inner training observations. Configurations minimize mean validation MSE on the annualized-volatility scale. MSE is used as the common tuning criterion because it is defined uniformly for all linear and tree-based candidates on the modelled volatility scale; sensitivity to alternative tuning losses was not evaluated. The grid order resolves exact ties deterministically. Table 4.13 lists the complete one-time search grids and fixed settings.

XGBoost early stopping uses the inner validation block, which is contained entirely within the initial 252-observation sample. Consequently, these scores are parameter-selection statistics, not independent estimates of generalization error. For the winning configuration, the rounded median of its three best iteration counts becomes the fixed number of boosting rounds in later refits. No later test or calibration outcome is used to revise it. All candidates, fold predictions, scores, stopping iterations, and selected configurations are archived. Randomized estimators use seed 42.

Table 4.13: One-time hyperparameter search on the initial 252 observations

| Model | Search and fixed settings |
|---|---|
| Elastic Net-Text | $\alpha \in \{10^{-5}, 10^{-4}, 10^{-3}, 10^{-2}, 10^{-1}\}$, $\rho \in \{0.1, 0.25, 0.5, 0.75, 0.9\}$; 25 configurations. Standardized predictors; maximum 100,000 iterations, tolerance $10^{-7}$. |
| RF-Text | 500 trees; maximum depth $\{4, 6, 8\}$; minimum split size $\{20, 40, 60\}$; minimum leaf size $\{10, 20, 30\}$; 27 configurations. All predictors eligible at a split; bootstrap sampling. |
| XGB-Text and XGB-Full | Maximum depth $\{4, 6, 8\}$, learning rate $\{0.01, 0.05, 0.10\}$, row and column fractions each $\{0.8, 0.9\}$, minimum child weight $\{1, 3, 5\}$; 108 configurations per information set. At most 1,000 rounds, early-stopping patience 50; squared-error objective, RMSE evaluation, histogram trees, L2 penalty 1. |
| Stacking | Nonnegative Ridge with fixed $\lambda = 1$; no penalty search. |

The small initial sample is a deliberate constraint of this design. In particular, the first inner training set has only 41 observations at $h = 22$, and parameters chosen in an early market environment need not remain optimal later. The results evaluate this frozen-selection procedure, not a continuously re-tuned forecasting system.

## 3.8 Walk-forward backtesting

Both estimation schemes use five chronological test blocks of approximately equal length and identical test dates. Index complete modeling rows from zero. The first 252 observations are reserved for parameter selection. After a purge of $h$, a further 63 observations permit stacking calibration independent of that initial selection; another purge precedes the first test block. For a dataset of length $n$, let

$$b_0 = 315 + 2h, \qquad b_k = b_0 + \left\lfloor \frac{k(n - b_0)}{5} \right\rfloor, \quad k = 1, \ldots, 5. \tag{4.28}$$

Test block $k$ consists of indices $b_{k-1}, \ldots, b_k - 1$. The expanding training set is $\{0, \ldots, b_{k-1} - h - 1\}$. The rolling set contains its final 252 observations. A model is refitted once before each test block; predictors continue to update daily within the block, but model parameters do not. Thus, rolling denotes a 252-observation estimation window at each of five refit dates, not daily parameter re-estimation.

The primary inference uses expanding-window forecasts; rolling-window forecasts provide a separately labeled robustness arm. There are $11 \times 5 = 55$ stock-fold comparisons per horizon and window, 165 per window, and 330 across both windows. The complete split manifest records all training, label-end, and test dates and sample sizes. All model comparisons use identical complete cases within a stock-horizon cell. No post-test embargo is needed because future observations never enter an earlier fit.

For example, AAPL at $h = 1$ has 316 admissible training observations before its first test block, which runs from 2002-06-17 to 2006-07-24 (1034 observations). Its five test-block sizes are 1034, 1034, 1035, 1034, 1035. The other 329 splits are recorded in the same manifest.

### 3.9 Evaluation metrics

Raw model forecasts are retained. A positive forecast for QLIKE is defined by

$$\widehat{y}_t^+ = \max(\widehat{y}_t, 10^{-6}).$$

Every application of this floor is recorded. The reported primary loss is the normalized QLIKE

$$L_t^Q = r_t - \log r_t - 1, \qquad r_t = \frac{y_t^2}{(\widehat{y}_t^+)^2}. \tag{4.29}$$

It is nonnegative and invariant to a common annualization factor. The alternative expression $\log((\widehat{y}_t^+)^2) + r_t$ is also stored, but differs by the observation-only term $\log(y_t^2) + 1$. Paired loss differences are unchanged; numerical mean losses and percentages must not be interchanged between these two normalizations.

Secondary DM comparisons use $(y_t - \widehat{y}_t)^2$ on the raw fitted annualized-volatility scale. MSE, RMSE, and MAE after the positivity floor are also archived, alongside raw MSE, so that the numerical safeguard is auditable. All models are fitted to the annualized-volatility level; no model uses a log-variance target. Primary horizon summaries give equal weight to the eleven stocks rather than weighting them by their different history lengths.

QLIKE is used as the primary comparison criterion, including nested HAR-X comparisons. Its proxy-robustness properties require assumptions such as those in Patton (2011); they do not prove that this smoothed OHLC target is an unbiased proxy for next-session integrated variance.

### 3.10 Statistical forecast comparison

For a competing model $m$, the DM differential is $d_t = L_{HAR,t} - L_{m,t}$, so positive values favor the competitor. The statistic is the mean differential divided by its Newey–West standard error, with Bartlett weights. The conventional forecast-horizon bandwidths $h = 1, 5, 22$ are retained as a protocol-comparability calculation. They capture overlap induced by an $h$-step forecast horizon, but they do not capture all serial dependence that can arise because the outcome itself averages future values of a trailing 252-session volatility estimator. For thesis-level inferential claims, the dependence-robust rule is therefore bandwidth $252 + h - 1$; the shorter $h$ rule is reported as a sensitivity/reference calculation. This hierarchy is fixed independently of which rule produces the more favorable result. Two-sided Student-$t$ reference probabilities use $T - 1$ degrees of freedom. These are asymptotic comparison procedures, not exact finite-sample tests for all model-selection settings.

For the four matched linear HAR-X extensions, the complementary one-sided Clark–West differential uses raw level forecasts:

$$f_t = (y_t - \widehat{y}_{HAR,t})^2 - (y_t - \widehat{y}_{m,t})^2 + (\widehat{y}_{HAR,t} - \widehat{y}_{m,t})^2. \tag{4.30}$$

CW addresses an adjusted squared-error comparison for nested models; it does not replace the primary QLIKE criterion. Different losses, the adjustment, and the one-sided alternative can lead to different outcomes. A favorable CW result does not by itself establish a QLIKE improvement.

The expanding-window family contains $10 \times 33 \times 2 = 660$ DM tests and $4 \times 33 = 132$ CW tests, or 792 tests in total. The two DM losses are QLIKE and raw squared error. Benjamini–Hochberg is applied once to the 792 raw $p$-values separately under each HAC rule; $q^{global} < 0.05$ denotes an adjusted rejection. Inferential statements in the thesis use the $252 + h - 1$ family, while the $h$-bandwidth family is retained for transparent comparison with the shorter convention. Neither selected pairwise tests nor SHAP rankings define hypothesis verdicts. Stacking contributes 66 of the 660 DM tests (33 stock–horizon cells times two losses). Because its near-constant forecasts generate many very small adverse $p$-values, including those 66 tests changes BH ranks and can make the adjusted threshold for other models less stringent. The family is nevertheless kept at 792 because it was defined at the model-universe level rather than after inspecting performance; all substantive matched-channel conclusions are therefore also checked under the longer HAC rule and interpreted without relying on Stacking.

Rolling results are labeled separately. The archive includes their own 792-test adjustment and a further joint BH correction over all 1,584 tests from both windows, making the additional window dimension explicit. A Benjamini–Yekutieli sensitivity adjustment is also retained because arbitrary dependence across the many related tests is not guaranteed to satisfy the usual BH conditions. These alternatives are not searched to select the most favorable significance statement.

Because the 252-session target construction can induce dependence beyond $h$, the complete $252 + h - 1$ test set is the dependence-robust inferential rule used for thesis conclusions. The shorter $h$ calculation remains visible as a protocol-comparability sensitivity and is never used to rescue a conclusion that fails under the longer rule. Broad all-pair DM comparisons and ex-post regime comparisons are diagnostic output; only strata with at least 30 matched observations receive test probabilities. Regime tests are not treated as evidence from an independent real-time classification exercise. Their HAC score sequence retains the original OOS time index, with zero scores outside the selected state; observations separated by another state are not compressed into adjacent sessions.

The separate 12,210-row pairwise diagnostic sheet is not part of the 792-test confirmatory family and is not used to support inferential claims. It is retained only as ancillary diagnostic material.

### 3.11 Regime-dependent evaluation

For each stock and horizon, a two-state Markov-switching model is fitted to the full available target series:

$$y_t = \mu_{s_t} + \epsilon_t, \qquad \epsilon_t \mid s_t \sim \mathcal{N}(0, \sigma^2_{s_t}), \qquad s_t \in \{0, 1\}. \tag{4.31}$$

The state follows a first-order Markov chain; both its intercept and variance switch. The lower-mean state is labeled low volatility. Maximum smoothed marginal probability assigns each date to a state. Parameter estimates, likelihoods, convergence diagnostics, state mappings, and both probability series are retained. A failed fit is not silently replaced by quantile labels.

These full-sample smoothed states are strictly ex-post descriptive strata. Their construction does not validate the two-state specification or remove within-state serial dependence in the smoothed target. The forecasting models do not receive them as predictors, tuning criteria, or trading signals. The same states are used for both window schemes on the same dates. Regime-specific losses and SHAP summaries describe retrospective heterogeneity; they do not demonstrate a deployable regime-timing strategy.

### 3.12 Explainability analysis

SHAP values are generated from the saved fitted models for all out-of-sample observations, separately by stock, horizon, model, window, and outer fold. For linear models, exact interventional attributions use the training mean in the standardized feature space. RF uses exact tree-path-dependent TreeSHAP; XGBoost uses exact tree contributions. Naive is a one-feature linear mapping. Stacking explanations allocate its prediction across the five base-model forecasts, not directly across raw news features.

Attributions explain raw annualized-volatility output before the positivity floor; a SHAP value of 0.01 corresponds to one percentage point of annualized volatility. Their sum plus the fold-specific reference prediction is checked against the corresponding forecast. Observation-level feature values, attributions, dates, folds, reference values, forecasts, and ex-post states are stored in compressed arrays. Pooling folds combines explanations from different fitted models and references; it is not one global response function. Individual AAII attributions are additionally sensitive to the algebraic dependence of shares and spreads.

Mean absolute SHAP values measure within-model attribution magnitude. Beeswarm plots show signs and dispersion, with feature values represented by color. They do not identify causal effects or establish incremental out-of-sample forecast value. That value is assessed through matched losses and adjusted tests. A broad distribution in the high-volatility stratum is not itself evidence that text improves forecasts more in that state.

The archive contains all 363 model-stock-horizon combinations per window, 726 in total. Appendix B.1 presents five selected views from the expanding-window models. All five selected views display every model input. The XGB-Full illustration therefore displays all forty inputs used by the model, while the archive retains the corresponding observation-level SHAP values and the complete set of model-stock-horizon combinations. No label is applied as a graphical

Table 4.14: Mean normalized QLIKE: primary expanding-window forecasts

| Model | $h = 1$ | $h = 5$ | $h = 22$ |
|---|---|---|---|
| HAR | 0.000292 | 0.000711 | 0.005700 |
| HAR-VIX | 0.000420 | 0.001083 | 0.007870 |
| HAR-AAII | 0.000385 | 0.001054 | 0.008681 |
| HAR-Text | 0.000446 | 0.001307 | 0.007835 |
| HAR-Full | 0.001236 | 0.003461 | 0.012130 |
| Elastic Net-Text | 0.000527 | 0.001327 | 0.009558 |
| RF-Text | 0.056176 | 0.056909 | 0.068661 |
| XGB-Text | 0.054099 | 0.058690 | 0.086695 |
| XGB-Full | 0.056628 | 0.067153 | 0.095420 |
| Naive | 0.000286 | 0.000676 | 0.003806 |
| Stacking | 0.437237 | 0.437242 | 0.438507 |

*Notes:* Equal-weight means across eleven stocks. Lower is better. Initial hyperparameters are frozen after selection on the first 252 complete observations. These are YZ-252 level-target results. Percentage differences reported in the text are computed from full-precision losses; displayed means are rounded to six decimals.

overlay. Posterior entropy and directionality retain distinct definitions and displayed names.

# 4 Empirical results

All results in this section and Appendix C come from the YZ-252, annualized-volatility-level calculation with initial-only parameter selection. The archived result identifier used for the cell-level winner counts is `YZ252_tune_once_initial252_v1`. Tables 4.14–4.18 and Appendix C report the primary expanding window. Rolling-window and longer-bandwidth results are retained and summarized separately.

## 4.1 Forecast performance and matched information channels

Naive has the lowest equal-weight mean QLIKE at all three horizons: approximately 0.000286, 0.000676, and 0.003806 for $h = 1, 5, 22$, respectively. Relative to HAR, the reductions are 2.12%, 4.86%, and 33.23%. These are descriptive equal-weight cross-stock averages, rather than dominance in every stock–horizon cell. At $h = 22$, the 33.23% reduction is concentrated in JNJ, JPM, and AAPL, which contribute 20.41, 10.67, and 5.91 percentage points, respectively. MA and LLY add only 0.50 and 0.11 percentage points, and the remaining six stocks offset the reduction by 4.37 percentage points. HAR has lower QLIKE than Naive in six of the eleven $h = 22$ stock cells, while Naive has lower QLIKE in 13 of the 33 stock–horizon cells overall.

Naive is the parameter-free forecast $\widehat{y}_{t+h|t} = v_{t-1}$. HAR, by contrast, estimates a larger and strongly collinear block of five volatility terms and five log-volume terms in every training window. The retained fitted models show material fold-specific variation for the three stocks that drive the aggregate $h = 22$ result. In the first expanding fold the training-sample has size of 337 observations. The summed raw volatility coefficients are 0.613, 0.852, and 0.943, respectively, and move closer to one in subsequent folds. The corresponding raw intercepts are -0.156, 0.400, and -1.147. Because the first-fold training-sample sizes are identical, differences in the estimated coefficients cannot be attributed to differences in the number of training observations

Table 4.15: Long-horizon Naive–HAR QLIKE decomposition under expanding estimation

| Stock | HAR QLIKE | Naive QLIKE | Contribution (pp) |
|---|---|---|---|
| JNJ | 0.016850 | 0.004055 | 20.41 |
| JPM | 0.012819 | 0.006127 | 10.67 |
| AAPL | 0.006773 | 0.003065 | 5.91 |
| MA | 0.005087 | 0.004774 | 0.50 |
| LLY | 0.002439 | 0.002370 | 0.11 |
| ABT | 0.003335 | 0.003382 | -0.07 |
| AMZN | 0.002303 | 0.002565 | -0.42 |
| WMT | 0.002491 | 0.002804 | -0.50 |
| UNH | 0.002524 | 0.002939 | -0.66 |
| ABBV | 0.003673 | 0.004133 | -0.73 |
| XOM | 0.004410 | 0.005652 | -1.98 |
| Equal-weight mean | 0.005700 | 0.003806 | 33.23 |

*Note:* Contribution equals $100(QLIKE_{HAR,i} - QLIKE_{Naive,i})/(11\,\overline{QLIKE}_{HAR})$. Positive values favor Naive. HAR is better in six of eleven $h = 22$ cells.

across these stocks. Instead, they reflect differences in the stock-specific data and fitted HAR relationships under the common estimation protocol. Thus, the fitted HAR system does not mechanically preserve the near-random-walk level behavior represented exactly by Naive. This is a diagnostic interpretation of finite-window OLS estimation, not a claim that Naive uniformly dominates HAR. Naive forecasts are identical across windows because the test dates and lagged volatility inputs are identical.

Table 4.16: Fold-specific raw HAR diagnostics for the concentrated $h = 22$ cases

| Stock | Fold | Intercept | $\sum \beta_{vol}$ | $\sum \beta_{volume}$ |
|---|---|---|---|---|
| JNJ | 1 | -0.155547 | 0.612599 | 0.016399 |
| JNJ | 2 | 0.021033 | 0.988049 | -0.001201 |
| JNJ | 3 | -0.037554 | 0.983101 | 0.002557 |
| JNJ | 4 | -0.024880 | 0.986091 | 0.001706 |
| JNJ | 5 | -0.018520 | 0.982657 | 0.001358 |
| JPM | 1 | 0.399551 | 0.851840 | -0.020801 |
| JPM | 2 | 0.055200 | 0.993053 | -0.003352 |
| JPM | 3 | -0.027824 | 0.988330 | 0.001947 |
| JPM | 4 | -0.016220 | 0.990635 | 0.001160 |
| JPM | 5 | -0.022716 | 0.989557 | 0.001562 |
| AAPL | 1 | -1.146752 | 0.942630 | 0.060154 |
| AAPL | 2 | -0.069461 | 0.968129 | 0.004073 |
| AAPL | 3 | -0.105833 | 0.974241 | 0.005761 |
| AAPL | 4 | -0.123624 | 0.977008 | 0.006558 |
| AAPL | 5 | -0.070034 | 0.972640 | 0.004000 |

*Note:* Expanding-window OLS. $\sum \beta_{vol}$ sums the coefficients of `ann_vol_lag1`, `ann_vol_lag5`, `ann_vol_lag22`, `ann_vol_mean5`, and `ann_vol_mean22`; $\sum \beta_{volume}$ analogously sums the five log-volume coefficients. Coefficients are transformed from the standardized fitting representation to the original regressor scale.

All four matched HAR-X extensions have higher mean QLIKE than HAR at each horizon. HAR-Text is worse by 52.62%, 83.95%, and 37.45%; HAR-VIX is worse by 43.77%, 52.44%, and 38.07%. HAR-VIX therefore does not have the lowest mean QLIKE. Frequent individual wins and the lowest cross-stock mean are different summaries.

Table 4.17: Matched HAR-X comparisons: expanding window

| Model | QLIKE better | DM better | DM worse | CW positive | CW significant |
|---|---|---|---|---|---|
| HAR-VIX | 18/33 | 8/33 | 11/33 | 22/33 | 18/33 |
| HAR-AAII | 2/33 | 0/33 | 25/33 | 15/33 | 7/33 |
| HAR-Text | 2/33 | 2/33 | 30/33 | 14/33 | 13/33 |
| HAR-Full | 2/33 | 1/33 | 30/33 | 14/33 | 11/33 |

*Notes:* DM significance refers to QLIKE; CW uses adjusted raw annualized-volatility squared error. Significant entries shown here use the short-bandwidth expanding-window BH family of 792 tests at $q^{global} < 0.05$, with bandwidth $h$; thesis-level inferential conclusions additionally require the dependence-robust $252 + h - 1$ rule. Counts do not compare identical loss criteria across DM and CW.

The tree-based specifications and Stacking have substantially larger mean losses under this protocol. This does not establish that nonlinear forecasting or forecast combination is intrinsically ineffective. The results concern initial-only selection, infrequent refits, and the specified calibration rule. The median sum of the five stacking coefficients across 165 fits is only 0.00477 under expanding estimation and 0.00434 under rolling estimation. A common shift in the base forecasts is therefore strongly attenuated in a typical fitted combination. This is consistent with the fixed Ridge penalty on unscaled volatility forecasts and the short calibration sample.

HAR-Text improves on HAR in only two of 33 expanding cells: JNJ and JPM at $h = 22$. Appendix C reports the directly auditable absolute QLIKE differences: HAR-Text has $\Delta = 0.008653$ for JNJ and $\Delta = 0.003374$ for JPM. In those same cells, Naive improves on HAR more strongly, with $\Delta = 0.012795$ and $\Delta = 0.006692$, respectively. The two favorable HAR-Text rejections therefore do not identify text-specific value; they are also consistent with weakness of the estimated HAR benchmark in those cells. Under the short bandwidth $h$, the reference family yields two favorable and 30 adverse adjusted QLIKE rejections. Under the thesis-level dependence-robust bandwidth $252 + h - 1$, the favorable count is zero and the adverse count is three. Hence the two favorable short-bandwidth results are not used as confirmatory evidence.

The complementary CW statistic is positive for HAR-Text in 14 cells and globally significant in 13. This is not evidence of thirteen QLIKE gains: CW adjusts a raw level-squared-error differential and uses a one-sided alternative. Its favorable results cannot replace the primary assessment.

VIX has the most recurrent matched-channel improvement: 18 cells, compared with two each for AAII, text, and HAR-Full. Nevertheless, HAR-VIX has eight favorable and eleven adverse primary QLIKE rejections, and its mean loss exceeds that of HAR at every horizon. For JNJ at $h = 1$, Appendix C reports a large adverse HAR-VIX loss difference relative to HAR. The evidence supports heterogeneity, not a universal VIX advantage. HAR-AAII has no favorable primary QLIKE rejection; HAR-Full has one.

## 4.2 Descriptive winner counts

Under expanding estimation, HAR-VIX has the lowest QLIKE in 17 cells, Naive in 12, and HAR in four. For raw MSE the counts are 16, 12, and five. Floored-MSE counts are identical because no expanding forecast requires the positivity floor. No other specification wins under these loss

Table 4.18: Cell-level winner counts across 33 tasks: expanding window

| Model | QLIKE | Raw MSE | Floored MSE |
|---|---|---|---|
| HAR | 4 | 5 | 5 |
| HAR-VIX | 17 | 16 | 16 |
| HAR-AAII | 0 | 0 | 0 |
| HAR-Text | 0 | 0 | 0 |
| HAR-Full | 0 | 0 | 0 |
| Elastic Net-Text | 0 | 0 | 0 |
| RF-Text | 0 | 0 | 0 |
| XGB-Text | 0 | 0 | 0 |
| XGB-Full | 0 | 0 | 0 |
| Naive | 12 | 12 | 12 |
| Stacking | 0 | 0 | 0 |

*Notes:* Each loss column sums to 33. MSE uses annualized-volatility levels, not log variance. Lowest point estimates are descriptive rankings, not inferential rejections.

definitions. In the rolling arm, Naive wins 31 QLIKE comparisons and 30 MSE comparisons; the remaining wins belong to HAR. Cell-level winners are descriptive. They do not establish a general model advantage without the corresponding benchmark loss differences and adjusted tests. Elastic Net-Text wins no expanding QLIKE or MSE cell in Table 4.18, and its mean QLIKE exceeds HAR at every horizon; there is therefore no cell-level or mean-loss evidence of an Elastic Net advantage here.

## 4.3 Stability, alternative windows, and ex-post regimes

For the active result set, aggregate fold- and regime-level diagnostics are available but the complete 165-row fold comparison export and the full cell-by-regime loss table are not. Accordingly, the quantities in this subsection are descriptive summaries and are not used as independently auditable confirmatory hypothesis tests. Appendix D reports every aggregate quantity used below. HAR-Text improves on HAR in 15, 14, and 23 of 55 expanding stock-fold comparisons at $h = 1, 5, 22$, respectively. Rolling counts are only five, three, and seven. No stock-horizon cell improves in *every* fold under either window. Rolling HAR-Text improves in two cells, with zero favorable and 29 adverse primary QLIKE rejections. JNJ at $h = 22$ is the only favorable point estimate common to both windows. The joint 1,584-test correction leaves the favorable expanding and rolling HAR-Text counts unchanged at two and zero.

With bandwidth $252 + h - 1$, HAR-Text has no favorable QLIKE rejection in either window. Adverse BH counts fall from 30 to three for expanding and from 29 to 24 for rolling. Under longer HAC and BY, the adverse counts are zero and six. HAR-VIX also has no favorable rejection under longer HAC. Very small primary probabilities are therefore not necessarily bandwidth-robust evidence.

Naive has nine favorable and eight adverse primary QLIKE rejections against expanding HAR; neither direction remains significant under longer HAC. Against rolling HAR, favorable counts are 29 under bandwidth $h$ and 17 under longer HAC, with no adverse rejections. Its lowest mean loss does not imply uniform inferential dominance over expanding HAR.

One rolling HAR-Full forecast for MA at $h = 5$ is negative. The declared positivity floor generates an extreme QLIKE observation. Appendix D reports its date, raw forecast, and a post-hoc floor sensitivity. The observation is retained, and the diagnostic does not alter the test family.

Regime results are relatively more favorable to text in the high state. HAR-Text beats HAR in ten high-state cells versus two low-state cells under expanding estimation, and in five versus two under rolling estimation. Its relative QLIKE improvement is larger in the high state in 22/33 expanding and 21/33 rolling cells. However, median improvements remain negative in both states: −8.73% versus −56.28% for expanding and −100.49% versus −194.89% for rolling. This indicates less adverse relative performance in high-volatility strata, not a generally positive textual gain or a formal common-state-effect test.

### 4.4 Conditional feature attribution: SHAP analysis

Across all OOS observations in RF-Text for ABT at $h = 5$, volatility inputs occupy ranks one to five. Lag-one volatility has mean absolute SHAP approximately 0.0551, compared with $2.00 \times 10^{-4}$ for the highest-ranked text input, the lagged 22-session mean directional sentiment.

Posterior entropy is directly present. Its lagged five-session mean ranks tenth, with mean absolute SHAP $1.90 \times 10^{-5}$; the highest-ranked directionality input ranks eleventh, at $8.78 \times 10^{-6}$. Upper-quartile entropy observations have mean SHAP $-2.21 \times 10^{-5}$, versus $2.03 \times 10^{-5}$ in the lower quartile. The lagged 22-session directionality mean instead has positive mean attribution in its upper quartile and negative mean attribution in its lower quartile. These are descriptive pooled comparisons, not monotonic or causal effects. Neither uniformly positive entropy attribution nor uniformly smaller entropy importance is supported.

In HAR-VIX for ABT at $h = 1$, volatility inputs also occupy the first five ranks, followed by lag-one VIX. In high-state XGB-Full for WMT at $h = 1$, the five volatility inputs lead the ranking, followed by the lagged 22-session VIX mean. Directionality and entropy at that averaging length rank fourteenth and fifteenth. Individual AAII ranks are not structural interpretations because shares and spreads are dependent.

HAR-Text's base forecast ranks first in low-state Stacking for WMT at $h = 22$; HAR-AAII and HAR-VIX rank first and second in high-state Stacking for UNH at $h = 5$. These rankings allocate a combination's forecast among correlated base forecasts and do not establish incremental accuracy. Different fold-specific references also prevent pooled colors from representing a single global response function. Appendix B.1 presents selected figures; the complete 726-combination manifest and observation-level arrays are in the accompanying archive. All five printed figures show every input of the respective specification, including all forty inputs for XGB-Full. The printed RF-Text and HAR-VIX figures show high-volatility subsets; the pooled statistics above use all OOS observations and are not subset-specific rankings. For the printed high-volatility RF-Text/ABT/$h = 5$ panel specifically, the lagged five-session entropy measure ranks tenth, the lagged 22-session news-volume measure ranks eleventh, and the highest-ranked directionality input (lagged 22-session mean) ranks thirteenth. These subset ranks are reported only to make the printed figure directly auditable; they do not replace the full-OOS ranking above. Attribution magnitudes and colors are not causal effects, and neither directionality nor posterior entropy is equated with economic uncertainty. The stacking plots describe base-forecast contributions rather than raw textual-variable effects.

### 4.5 Summary assessment of the empirical hypotheses

QLIKE comparisons and the primary global adjusted values determine the main assessment. CW is complementary adjusted squared-error evidence and may disagree with QLIKE; it does not override it. Stability and Markov-regime comparisons are descriptive. Conclusions are conditional on the disclosed initial-selection protocol and target, not universal statements about news information or nonlinear learning. Table 4.19 summarizes the resulting hypothesis assessments and their limitations.

Table 4.19: Assessment of the Article 4 hypotheses under the YZ-252 protocol

| Hypothesis | Assessment | Evidence and limitation |
|---|---|---|
| H4.1 | Not generally supported | HAR-Text improves in 2/33 expanding cells; under bandwidth $h$, primary QLIKE tests give two favorable and 30 adverse rejections, versus zero favorable and three adverse rejections under bandwidth $252 + h - 1$. In both favorable cells Naive improves on HAR by a larger absolute QLIKE difference than HAR-Text, so the two short-bandwidth rejections do not identify text-specific value. Thirteen favorable CW rejections concern a different adjusted loss. |
| H4.2 | Not supported | Text has fewer matched QLIKE improvements than VIX (2/33 versus 18/33); AAII also has 2/33. The requirement of at least equally recurrent gains across both conventional channels is not met. VIX does not have the lowest mean QLIKE. |
| H4.3 | Descriptive sensitivity only | HAR–Text wins 15/55, 14/55, and 23/55 expanding fold comparisons and 5/55, 3/55, and 7/55 under rolling estimation at $h = 1, 5, 22$. The full active fold table is retained in the replication archive. |
| H4.4 | Descriptive heterogeneity only | HAR–Text has more high- than low-state QLIKE wins (10 versus 2 expanding; 5 versus 2 rolling), but the median relative improvement is negative in both states. The full active regime table is retained in the replication archive. |

## 5 Conclusion

The analysis does not establish a general incremental forecasting gain from market-wide transformer-derived news. HAR-Text lowers QLIKE in two of 33 expanding tasks. Under the short bandwidth $h$, there are two favorable and 30 adverse adjusted QLIKE rejections; under $252 + h - 1$, there are zero favorable and three adverse rejections. The positive CW findings concern a different adjusted loss and cannot substitute for QLIKE evidence.

Naive has the lowest mean QLIKE at every horizon, with reductions relative to HAR of 2.12% at $h = 1$, 4.86% at $h = 5$, and 33.23% at $h = 22$. The long-horizon average is concentrated:

JNJ, JPM, and AAPL account for 36.99 percentage points of the 33.23% reduction, whereas HAR has lower QLIKE in six of the eleven $h = 22$ stocks. The result is therefore a concentrated descriptive loss pattern under the retained estimation design, not uniform statistical dominance of the simple benchmark across the panel. The fold-specific HAR diagnostics identify a plausible finite-window OLS mechanism: the larger persistence and volume system has unstable intercept and persistence estimates in the influential first folds.

HAR-VIX wins the most expanding cells but has higher mean QLIKE than HAR at every horizon. VIX therefore has no broad mean-loss advantage. The evidence does not support H4.1 as a general proposition or H4.2. H4.3 and H4.4 remain descriptive: the complete active fold and regime exports document instability across windows and retrospective states, but they do not establish a deployable real-time regime-timing rule. Text is relatively less adverse in high-volatility strata in a majority of cells, while median gains remain negative in both states. SHAP documents conditional attribution without overriding the loss assessment.

The long-window target, sparse refits, fixed early-sample settings, and strongly shrunk stacking combination are integral to the result. The rolling HAR-Full floor event and dependence sensitivity further limit broad rankings. Additional information does not reliably improve this particular highly persistent target under this procedure. The study remains an ex-post measurement experiment. ModernFinBERT was not historically available over the evaluation period, its training-data cutoff is unknown, and the semantic agreement audit is not expert ground truth. The equity universe is an ex-post selected set of surviving large firms. Moreover, predicting a future average of long trailing-window variance estimates is not the same task as predicting the variance generated only by the next few returns. Initial-only tuning, sparse block refits, and small initial validation samples limit claims about adaptive forecasting systems. These restrictions remain substantive even when all computational timing and multiple-testing checks pass.

# Appendix A

## A.1 Final empirical panel and diagnostic corpus partitions

The ModernFinBERT layer is external and is not trained or fine-tuned in this study. The fixed 80/10/10 corpus subsets are treated exclusively as diagnostic corpus partitions. Their only role is to summarize predicted-class frequencies, vocabulary overlap, and duplicate content. They do not enter model estimation or out-of-sample forecast evaluation. Table A.1 reports the predicted-class distributions across the three diagnostic partitions.

Table A.1: Distribution of ModernFinBERT-predicted classes across diagnostic corpus partitions

| Predicted class | 80% partition | 10% partition A | 10% partition B |
|---|---|---|---|
| Neutral | 187,338 | 23,418 | 23,417 |
| Bearish | 68,810 | 8,601 | 8,601 |
| Bullish | 34,276 | 4,284 | 4,285 |

Overlap and deduplication diagnostics are reported once in Table 4.4.

### A.1.1 Model-panel coverage

Table A.2 presents the common-case sample and OOS coverage.

Table A.2: Common-case sample and OOS coverage

| Stock | $h$ | Model rows | OOS rows | First OOS | Last OOS |
|---|---|---|---|---|---|
| AAPL | 1 | 5489 | 5172 | 2002-06-17 | 2022-12-29 |
| AAPL | 5 | 5485 | 5160 | 2002-06-27 | 2022-12-22 |
| AAPL | 22 | 5468 | 5109 | 2002-08-15 | 2022-11-29 |
| ABBV | 1 | 2220 | 1903 | 2015-06-11 | 2022-12-29 |
| ABBV | 5 | 2216 | 1891 | 2015-06-23 | 2022-12-22 |
| ABBV | 22 | 2199 | 1840 | 2015-08-11 | 2022-11-29 |
| ABT | 1 | 5489 | 5172 | 2002-06-17 | 2022-12-29 |
| ABT | 5 | 5485 | 5160 | 2002-06-27 | 2022-12-22 |
| ABT | 22 | 5468 | 5109 | 2002-08-15 | 2022-11-29 |
| AMZN | 1 | 5489 | 5172 | 2002-06-17 | 2022-12-29 |
| AMZN | 5 | 5485 | 5160 | 2002-06-27 | 2022-12-22 |
| AMZN | 22 | 5468 | 5109 | 2002-08-15 | 2022-11-29 |
| JNJ | 1 | 5489 | 5172 | 2002-06-17 | 2022-12-29 |
| JNJ | 5 | 5485 | 5160 | 2002-06-27 | 2022-12-22 |
| JNJ | 22 | 5468 | 5109 | 2002-08-15 | 2022-11-29 |
| JPM | 1 | 5489 | 5172 | 2002-06-17 | 2022-12-29 |
| JPM | 5 | 5485 | 5160 | 2002-06-27 | 2022-12-22 |
| JPM | 22 | 5468 | 5109 | 2002-08-15 | 2022-11-29 |
| LLY | 1 | 5489 | 5172 | 2002-06-17 | 2022-12-29 |
| LLY | 5 | 5485 | 5160 | 2002-06-27 | 2022-12-22 |
| LLY | 22 | 5468 | 5109 | 2002-08-15 | 2022-11-29 |
| MA | 1 | 3882 | 3565 | 2008-10-31 | 2022-12-29 |

| Stock | $h$ | Model rows | OOS rows | First OOS | Last OOS |
|---|---|---|---|---|---|
| MA | 5 | 3878 | 3553 | 2008-11-12 | 2022-12-22 |
| MA | 22 | 3861 | 3502 | 2009-01-02 | 2022-11-29 |
| UNH | 1 | 5489 | 5172 | 2002-06-17 | 2022-12-29 |
| UNH | 5 | 5485 | 5160 | 2002-06-27 | 2022-12-22 |
| UNH | 22 | 5468 | 5109 | 2002-08-15 | 2022-11-29 |
| WMT | 1 | 5488 | 5171 | 2002-06-17 | 2022-12-28 |
| WMT | 5 | 5484 | 5159 | 2002-06-27 | 2022-12-21 |
| WMT | 22 | 5467 | 5108 | 2002-08-15 | 2022-11-28 |
| XOM | 1 | 5489 | 5172 | 2002-06-17 | 2022-12-29 |
| XOM | 5 | 5485 | 5160 | 2002-06-27 | 2022-12-22 |
| XOM | 22 | 5468 | 5109 | 2002-08-15 | 2022-11-29 |

# Appendix B

## B.1 Selected SHAP figures

These figures use the same model names, feature definitions, horizons, and expanding-window forecasts as the results section. News volume is a session article count; directionality and posterior entropy are distinct posterior-derived measures. All dates, values, and fold-specific references are preserved in the array archive. Full-feature plots for every combination are also supplied electronically. Figures B.1–B.5 show the five selected full-input views discussed in the main text.

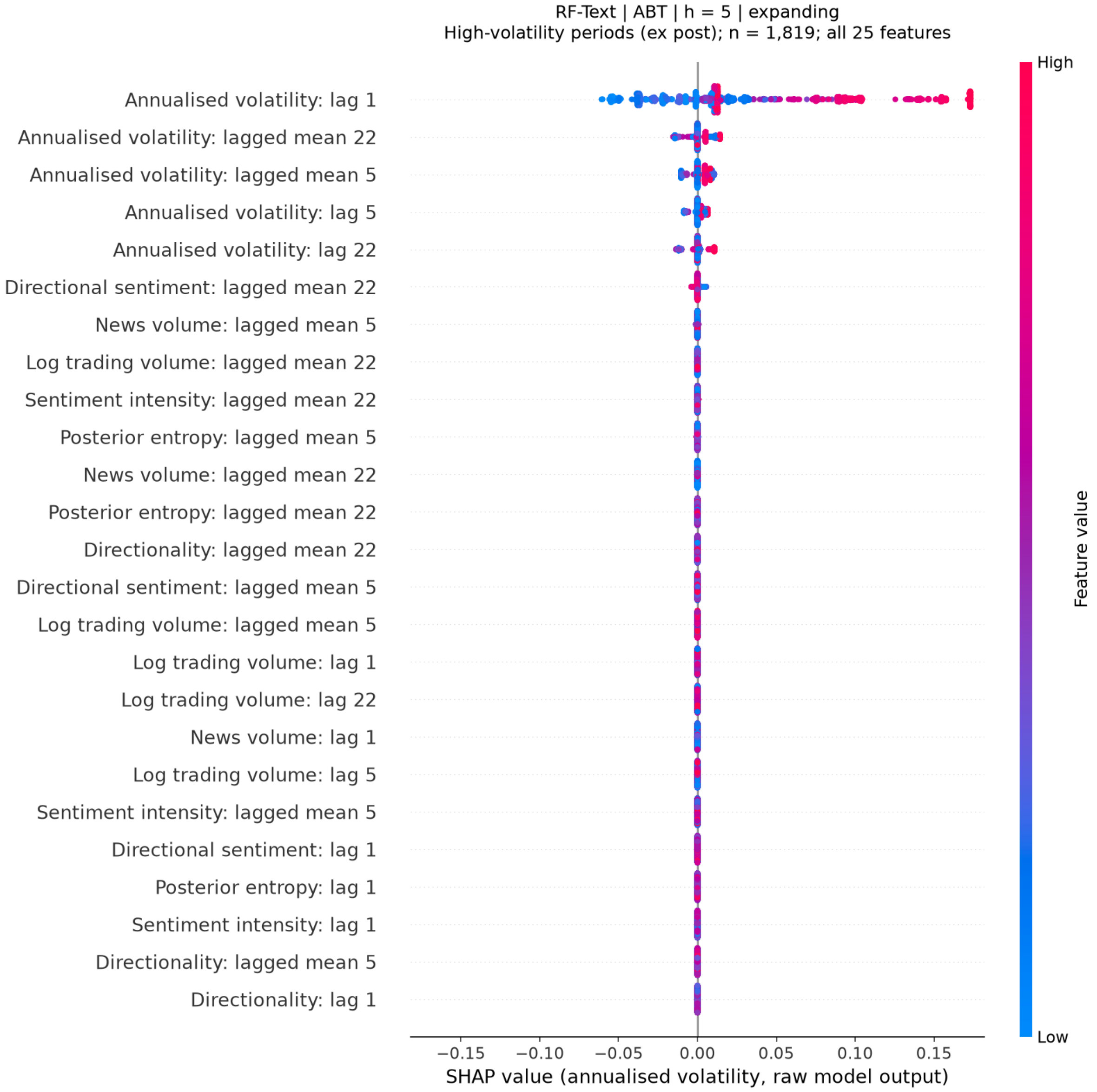

Figure B.1: RF-Text, ABT, $h = 5$, expanding window; high-volatility ex-post subset, n = 1,819. All 25 inputs are displayed. Values explain raw annualized-volatility output.

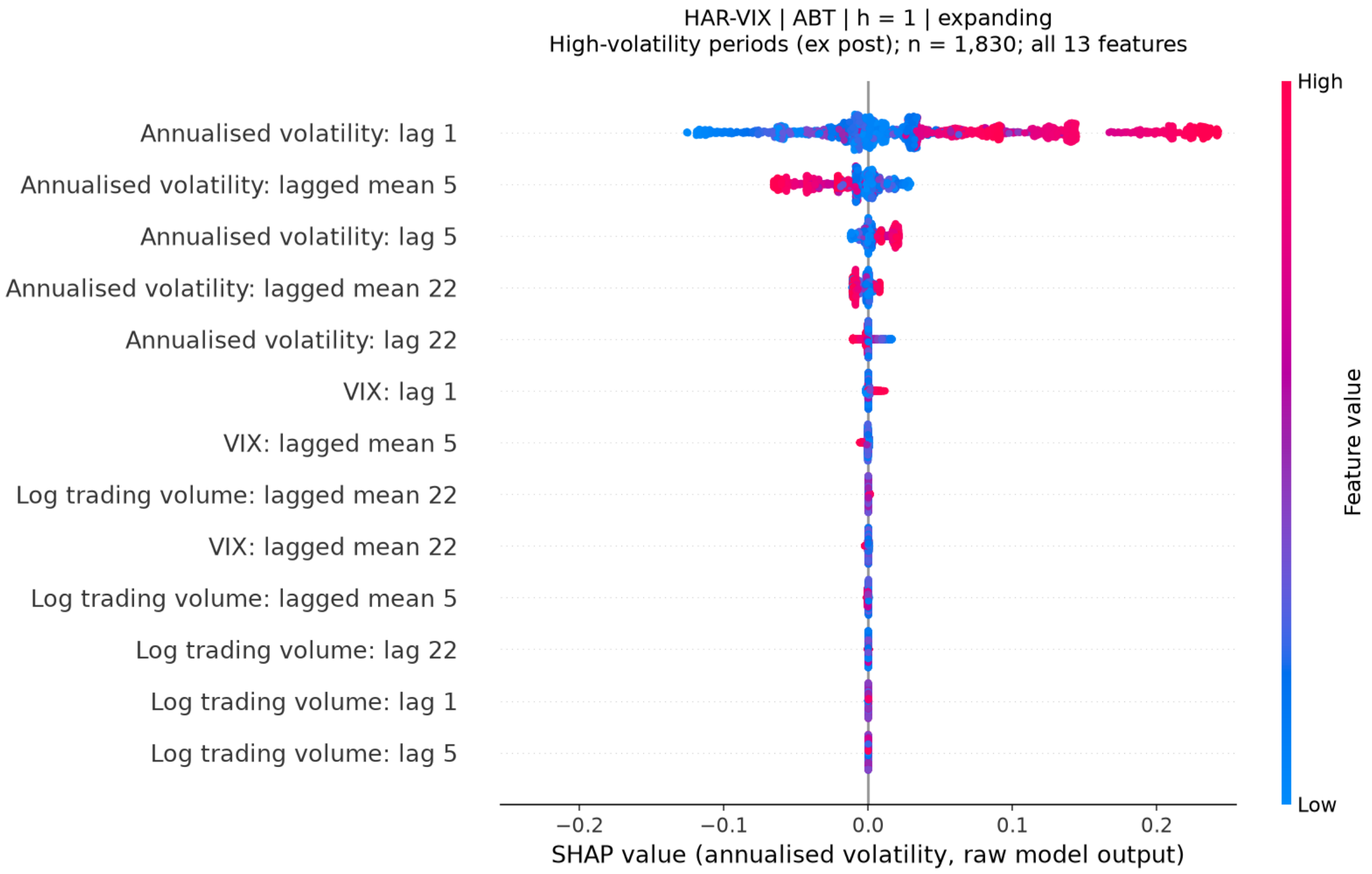

Figure B.2: HAR-VIX, ABT, $h$ = 1, expanding window; high-volatility ex-post subset, n = 1,830. All 13 inputs are displayed. Values explain raw annualized-volatility output.

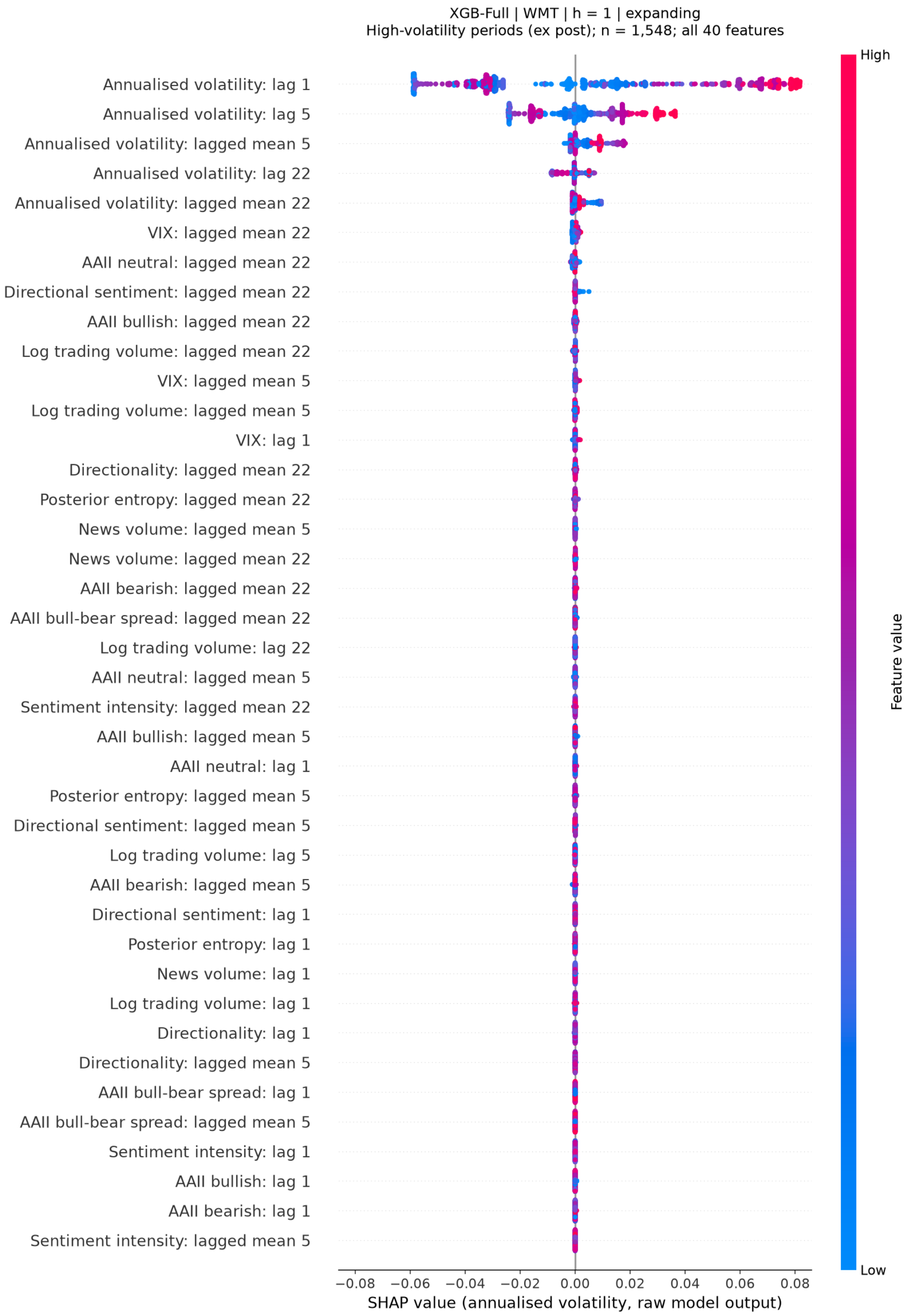

Figure B.3: XGB-Full, WMT, $h = 1$, expanding window; high-volatility ex-post subset, n = 1,548. All 40 inputs are displayed. Values explain raw annualized-volatility output.

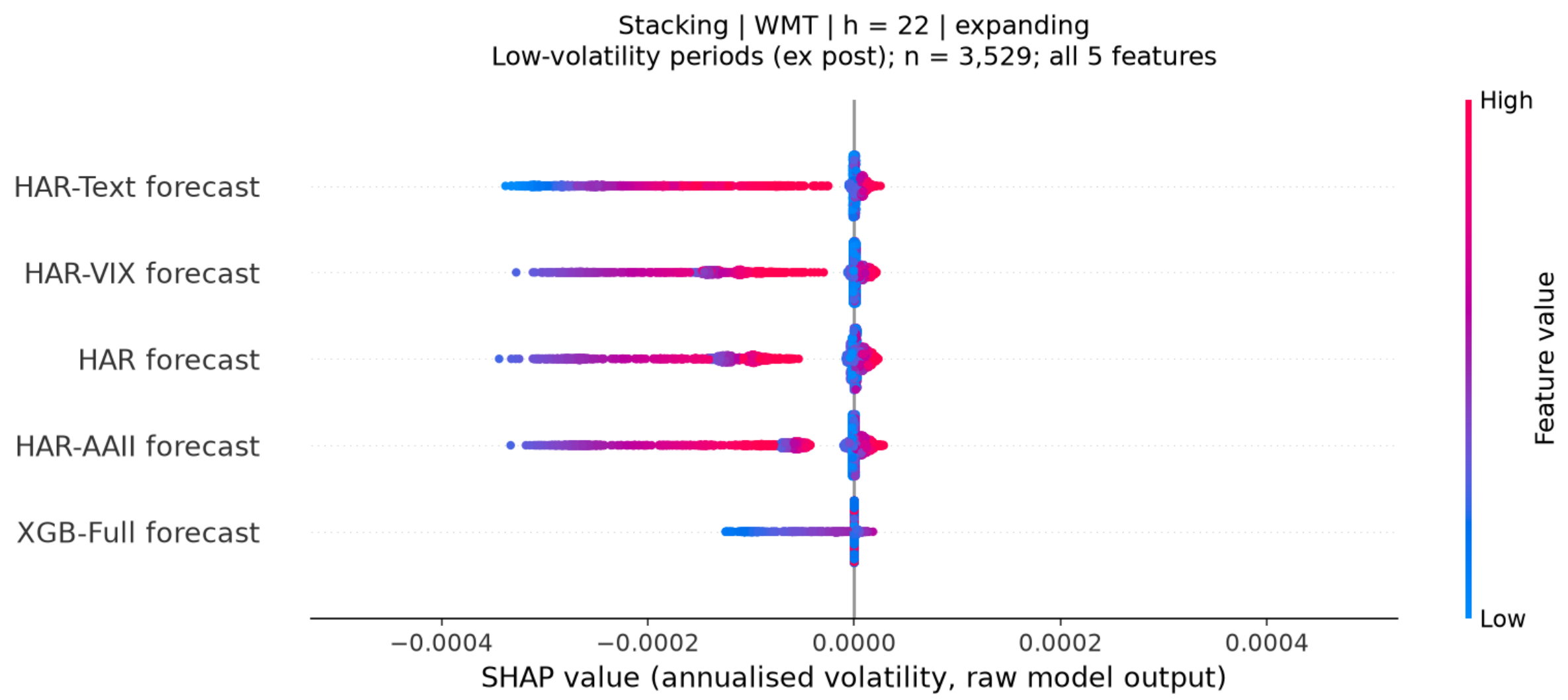

Figure B.4: Stacking, WMT, $h = 22$, expanding window; low-volatility ex-post subset, n = 3,529. All 5 inputs are displayed. Values explain raw annualized-volatility output.

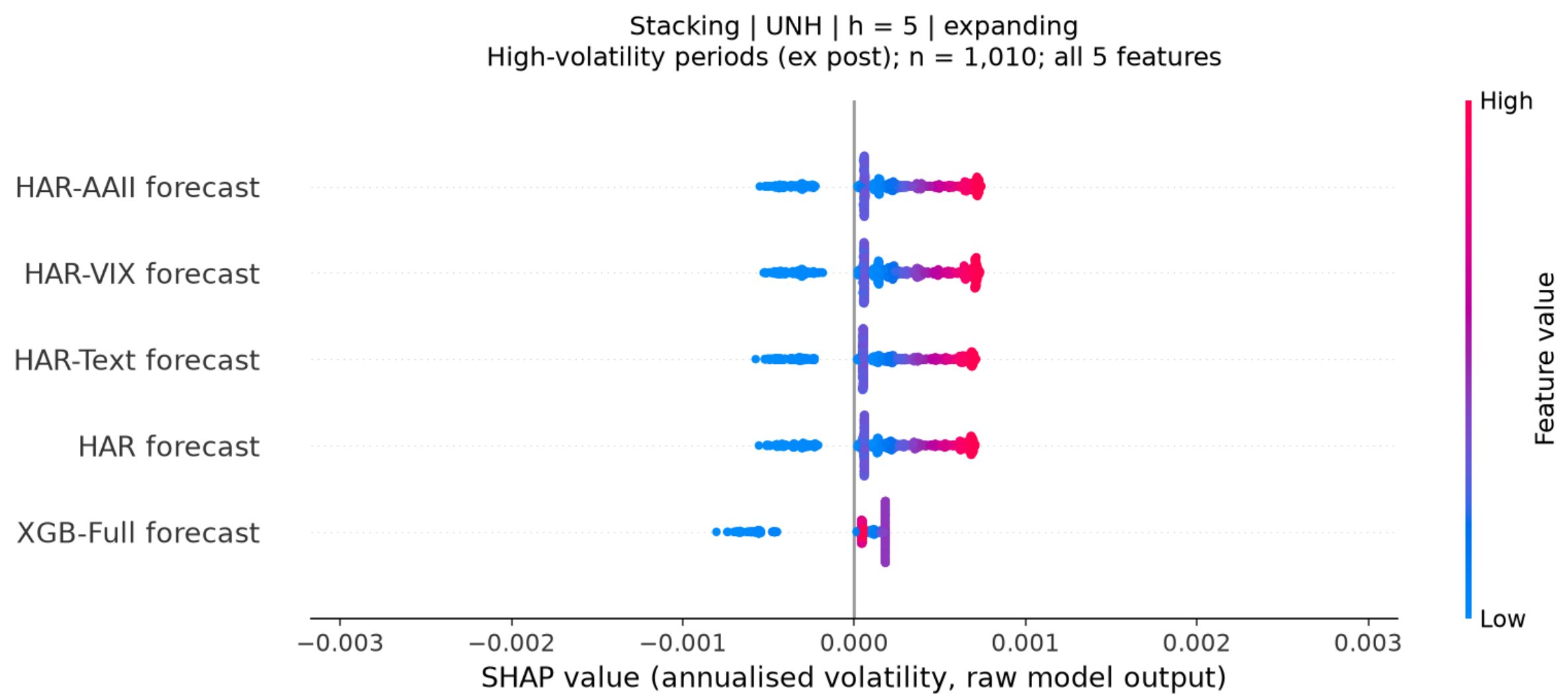

Figure B.5: Stacking, UNH, $h = 5$, expanding window; high-volatility ex-post subset, n = 1,010. All 5 inputs are displayed. Values explain raw annualized-volatility output.

## B.2 Output inventory and audit scope

The study contains 726 aggregate model-window cells, 3,630 model-window-fold rows, 330 split records, the complete initial-search predictions and scores, all fitted forecasting and calibration models, 726 SHAP arrays and full-feature plots, and all raw forecast errors. The main test family has 792 entries; both windows together have 1,584. Repeating the calculations under the longer HAC bandwidth yields 3,168 rows in the complete sensitivity table. These are not 3,168 independent economic hypotheses. The all-pair and regime tables are diagnostic. Table B.1 summarizes the retained output inventory.

Table B.1: Output inventory

| Output | Rows or combinations |
|---|---:|
| Aggregate model-window cells | 726 |
| Outer model-window-fold rows | 3,630 |
| Temporal splits | 330 |
| Primary expanding tests | 792 |
| Both-window tests, one HAC rule | 1,584 |
| All tests, two HAC rules | 3,168 |
| Full SHAP arrays and plots | 726 |
| Pairwise diagnostic records (excluded from confirmatory family) | 12,210 |

# Appendix C

## C.1 Complete short-bandwidth benchmark-comparison reference family

All tables in this appendix use expanding-window forecasts, one-time initial parameter selection, the short forecast-horizon HAC bandwidth $h$, and the same combined 792-test BH adjustment. They provide the complete row-level reference for the short-bandwidth convention. Thesis-level inferential conclusions use the dependence-robust $252+h-1$ rule summarized in Appendix D.1. Positive statistics favor the indicated competitor over HAR. DM tests are two-sided; CW tests are one-sided and use adjusted raw level squared error. $\Delta$ is the mean loss improvement (adjusted for CW). Rolling and longer-bandwidth results are supplied in the electronic archive. Probabilities are displayed with rounding, while rejection counts use their full stored precision. Tables C.1–C.3 report QLIKE DM tests; Tables C.4–C.6 report squared-error DM tests; and Tables C.7–C.9 report the matched Clark–West comparisons. Numerical underflow is shown as $< 10^{-300}$, not as an exact zero probability.

Table C.1: DM, QLIKE, $h = 1$: expanding window

| Stock | Model | $n$ | $\Delta$ | Statistic | $p$ | $q^{global}$ |
|---|---|---|---|---|---|---|
| AAPL | Elastic Net-Text | 5172 | -0.000697 | -15.476 | $7.59 \times 10^{-53}$ | $5.67 \times 10^{-52}$ |
| AAPL | HAR-AAII | 5172 | -0.000101 | -12.919 | $1.33 \times 10^{-37}$ | $6.66 \times 10^{-37}$ |
| AAPL | HAR-Full | 5172 | -0.003346 | -14.927 | $2.33 \times 10^{-49}$ | $1.55 \times 10^{-48}$ |
| AAPL | HAR-Text | 5172 | -0.000653 | -14.166 | $1.01 \times 10^{-44}$ | $6.04 \times 10^{-44}$ |
| AAPL | HAR-VIX | 5172 | -0.000305 | -15.274 | $1.50 \times 10^{-51}$ | $1.05 \times 10^{-50}$ |
| AAPL | Naive | 5172 | $2.00 \times 10^{-5}$ | 4.918 | $9.00 \times 10^{-7}$ | $1.56 \times 10^{-6}$ |
| AAPL | RF-Text | 5172 | -0.041805 | -29.959 | $5.51 \times 10^{-182}$ | $7.28 \times 10^{-180}$ |
| AAPL | Stacking | 5172 | -0.211011 | -34.090 | $5.92 \times 10^{-230}$ | $1.56 \times 10^{-227}$ |
| AAPL | XGB-Full | 5172 | -0.038760 | -25.968 | $7.44 \times 10^{-140}$ | $2.68 \times 10^{-138}$ |
| AAPL | XGB-Text | 5172 | -0.036735 | -26.838 | $1.18 \times 10^{-148}$ | $5.84 \times 10^{-147}$ |
| ABBV | Elastic Net-Text | 1903 | -0.000110 | -10.611 | $1.35 \times 10^{-25}$ | $4.84 \times 10^{-25}$ |
| ABBV | HAR-AAII | 1903 | $-3.86 \times 10^{-5}$ | -3.106 | 0.0019 | 0.0027 |
| ABBV | HAR-Full | 1903 | -0.000301 | -9.593 | $2.56 \times 10^{-21}$ | $8.22 \times 10^{-21}$ |
| ABBV | HAR-Text | 1903 | $-4.73 \times 10^{-5}$ | -8.771 | $3.84 \times 10^{-18}$ | $1.11 \times 10^{-17}$ |
| ABBV | HAR-VIX | 1903 | $-6.83 \times 10^{-7}$ | -0.067 | 0.9469 | 1.0000 |
| ABBV | Naive | 1903 | $-2.63 \times 10^{-5}$ | -3.212 | 0.0013 | 0.0019 |
| ABBV | RF-Text | 1903 | -0.030898 | -14.324 | $2.85 \times 10^{-44}$ | $1.69 \times 10^{-43}$ |
| ABBV | Stacking | 1903 | -0.087203 | -27.668 | $6.89 \times 10^{-142}$ | $2.60 \times 10^{-140}$ |
| ABBV | XGB-Full | 1903 | -0.034291 | -15.164 | $4.06 \times 10^{-49}$ | $2.65 \times 10^{-48}$ |
| ABBV | XGB-Text | 1903 | -0.034660 | -15.147 | $5.07 \times 10^{-49}$ | $3.29 \times 10^{-48}$ |
| ABT | Elastic Net-Text | 5172 | $-3.43 \times 10^{-5}$ | -3.845 | 0.0001 | 0.0002 |
| ABT | HAR-AAII | 5172 | $-2.25 \times 10^{-5}$ | -8.070 | $8.70 \times 10^{-16}$ | $2.28 \times 10^{-15}$ |
| ABT | HAR-Full | 5172 | -0.000145 | -10.824 | $5.16 \times 10^{-27}$ | $1.92 \times 10^{-26}$ |
| ABT | HAR-Text | 5172 | $-7.60 \times 10^{-5}$ | -12.545 | $1.39 \times 10^{-35}$ | $6.57 \times 10^{-35}$ |
| ABT | HAR-VIX | 5172 | $8.18 \times 10^{-6}$ | 1.745 | 0.0811 | 0.0991 |
| ABT | Naive | 5172 | $-2.59 \times 10^{-5}$ | -2.238 | 0.0253 | 0.0323 |
| ABT | RF-Text | 5172 | -0.019487 | -19.766 | $6.81 \times 10^{-84}$ | $8.42 \times 10^{-83}$ |
| ABT | Stacking | 5172 | -0.240219 | -26.919 | $1.73 \times 10^{-149}$ | $9.11 \times 10^{-148}$ |
| ABT | XGB-Full | 5172 | -0.021449 | -18.689 | $1.79 \times 10^{-75}$ | $1.81 \times 10^{-74}$ |
| ABT | XGB-Text | 5172 | -0.021636 | -18.651 | $3.44 \times 10^{-75}$ | $3.45 \times 10^{-74}$ |
| AMZN | Elastic Net-Text | 5172 | $-7.73 \times 10^{-5}$ | -9.886 | $7.66 \times 10^{-23}$ | $2.54 \times 10^{-22}$ |

| Stock | Model | $n$ | $\Delta$ | Statistic | $p$ | $q^{global}$ |
|---|---|---|---|---|---|---|
| AMZN | HAR-AAII | 5172 | -0.000664 | -19.166 | $3.77 \times 10^{-79}$ | $4.15 \times 10^{-78}$ |
| AMZN | HAR-Full | 5172 | -0.003665 | -13.998 | $1.00 \times 10^{-43}$ | $5.85 \times 10^{-43}$ |
| AMZN | HAR-Text | 5172 | $-6.97 \times 10^{-5}$ | -7.347 | $2.34 \times 10^{-13}$ | $5.65 \times 10^{-13}$ |
| AMZN | HAR-VIX | 5172 | $-5.93 \times 10^{-5}$ | -10.635 | $3.83 \times 10^{-26}$ | $1.40 \times 10^{-25}$ |
| AMZN | Naive | 5172 | $4.19 \times 10^{-5}$ | 7.702 | $1.60 \times 10^{-14}$ | $4.01 \times 10^{-14}$ |
| AMZN | RF-Text | 5172 | -0.156901 | -26.539 | $1.32 \times 10^{-145}$ | $5.81 \times 10^{-144}$ |
| AMZN | Stacking | 5172 | -0.225694 | -34.114 | $3.07 \times 10^{-230}$ | $1.22 \times 10^{-227}$ |
| AMZN | XGB-Full | 5172 | -0.179197 | -29.070 | $3.16 \times 10^{-172}$ | $3.13 \times 10^{-170}$ |
| AMZN | XGB-Text | 5172 | -0.151392 | -25.847 | $1.20 \times 10^{-138}$ | $3.66 \times 10^{-137}$ |
| JNJ | Elastic Net-Text | 5172 | -0.000800 | -20.412 | $3.95 \times 10^{-89}$ | $5.58 \times 10^{-88}$ |
| JNJ | HAR-AAII | 5172 | -0.000114 | -14.859 | $6.18 \times 10^{-49}$ | $3.95 \times 10^{-48}$ |
| JNJ | HAR-Full | 5172 | -0.001267 | -18.499 | $4.92 \times 10^{-74}$ | $4.76 \times 10^{-73}$ |
| JNJ | HAR-Text | 5172 | -0.000200 | -6.229 | $5.06 \times 10^{-10}$ | $1.03 \times 10^{-9}$ |
| JNJ | HAR-VIX | 5172 | -0.000941 | -18.570 | $1.43 \times 10^{-74}$ | $1.39 \times 10^{-73}$ |
| JNJ | Naive | 5172 | 0.000213 | 12.340 | $1.68 \times 10^{-34}$ | $7.52 \times 10^{-34}$ |
| JNJ | RF-Text | 5172 | -0.044908 | -20.797 | $2.50 \times 10^{-92}$ | $3.74 \times 10^{-91}$ |
| JNJ | Stacking | 5172 | -0.265688 | -29.200 | $1.24 \times 10^{-173}$ | $1.40 \times 10^{-171}$ |
| JNJ | XGB-Full | 5172 | -0.044954 | -20.867 | $6.54 \times 10^{-93}$ | $9.96 \times 10^{-92}$ |
| JNJ | XGB-Text | 5172 | -0.044095 | -20.364 | $9.64 \times 10^{-89}$ | $1.32 \times 10^{-87}$ |
| JPM | Elastic Net-Text | 5172 | -0.000320 | -12.933 | $1.12 \times 10^{-37}$ | $5.68 \times 10^{-37}$ |
| JPM | HAR-AAII | 5172 | $-9.85 \times 10^{-6}$ | -4.903 | $9.75 \times 10^{-7}$ | $1.68 \times 10^{-6}$ |
| JPM | HAR-Full | 5172 | -0.000530 | -9.302 | $1.99 \times 10^{-20}$ | $6.21 \times 10^{-20}$ |
| JPM | HAR-Text | 5172 | $-4.66 \times 10^{-5}$ | -7.383 | $1.79 \times 10^{-13}$ | $4.37 \times 10^{-13}$ |
| JPM | HAR-VIX | 5172 | $3.13 \times 10^{-5}$ | 3.153 | 0.0016 | 0.0023 |
| JPM | Naive | 5172 | $-4.85 \times 10^{-5}$ | -2.358 | 0.0184 | 0.0238 |
| JPM | RF-Text | 5172 | -0.121255 | -22.299 | $3.04 \times 10^{-105}$ | $6.17 \times 10^{-104}$ |
| JPM | Stacking | 5172 | -1.862345 | -18.753 | $5.79 \times 10^{-76}$ | $5.95 \times 10^{-75}$ |
| JPM | XGB-Full | 5172 | -0.119268 | -21.876 | $1.55 \times 10^{-101}$ | $2.93 \times 10^{-100}$ |
| JPM | XGB-Text | 5172 | -0.119123 | -21.892 | $1.14 \times 10^{-101}$ | $2.21 \times 10^{-100}$ |
| LLY | Elastic Net-Text | 5172 | $-3.62 \times 10^{-5}$ | -8.304 | $1.27 \times 10^{-16}$ | $3.51 \times 10^{-16}$ |
| LLY | HAR-AAII | 5172 | $-7.06 \times 10^{-6}$ | -4.384 | $1.19 \times 10^{-5}$ | $1.92 \times 10^{-5}$ |
| LLY | HAR-Full | 5172 | -0.000182 | -8.286 | $1.48 \times 10^{-16}$ | $4.08 \times 10^{-16}$ |
| LLY | HAR-Text | 5172 | -0.000123 | -6.772 | $1.41 \times 10^{-11}$ | $3.14 \times 10^{-11}$ |
| LLY | HAR-VIX | 5172 | $5.81 \times 10^{-6}$ | 1.743 | 0.0814 | 0.0993 |
| LLY | Naive | 5172 | $-3.46 \times 10^{-6}$ | -0.782 | 0.4343 | 0.4935 |
| LLY | RF-Text | 5172 | -0.006306 | -14.824 | $1.02 \times 10^{-48}$ | $6.41 \times 10^{-48}$ |
| LLY | Stacking | 5172 | -0.343629 | -27.924 | $5.96 \times 10^{-160}$ | $4.72 \times 10^{-158}$ |
| LLY | XGB-Full | 5172 | -0.006879 | -14.891 | $3.88 \times 10^{-49}$ | $2.56 \times 10^{-48}$ |
| LLY | XGB-Text | 5172 | -0.009576 | -17.137 | $4.51 \times 10^{-64}$ | $4.11 \times 10^{-63}$ |
| MA | Elastic Net-Text | 3565 | -0.000251 | -9.421 | $7.76 \times 10^{-21}$ | $2.45 \times 10^{-20}$ |
| MA | HAR-AAII | 3565 | $-4.76 \times 10^{-5}$ | -10.389 | $6.32 \times 10^{-25}$ | $2.21 \times 10^{-24}$ |
| MA | HAR-Full | 3565 | -0.000696 | -12.626 | $8.73 \times 10^{-36}$ | $4.14 \times 10^{-35}$ |
| MA | HAR-Text | 3565 | -0.000313 | -13.691 | $1.27 \times 10^{-41}$ | $7.13 \times 10^{-41}$ |
| MA | HAR-VIX | 3565 | -0.000173 | -9.677 | $7.01 \times 10^{-22}$ | $2.28 \times 10^{-21}$ |
| MA | Naive | 3565 | $9.76 \times 10^{-6}$ | 0.694 | 0.4875 | 0.5499 |
| MA | RF-Text | 3565 | -0.039920 | -26.322 | $1.08 \times 10^{-139}$ | $3.72 \times 10^{-138}$ |
| MA | Stacking | 3565 | -0.275257 | -26.309 | $1.47 \times 10^{-139}$ | $4.84 \times 10^{-138}$ |
| MA | XGB-Full | 3565 | -0.029953 | -22.206 | $1.98 \times 10^{-102}$ | $3.91 \times 10^{-101}$ |
| MA | XGB-Text | 3565 | -0.030502 | -22.742 | $5.01 \times 10^{-107}$ | $1.07 \times 10^{-105}$ |
| UNH | Elastic Net-Text | 5172 | -0.000103 | -9.520 | $2.58 \times 10^{-21}$ | $8.24 \times 10^{-21}$ |
| UNH | HAR-AAII | 5172 | $-6.52 \times 10^{-6}$ | -9.624 | $9.64 \times 10^{-22}$ | $3.12 \times 10^{-21}$ |

| Stock | Model | $n$ | $\Delta$ | Statistic | $p$ | $q^{global}$ |
|---|---|---|---|---|---|---|
| UNH | HAR-Full | 5172 | -0.000234 | -13.167 | $5.71 \times 10^{-39}$ | $2.97 \times 10^{-38}$ |
| UNH | HAR-Text | 5172 | $-9.68 \times 10^{-5}$ | -12.876 | $2.30 \times 10^{-37}$ | $1.15 \times 10^{-36}$ |
| UNH | HAR-VIX | 5172 | $6.82 \times 10^{-6}$ | 1.888 | 0.0590 | 0.0730 |
| UNH | Naive | 5172 | $-2.48 \times 10^{-5}$ | -4.456 | $8.51 \times 10^{-6}$ | $1.40 \times 10^{-5}$ |
| UNH | RF-Text | 5172 | -0.069190 | -14.220 | $4.81 \times 10^{-45}$ | $2.91 \times 10^{-44}$ |
| UNH | Stacking | 5172 | -0.468932 | -20.667 | $3.04 \times 10^{-91}$ | $4.46 \times 10^{-90}$ |
| UNH | XGB-Full | 5172 | -0.062967 | -13.199 | $3.78 \times 10^{-39}$ | $1.98 \times 10^{-38}$ |
| UNH | XGB-Text | 5172 | -0.062565 | -13.068 | $2.02 \times 10^{-38}$ | $1.05 \times 10^{-37}$ |
| WMT | Elastic Net-Text | 5171 | $-8.50 \times 10^{-5}$ | -5.872 | $4.59 \times 10^{-9}$ | $8.92 \times 10^{-9}$ |
| WMT | HAR-AAII | 5171 | $-1.49 \times 10^{-5}$ | -10.326 | $9.36 \times 10^{-25}$ | $3.24 \times 10^{-24}$ |
| WMT | HAR-Full | 5171 | $-1.61 \times 10^{-5}$ | -3.794 | 0.0002 | 0.0002 |
| WMT | HAR-Text | 5171 | $-4.87 \times 10^{-5}$ | -11.692 | $3.44 \times 10^{-31}$ | $1.45 \times 10^{-30}$ |
| WMT | HAR-VIX | 5171 | $8.58 \times 10^{-6}$ | 2.701 | 0.0069 | 0.0093 |
| WMT | Naive | 5171 | $-2.09 \times 10^{-5}$ | -2.427 | 0.0153 | 0.0199 |
| WMT | RF-Text | 5171 | -0.044570 | -21.146 | $2.91 \times 10^{-95}$ | $4.52 \times 10^{-94}$ |
| WMT | Stacking | 5171 | -0.274440 | -25.847 | $1.20 \times 10^{-138}$ | $3.66 \times 10^{-137}$ |
| WMT | XGB-Full | 5171 | -0.043164 | -20.402 | $4.78 \times 10^{-89}$ | $6.64 \times 10^{-88}$ |
| WMT | XGB-Text | 5171 | -0.043088 | -20.251 | $8.17 \times 10^{-88}$ | $1.10 \times 10^{-86}$ |
| XOM | Elastic Net-Text | 5172 | $-7.39 \times 10^{-5}$ | -7.411 | $1.46 \times 10^{-13}$ | $3.59 \times 10^{-13}$ |
| XOM | HAR-AAII | 5172 | $-1.57 \times 10^{-7}$ | -0.374 | 0.7085 | 0.7826 |
| XOM | HAR-Full | 5172 | $-6.76 \times 10^{-6}$ | -1.654 | 0.0983 | 0.1190 |
| XOM | HAR-Text | 5172 | $-1.58 \times 10^{-5}$ | -6.540 | $6.75 \times 10^{-11}$ | $1.44 \times 10^{-10}$ |
| XOM | HAR-VIX | 5172 | $1.31 \times 10^{-5}$ | 3.647 | 0.0003 | 0.0004 |
| XOM | Naive | 5172 | $-6.65 \times 10^{-5}$ | -3.504 | 0.0005 | 0.0007 |
| XOM | RF-Text | 5172 | -0.039488 | -26.635 | $1.39 \times 10^{-146}$ | $6.49 \times 10^{-145}$ |
| XOM | Stacking | 5172 | -0.551977 | -27.897 | $1.16 \times 10^{-159}$ | $8.36 \times 10^{-158}$ |
| XOM | XGB-Full | 5172 | -0.038813 | -26.430 | $1.69 \times 10^{-144}$ | $7.06 \times 10^{-143}$ |
| XOM | XGB-Text | 5172 | -0.038507 | -26.356 | $9.54 \times 10^{-144}$ | $3.78 \times 10^{-142}$ |

Table C.2: DM, QLIKE, $h = 5$: expanding window

| Stock | Model | $n$ | $\Delta$ | Statistic | $p$ | $q^{global}$ |
|---|---|---|---|---|---|---|
| AAPL | Elastic Net-Text | 5160 | -0.002636 | -9.452 | $4.92 \times 10^{-21}$ | $1.56 \times 10^{-20}$ |
| AAPL | HAR-AAII | 5160 | -0.000372 | -8.229 | $2.37 \times 10^{-16}$ | $6.44 \times 10^{-16}$ |
| AAPL | HAR-Full | 5160 | -0.007296 | -8.983 | $3.64 \times 10^{-19}$ | $1.09 \times 10^{-18}$ |
| AAPL | HAR-Text | 5160 | -0.002466 | -8.954 | $4.72 \times 10^{-19}$ | $1.41 \times 10^{-18}$ |
| AAPL | HAR-VIX | 5160 | -0.001154 | -9.706 | $4.39 \times 10^{-22}$ | $1.44 \times 10^{-21}$ |
| AAPL | Naive | 5160 | 0.000120 | 5.080 | $3.91 \times 10^{-7}$ | $6.89 \times 10^{-7}$ |
| AAPL | RF-Text | 5160 | -0.037798 | -15.363 | $4.07 \times 10^{-52}$ | $2.90 \times 10^{-51}$ |
| AAPL | Stacking | 5160 | -0.217876 | -19.396 | $6.02 \times 10^{-81}$ | $7.01 \times 10^{-80}$ |
| AAPL | XGB-Full | 5160 | -0.040506 | -15.725 | $1.85 \times 10^{-54}$ | $1.48 \times 10^{-53}$ |
| AAPL | XGB-Text | 5160 | -0.040968 | -16.029 | $1.81 \times 10^{-56}$ | $1.50 \times 10^{-55}$ |
| ABBV | Elastic Net-Text | 1891 | -0.000678 | -7.407 | $1.94 \times 10^{-13}$ | $4.71 \times 10^{-13}$ |
| ABBV | HAR-AAII | 1891 | $-5.46 \times 10^{-5}$ | -4.404 | $1.12 \times 10^{-5}$ | $1.82 \times 10^{-5}$ |
| ABBV | HAR-Full | 1891 | -0.000824 | -7.245 | $6.28 \times 10^{-13}$ | $1.47 \times 10^{-12}$ |
| ABBV | HAR-Text | 1891 | -0.000141 | -6.679 | $3.15 \times 10^{-11}$ | $6.87 \times 10^{-11}$ |
| ABBV | HAR-VIX | 1891 | $-8.89 \times 10^{-6}$ | -0.187 | 0.8520 | 0.9282 |
| ABBV | Naive | 1891 | $-8.72 \times 10^{-5}$ | -2.163 | 0.0306 | 0.0389 |
| ABBV | RF-Text | 1891 | -0.035780 | -9.310 | $3.43 \times 10^{-20}$ | $1.05 \times 10^{-19}$ |

| Stock | Model | $n$ | $\Delta$ | Statistic | $p$ | $q^{global}$ |
|---|---|---|---|---|---|---|
| ABBV | Stacking | 1891 | -0.088359 | -16.127 | $6.44 \times 10^{-55}$ | $5.20 \times 10^{-54}$ |
| ABBV | XGB-Full | 1891 | -0.039297 | -9.660 | $1.39 \times 10^{-21}$ | $4.47 \times 10^{-21}$ |
| ABBV | XGB-Text | 1891 | -0.042089 | -10.267 | $4.20 \times 10^{-24}$ | $1.41 \times 10^{-23}$ |
| ABT | Elastic Net-Text | 5160 | -0.000159 | -3.376 | 0.0007 | 0.0011 |
| ABT | HAR-AAII | 5160 | $-8.75 \times 10^{-5}$ | -4.964 | $7.12 \times 10^{-7}$ | $1.24 \times 10^{-6}$ |
| ABT | HAR-Full | 5160 | -0.000245 | -5.887 | $4.19 \times 10^{-9}$ | $8.21 \times 10^{-9}$ |
| ABT | HAR-Text | 5160 | -0.000604 | -7.793 | $7.86 \times 10^{-15}$ | $1.99 \times 10^{-14}$ |
| ABT | HAR-VIX | 5160 | $2.86 \times 10^{-5}$ | 1.462 | 0.1437 | 0.1699 |
| ABT | Naive | 5160 | $-6.23 \times 10^{-5}$ | -1.221 | 0.2221 | 0.2572 |
| ABT | RF-Text | 5160 | -0.020457 | -11.475 | $4.09 \times 10^{-30}$ | $1.65 \times 10^{-29}$ |
| ABT | Stacking | 5160 | -0.240654 | -15.439 | $1.32 \times 10^{-52}$ | $9.70 \times 10^{-52}$ |
| ABT | XGB-Full | 5160 | -0.020866 | -11.164 | $1.30 \times 10^{-28}$ | $5.02 \times 10^{-28}$ |
| ABT | XGB-Text | 5160 | -0.022153 | -10.944 | $1.42 \times 10^{-27}$ | $5.32 \times 10^{-27}$ |
| AMZN | Elastic Net-Text | 5160 | -0.000238 | -8.766 | $2.47 \times 10^{-18}$ | $7.16 \times 10^{-18}$ |
| AMZN | HAR-AAII | 5160 | -0.002611 | -11.206 | $8.22 \times 10^{-29}$ | $3.19 \times 10^{-28}$ |
| AMZN | HAR-Full | 5160 | -0.014554 | -8.027 | $1.22 \times 10^{-15}$ | $3.19 \times 10^{-15}$ |
| AMZN | HAR-Text | 5160 | -0.000529 | -6.337 | $2.54 \times 10^{-10}$ | $5.30 \times 10^{-10}$ |
| AMZN | HAR-VIX | 5160 | -0.000278 | -8.613 | $9.34 \times 10^{-18}$ | $2.68 \times 10^{-17}$ |
| AMZN | Naive | 5160 | $5.41 \times 10^{-5}$ | 2.568 | 0.0102 | 0.0135 |
| AMZN | RF-Text | 5160 | -0.156276 | -15.427 | $1.58 \times 10^{-52}$ | $1.14 \times 10^{-51}$ |
| AMZN | Stacking | 5160 | -0.234569 | -20.431 | $2.80 \times 10^{-89}$ | $4.04 \times 10^{-88}$ |
| AMZN | XGB-Full | 5160 | -0.268334 | -23.134 | $9.64 \times 10^{-113}$ | $2.25 \times 10^{-111}$ |
| AMZN | XGB-Text | 5160 | -0.172981 | -16.357 | $1.14 \times 10^{-58}$ | $1.00 \times 10^{-57}$ |
| JNJ | Elastic Net-Text | 5160 | -0.001573 | -12.340 | $1.69 \times 10^{-34}$ | $7.52 \times 10^{-34}$ |
| JNJ | HAR-AAII | 5160 | -0.000394 | -8.903 | $7.42 \times 10^{-19}$ | $2.19 \times 10^{-18}$ |
| JNJ | HAR-Full | 5160 | -0.002401 | -10.776 | $8.61 \times 10^{-27}$ | $3.16 \times 10^{-26}$ |
| JNJ | HAR-Text | 5160 | -0.000496 | -3.928 | $8.68 \times 10^{-5}$ | 0.0001 |
| JNJ | HAR-VIX | 5160 | -0.002579 | -11.030 | $5.59 \times 10^{-28}$ | $2.12 \times 10^{-27}$ |
| JNJ | Naive | 5160 | 0.000858 | 8.193 | $3.18 \times 10^{-16}$ | $8.51 \times 10^{-16}$ |
| JNJ | RF-Text | 5160 | -0.046956 | -12.338 | $1.73 \times 10^{-34}$ | $7.68 \times 10^{-34}$ |
| JNJ | Stacking | 5160 | -0.265900 | -16.870 | $3.36 \times 10^{-62}$ | $3.03 \times 10^{-61}$ |
| JNJ | XGB-Full | 5160 | -0.046305 | -12.352 | $1.45 \times 10^{-34}$ | $6.57 \times 10^{-34}$ |
| JNJ | XGB-Text | 5160 | -0.046004 | -12.174 | $1.23 \times 10^{-33}$ | $5.41 \times 10^{-33}$ |
| JPM | Elastic Net-Text | 5160 | -0.000521 | -6.443 | $1.27 \times 10^{-10}$ | $2.71 \times 10^{-10}$ |
| JPM | HAR-AAII | 5160 | $-2.41 \times 10^{-5}$ | -1.699 | 0.0894 | 0.1089 |
| JPM | HAR-Full | 5160 | -0.001890 | -5.175 | $2.36 \times 10^{-7}$ | $4.21 \times 10^{-7}$ |
| JPM | HAR-Text | 5160 | -0.000409 | -5.883 | $4.29 \times 10^{-9}$ | $8.39 \times 10^{-9}$ |
| JPM | HAR-VIX | 5160 | 0.000110 | 3.894 | $9.98 \times 10^{-5}$ | 0.0002 |
| JPM | Naive | 5160 | -0.000155 | -1.459 | 0.1446 | 0.1707 |
| JPM | RF-Text | 5160 | -0.123528 | -12.995 | $5.14 \times 10^{-38}$ | $2.61 \times 10^{-37}$ |
| JPM | Stacking | 5160 | -1.836307 | -10.908 | $2.10 \times 10^{-27}$ | $7.84 \times 10^{-27}$ |
| JPM | XGB-Full | 5160 | -0.122145 | -12.782 | $7.49 \times 10^{-37}$ | $3.62 \times 10^{-36}$ |
| JPM | XGB-Text | 5160 | -0.122629 | -12.783 | $7.39 \times 10^{-37}$ | $3.59 \times 10^{-36}$ |
| LLY | Elastic Net-Text | 5160 | $-7.41 \times 10^{-5}$ | -3.845 | 0.0001 | 0.0002 |
| LLY | HAR-AAII | 5160 | $-2.54 \times 10^{-5}$ | -3.378 | 0.0007 | 0.0011 |
| LLY | HAR-Full | 5160 | -0.000848 | -5.908 | $3.69 \times 10^{-9}$ | $7.29 \times 10^{-9}$ |
| LLY | HAR-Text | 5160 | -0.000491 | -4.303 | $1.71 \times 10^{-5}$ | $2.73 \times 10^{-5}$ |
| LLY | HAR-VIX | 5160 | $2.03 \times 10^{-5}$ | 1.291 | 0.1967 | 0.2294 |
| LLY | Naive | 5160 | $-1.81 \times 10^{-5}$ | -0.853 | 0.3937 | 0.4480 |
| LLY | RF-Text | 5160 | -0.007882 | -9.776 | $2.24 \times 10^{-22}$ | $7.39 \times 10^{-22}$ |
| LLY | Stacking | 5160 | -0.347130 | -16.210 | $1.11 \times 10^{-57}$ | $9.57 \times 10^{-57}$ |

| Stock | Model | $n$ | $\Delta$ | Statistic | $p$ | $q^{global}$ |
|---|---|---|---|---|---|---|
| LLY | XGB-Full | 5160 | -0.009104 | -10.323 | $9.60 \times 10^{-25}$ | $3.31 \times 10^{-24}$ |
| LLY | XGB-Text | 5160 | -0.010290 | -10.461 | $2.33 \times 10^{-25}$ | $8.31 \times 10^{-25}$ |
| MA | Elastic Net-Text | 3553 | -0.000573 | -6.327 | $2.82 \times 10^{-10}$ | $5.86 \times 10^{-10}$ |
| MA | HAR-AAII | 3553 | -0.000135 | -6.864 | $7.89 \times 10^{-12}$ | $1.78 \times 10^{-11}$ |
| MA | HAR-Full | 3553 | -0.001495 | -8.147 | $5.10 \times 10^{-16}$ | $1.35 \times 10^{-15}$ |
| MA | HAR-Text | 3553 | -0.000913 | -8.210 | $3.06 \times 10^{-16}$ | $8.20 \times 10^{-16}$ |
| MA | HAR-VIX | 3553 | -0.000329 | -5.836 | $5.82 \times 10^{-9}$ | $1.13 \times 10^{-8}$ |
| MA | Naive | 3553 | $3.37 \times 10^{-5}$ | 0.449 | 0.6532 | 0.7277 |
| MA | RF-Text | 3553 | -0.034184 | -13.046 | $4.96 \times 10^{-38}$ | $2.54 \times 10^{-37}$ |
| MA | Stacking | 3553 | -0.272290 | -15.055 | $1.07 \times 10^{-49}$ | $7.34 \times 10^{-49}$ |
| MA | XGB-Full | 3553 | -0.032498 | -13.590 | $4.81 \times 10^{-41}$ | $2.61 \times 10^{-40}$ |
| MA | XGB-Text | 3553 | -0.031952 | -13.499 | $1.55 \times 10^{-40}$ | $8.37 \times 10^{-40}$ |
| UNH | Elastic Net-Text | 5160 | -0.000166 | -5.395 | $7.17 \times 10^{-8}$ | $1.31 \times 10^{-7}$ |
| UNH | HAR-AAII | 5160 | $-5.15 \times 10^{-6}$ | -2.020 | 0.0434 | 0.0543 |
| UNH | HAR-Full | 5160 | -0.000336 | -7.246 | $4.93 \times 10^{-13}$ | $1.17 \times 10^{-12}$ |
| UNH | HAR-Text | 5160 | -0.000241 | -7.552 | $5.06 \times 10^{-14}$ | $1.25 \times 10^{-13}$ |
| UNH | HAR-VIX | 5160 | $1.86 \times 10^{-5}$ | 1.782 | 0.0748 | 0.0916 |
| UNH | Naive | 5160 | $-7.96 \times 10^{-5}$ | -2.765 | 0.0057 | 0.0077 |
| UNH | RF-Text | 5160 | -0.069533 | -8.212 | $2.73 \times 10^{-16}$ | $7.34 \times 10^{-16}$ |
| UNH | Stacking | 5160 | -0.467005 | -11.928 | $2.25 \times 10^{-32}$ | $9.61 \times 10^{-32}$ |
| UNH | XGB-Full | 5160 | -0.064969 | -7.717 | $1.42 \times 10^{-14}$ | $3.57 \times 10^{-14}$ |
| UNH | XGB-Text | 5160 | -0.064610 | -7.700 | $1.63 \times 10^{-14}$ | $4.06 \times 10^{-14}$ |
| WMT | Elastic Net-Text | 5159 | -0.000129 | -3.174 | 0.0015 | 0.0021 |
| WMT | HAR-AAII | 5159 | $-6.34 \times 10^{-5}$ | -7.251 | $4.74 \times 10^{-13}$ | $1.13 \times 10^{-12}$ |
| WMT | HAR-Full | 5159 | -0.000375 | -6.439 | $1.31 \times 10^{-10}$ | $2.78 \times 10^{-10}$ |
| WMT | HAR-Text | 5159 | -0.000250 | -7.166 | $8.80 \times 10^{-13}$ | $2.04 \times 10^{-12}$ |
| WMT | HAR-VIX | 5159 | $3.85 \times 10^{-5}$ | 2.374 | 0.0176 | 0.0228 |
| WMT | Naive | 5159 | $-6.80 \times 10^{-5}$ | -1.647 | 0.0997 | 0.1205 |
| WMT | RF-Text | 5159 | -0.045632 | -12.382 | $1.02 \times 10^{-34}$ | $4.62 \times 10^{-34}$ |
| WMT | Stacking | 5159 | -0.273987 | -14.959 | $1.48 \times 10^{-49}$ | $1.01 \times 10^{-48}$ |
| WMT | XGB-Full | 5159 | -0.045782 | -12.111 | $2.61 \times 10^{-33}$ | $1.13 \times 10^{-32}$ |
| WMT | XGB-Text | 5159 | -0.044416 | -11.903 | $2.99 \times 10^{-32}$ | $1.28 \times 10^{-31}$ |
| XOM | Elastic Net-Text | 5160 | $-2.86 \times 10^{-5}$ | -3.820 | 0.0001 | 0.0002 |
| XOM | HAR-AAII | 5160 | $9.94 \times 10^{-7}$ | 0.352 | 0.7246 | 0.7982 |
| XOM | HAR-Full | 5160 | $8.10 \times 10^{-6}$ | 0.632 | 0.5276 | 0.5919 |
| XOM | HAR-Text | 5160 | $-2.16 \times 10^{-5}$ | -2.680 | 0.0074 | 0.0098 |
| XOM | HAR-VIX | 5160 | $3.48 \times 10^{-5}$ | 2.986 | 0.0028 | 0.0039 |
| XOM | Naive | 5160 | -0.000216 | -1.981 | 0.0477 | 0.0595 |
| XOM | RF-Text | 5160 | -0.040149 | -15.569 | $1.91 \times 10^{-53}$ | $1.47 \times 10^{-52}$ |
| XOM | Stacking | 5160 | -0.557770 | -16.051 | $1.30 \times 10^{-56}$ | $1.09 \times 10^{-55}$ |
| XOM | XGB-Full | 5160 | -0.041059 | -15.683 | $3.49 \times 10^{-54}$ | $2.76 \times 10^{-53}$ |
| XOM | XGB-Text | 5160 | -0.039676 | -15.316 | $8.13 \times 10^{-52}$ | $5.75 \times 10^{-51}$ |

Table C.3: DM, QLIKE, $h = 22$: expanding window

| Stock | Model | $n$ | $\Delta$ | Statistic | $p$ | $q^{global}$ |
|---|---|---|---|---|---|---|
| AAPL | Elastic Net-Text | 5109 | -0.013167 | -4.779 | $1.81 \times 10^{-6}$ | $3.09 \times 10^{-6}$ |
| AAPL | HAR-AAII | 5109 | -0.002319 | -4.477 | $7.73 \times 10^{-6}$ | $1.28 \times 10^{-5}$ |
| AAPL | HAR-Full | 5109 | -0.017063 | -5.889 | $4.13 \times 10^{-9}$ | $8.12 \times 10^{-9}$ |

| Stock | Model | $n$ | $\Delta$ | Statistic | $p$ | $q^{global}$ |
|---|---|---|---|---|---|---|
| AAPL | HAR-Text | 5109 | -0.012756 | -4.745 | $2.14 \times 10^{-6}$ | $3.65 \times 10^{-6}$ |
| AAPL | HAR-VIX | 5109 | -0.006462 | -5.760 | $8.93 \times 10^{-9}$ | $1.71 \times 10^{-8}$ |
| AAPL | Naive | 5109 | 0.003709 | 4.738 | $2.22 \times 10^{-6}$ | $3.77 \times 10^{-6}$ |
| AAPL | RF-Text | 5109 | -0.046223 | -9.931 | $4.95 \times 10^{-23}$ | $1.65 \times 10^{-22}$ |
| AAPL | Stacking | 5109 | -0.237732 | -9.293 | $2.16 \times 10^{-20}$ | $6.72 \times 10^{-20}$ |
| AAPL | XGB-Full | 5109 | -0.244230 | -18.897 | $4.90 \times 10^{-77}$ | $5.31 \times 10^{-76}$ |
| AAPL | XGB-Text | 5109 | -0.045592 | -9.252 | $3.16 \times 10^{-20}$ | $9.70 \times 10^{-20}$ |
| ABBV | Elastic Net-Text | 1840 | -0.001997 | -5.489 | $4.61 \times 10^{-8}$ | $8.42 \times 10^{-8}$ |
| ABBV | HAR-AAII | 1840 | -0.001366 | -1.309 | 0.1906 | 0.2226 |
| ABBV | HAR-Full | 1840 | -0.004245 | -4.743 | $2.27 \times 10^{-6}$ | $3.84 \times 10^{-6}$ |
| ABBV | HAR-Text | 1840 | -0.000711 | -3.725 | 0.0002 | 0.0003 |
| ABBV | HAR-VIX | 1840 | -0.000377 | -0.957 | 0.3389 | 0.3873 |
| ABBV | Naive | 1840 | -0.000460 | -1.317 | 0.1880 | 0.2200 |
| ABBV | RF-Text | 1840 | -0.056040 | -5.651 | $1.85 \times 10^{-8}$ | $3.44 \times 10^{-8}$ |
| ABBV | Stacking | 1840 | -0.090221 | -8.302 | $1.96 \times 10^{-16}$ | $5.37 \times 10^{-16}$ |
| ABBV | XGB-Full | 1840 | -0.065192 | -6.583 | $5.98 \times 10^{-11}$ | $1.28 \times 10^{-10}$ |
| ABBV | XGB-Text | 1840 | -0.060514 | -6.110 | $1.21 \times 10^{-9}$ | $2.42 \times 10^{-9}$ |
| ABT | Elastic Net-Text | 5109 | -0.000973 | -4.300 | $1.74 \times 10^{-5}$ | $2.77 \times 10^{-5}$ |
| ABT | HAR-AAII | 5109 | -0.000396 | -2.082 | 0.0374 | 0.0472 |
| ABT | HAR-Full | 5109 | -0.001706 | -2.738 | 0.0062 | 0.0083 |
| ABT | HAR-Text | 5109 | -0.003484 | -4.447 | $8.88 \times 10^{-6}$ | $1.45 \times 10^{-5}$ |
| ABT | HAR-VIX | 5109 | 0.000213 | 1.073 | 0.2834 | 0.3257 |
| ABT | Naive | 5109 | $-4.67 \times 10^{-5}$ | -0.126 | 0.8999 | 0.9777 |
| ABT | RF-Text | 5109 | -0.029474 | -5.936 | $3.11 \times 10^{-9}$ | $6.15 \times 10^{-9}$ |
| ABT | Stacking | 5109 | -0.240105 | -7.745 | $1.14 \times 10^{-14}$ | $2.88 \times 10^{-14}$ |
| ABT | XGB-Full | 5109 | -0.026057 | -6.104 | $1.11 \times 10^{-9}$ | $2.22 \times 10^{-9}$ |
| ABT | XGB-Text | 5109 | -0.031814 | -6.025 | $1.81 \times 10^{-9}$ | $3.61 \times 10^{-9}$ |
| AMZN | Elastic Net-Text | 5109 | -0.012524 | -5.438 | $5.65 \times 10^{-8}$ | $1.03 \times 10^{-7}$ |
| AMZN | HAR-AAII | 5109 | -0.005361 | -4.909 | $9.42 \times 10^{-7}$ | $1.63 \times 10^{-6}$ |
| AMZN | HAR-Full | 5109 | -0.019073 | -3.963 | $7.51 \times 10^{-5}$ | 0.0001 |
| AMZN | HAR-Text | 5109 | -0.011514 | -4.968 | $6.97 \times 10^{-7}$ | $1.22 \times 10^{-6}$ |
| AMZN | HAR-VIX | 5109 | $-2.00 \times 10^{-5}$ | -0.138 | 0.8903 | 0.9685 |
| AMZN | Naive | 5109 | -0.000262 | -1.565 | 0.1176 | 0.1410 |
| AMZN | RF-Text | 5109 | -0.162398 | -8.149 | $4.58 \times 10^{-16}$ | $1.22 \times 10^{-15}$ |
| AMZN | Stacking | 5109 | -0.281791 | -11.433 | $6.61 \times 10^{-30}$ | $2.64 \times 10^{-29}$ |
| AMZN | XGB-Full | 5109 | -0.255420 | -11.411 | $8.47 \times 10^{-30}$ | $3.35 \times 10^{-29}$ |
| AMZN | XGB-Text | 5109 | -0.341634 | -14.385 | $5.06 \times 10^{-46}$ | $3.11 \times 10^{-45}$ |
| JNJ | Elastic Net-Text | 5109 | 0.001263 | 2.213 | 0.0269 | 0.0343 |
| JNJ | HAR-AAII | 5109 | -0.020136 | -5.632 | $1.87 \times 10^{-8}$ | $3.48 \times 10^{-8}$ |
| JNJ | HAR-Full | 5109 | -0.006377 | -3.017 | 0.0026 | 0.0035 |
| JNJ | HAR-Text | 5109 | 0.008653 | 5.099 | $3.55 \times 10^{-7}$ | $6.26 \times 10^{-7}$ |
| JNJ | HAR-VIX | 5109 | -0.017183 | -6.136 | $9.09 \times 10^{-10}$ | $1.84 \times 10^{-9}$ |
| JNJ | Naive | 5109 | 0.012795 | 5.315 | $1.11 \times 10^{-7}$ | $2.00 \times 10^{-7}$ |
| JNJ | RF-Text | 5109 | -0.045159 | -7.388 | $1.73 \times 10^{-13}$ | $4.25 \times 10^{-13}$ |
| JNJ | Stacking | 5109 | -0.258156 | -8.360 | $8.02 \times 10^{-17}$ | $2.24 \times 10^{-16}$ |
| JNJ | XGB-Full | 5109 | -0.043315 | -6.872 | $7.09 \times 10^{-12}$ | $1.61 \times 10^{-11}$ |
| JNJ | XGB-Text | 5109 | -0.045693 | -6.868 | $7.28 \times 10^{-12}$ | $1.64 \times 10^{-11}$ |
| JPM | Elastic Net-Text | 5109 | -0.008040 | -7.335 | $2.56 \times 10^{-13}$ | $6.17 \times 10^{-13}$ |
| JPM | HAR-AAII | 5109 | -0.002077 | -2.880 | 0.0040 | 0.0055 |
| JPM | HAR-Full | 5109 | 0.003573 | 2.619 | 0.0088 | 0.0117 |
| JPM | HAR-Text | 5109 | 0.003374 | 4.353 | $1.37 \times 10^{-5}$ | $2.20 \times 10^{-5}$ |

| Stock | Model | $n$ | $\Delta$ | Statistic | $p$ | $q^{global}$ |
|---|---|---|---|---|---|---|
| JPM | HAR-VIX | 5109 | 0.001238 | 4.154 | $3.32 \times 10^{-5}$ | $5.22 \times 10^{-5}$ |
| JPM | Naive | 5109 | 0.006692 | 4.024 | $5.81 \times 10^{-5}$ | $9.01 \times 10^{-5}$ |
| JPM | RF-Text | 5109 | -0.125840 | -6.919 | $5.11 \times 10^{-12}$ | $1.17 \times 10^{-11}$ |
| JPM | Stacking | 5109 | -1.751174 | -5.677 | $1.45 \times 10^{-8}$ | $2.72 \times 10^{-8}$ |
| JPM | XGB-Full | 5109 | -0.123999 | -6.839 | $8.90 \times 10^{-12}$ | $2.00 \times 10^{-11}$ |
| JPM | XGB-Text | 5109 | -0.131240 | -6.868 | $7.28 \times 10^{-12}$ | $1.64 \times 10^{-11}$ |
| LLY | Elastic Net-Text | 5109 | -0.000826 | -2.580 | 0.0099 | 0.0131 |
| LLY | HAR-AAII | 5109 | -0.000792 | -2.857 | 0.0043 | 0.0058 |
| LLY | HAR-Full | 5109 | -0.009827 | -3.684 | 0.0002 | 0.0003 |
| LLY | HAR-Text | 5109 | -0.001228 | -3.239 | 0.0012 | 0.0017 |
| LLY | HAR-VIX | 5109 | 0.000419 | 2.507 | 0.0122 | 0.0160 |
| LLY | Naive | 5109 | $6.87 \times 10^{-5}$ | 0.308 | 0.7582 | 0.8317 |
| LLY | RF-Text | 5109 | -0.011770 | -5.358 | $8.76 \times 10^{-8}$ | $1.58 \times 10^{-7}$ |
| LLY | Stacking | 5109 | -0.355208 | -8.333 | $1.00 \times 10^{-16}$ | $2.79 \times 10^{-16}$ |
| LLY | XGB-Full | 5109 | -0.011105 | -4.942 | $8.00 \times 10^{-7}$ | $1.39 \times 10^{-6}$ |
| LLY | XGB-Text | 5109 | -0.014809 | -5.681 | $1.41 \times 10^{-8}$ | $2.66 \times 10^{-8}$ |
| MA | Elastic Net-Text | 3502 | -0.004640 | -4.030 | $5.69 \times 10^{-5}$ | $8.83 \times 10^{-5}$ |
| MA | HAR-AAII | 3502 | -0.000131 | -1.231 | 0.2184 | 0.2533 |
| MA | HAR-Full | 3502 | -0.006297 | -4.360 | $1.34 \times 10^{-5}$ | $2.16 \times 10^{-5}$ |
| MA | HAR-Text | 3502 | -0.003150 | -3.618 | 0.0003 | 0.0004 |
| MA | HAR-VIX | 3502 | -0.002262 | -4.037 | $5.52 \times 10^{-5}$ | $8.61 \times 10^{-5}$ |
| MA | Naive | 3502 | 0.000313 | 0.461 | 0.6446 | 0.7190 |
| MA | RF-Text | 3502 | -0.047372 | -7.596 | $3.90 \times 10^{-14}$ | $9.69 \times 10^{-14}$ |
| MA | Stacking | 3502 | -0.244383 | -9.298 | $2.45 \times 10^{-20}$ | $7.55 \times 10^{-20}$ |
| MA | XGB-Full | 3502 | -0.049675 | -6.661 | $3.15 \times 10^{-11}$ | $6.87 \times 10^{-11}$ |
| MA | XGB-Text | 3502 | -0.045609 | -8.019 | $1.44 \times 10^{-15}$ | $3.73 \times 10^{-15}$ |
| UNH | Elastic Net-Text | 5109 | -0.000697 | -3.973 | $7.20 \times 10^{-5}$ | 0.0001 |
| UNH | HAR-AAII | 5109 | $-5.25 \times 10^{-5}$ | -1.467 | 0.1423 | 0.1685 |
| UNH | HAR-Full | 5109 | -0.000843 | -4.612 | $4.08 \times 10^{-6}$ | $6.81 \times 10^{-6}$ |
| UNH | HAR-Text | 5109 | -0.000459 | -3.253 | 0.0012 | 0.0016 |
| UNH | HAR-VIX | 5109 | 0.000121 | 1.830 | 0.0672 | 0.0828 |
| UNH | Naive | 5109 | -0.000415 | -1.670 | 0.0950 | 0.1152 |
| UNH | RF-Text | 5109 | -0.071189 | -4.140 | $3.53 \times 10^{-5}$ | $5.53 \times 10^{-5}$ |
| UNH | Stacking | 5109 | -0.453785 | -6.141 | $8.83 \times 10^{-10}$ | $1.79 \times 10^{-9}$ |
| UNH | XGB-Full | 5109 | -0.070740 | -4.203 | $2.68 \times 10^{-5}$ | $4.24 \times 10^{-5}$ |
| UNH | XGB-Text | 5109 | -0.073266 | -4.335 | $1.49 \times 10^{-5}$ | $2.38 \times 10^{-5}$ |
| WMT | Elastic Net-Text | 5108 | $-8.89 \times 10^{-5}$ | -0.497 | 0.6193 | 0.6918 |
| WMT | HAR-AAII | 5108 | -0.000321 | -3.643 | 0.0003 | 0.0004 |
| WMT | HAR-Full | 5108 | -0.007191 | -3.425 | 0.0006 | 0.0009 |
| WMT | HAR-Text | 5108 | -0.001816 | -3.405 | 0.0007 | 0.0010 |
| WMT | HAR-VIX | 5108 | 0.000117 | 0.764 | 0.4451 | 0.5050 |
| WMT | Naive | 5108 | -0.000313 | -1.391 | 0.1642 | 0.1932 |
| WMT | RF-Text | 5108 | -0.054074 | -6.714 | $2.10 \times 10^{-11}$ | $4.64 \times 10^{-11}$ |
| WMT | Stacking | 5108 | -0.269833 | -7.791 | $7.99 \times 10^{-15}$ | $2.02 \times 10^{-14}$ |
| WMT | XGB-Full | 5108 | -0.052820 | -6.600 | $4.53 \times 10^{-11}$ | $9.78 \times 10^{-11}$ |
| WMT | XGB-Text | 5108 | -0.059171 | -6.402 | $1.67 \times 10^{-10}$ | $3.51 \times 10^{-10}$ |
| XOM | Elastic Net-Text | 5109 | -0.000744 | -2.257 | 0.0240 | 0.0309 |
| XOM | HAR-AAII | 5109 | 0.000167 | 1.791 | 0.0733 | 0.0899 |
| XOM | HAR-Full | 5109 | -0.001671 | -2.808 | 0.0050 | 0.0068 |
| XOM | HAR-Text | 5109 | -0.000392 | -1.550 | 0.1212 | 0.1448 |
| XOM | HAR-VIX | 5109 | 0.000324 | 1.985 | 0.0472 | 0.0589 |

| Stock | Model | $n$ | $\Delta$ | Statistic | $p$ | $q^{global}$ |
|---|---|---|---|---|---|---|
| XOM | Naive | 5109 | -0.001242 | -1.357 | 0.1749 | 0.2052 |
| XOM | RF-Text | 5109 | -0.043031 | -8.007 | $1.44 \times 10^{-15}$ | $3.74 \times 10^{-15}$ |
| XOM | Stacking | 5109 | -0.578486 | -8.129 | $5.36 \times 10^{-16}$ | $1.42 \times 10^{-15}$ |
| XOM | XGB-Full | 5109 | -0.044364 | -8.367 | $7.57 \times 10^{-17}$ | $2.12 \times 10^{-16}$ |
| XOM | XGB-Text | 5109 | -0.041597 | -8.080 | $7.99 \times 10^{-16}$ | $2.10 \times 10^{-15}$ |

Table C.4: DM, annualized-volatility squared error, $h = 1$: expanding window

| Stock | Model | $n$ | $\Delta$ | Statistic | $p$ | $q^{global}$ |
|---|---|---|---|---|---|---|
| AAPL | Elastic Net-Text | 5172 | $-6.11 \times 10^{-5}$ | -15.483 | $6.82 \times 10^{-53}$ | $5.14 \times 10^{-52}$ |
| AAPL | HAR-AAII | 5172 | $-8.66 \times 10^{-6}$ | -13.245 | $2.10 \times 10^{-39}$ | $1.11 \times 10^{-38}$ |
| AAPL | HAR-Full | 5172 | -0.000319 | -14.532 | $6.33 \times 10^{-47}$ | $3.92 \times 10^{-46}$ |
| AAPL | HAR-Text | 5172 | $-5.75 \times 10^{-5}$ | -14.120 | $1.90 \times 10^{-44}$ | $1.13 \times 10^{-43}$ |
| AAPL | HAR-VIX | 5172 | $-2.88 \times 10^{-5}$ | -16.722 | $3.57 \times 10^{-61}$ | $3.17 \times 10^{-60}$ |
| AAPL | Naive | 5172 | $1.67 \times 10^{-6}$ | 6.237 | $4.81 \times 10^{-10}$ | $9.84 \times 10^{-10}$ |
| AAPL | RF-Text | 5172 | -0.004504 | -24.356 | $3.88 \times 10^{-124}$ | $1.06 \times 10^{-122}$ |
| AAPL | Stacking | 5172 | -0.011979 | -39.380 | $7.09 \times 10^{-297}$ | $5.61 \times 10^{-294}$ |
| AAPL | XGB-Full | 5172 | -0.004843 | -22.963 | $3.40 \times 10^{-111}$ | $7.69 \times 10^{-110}$ |
| AAPL | XGB-Text | 5172 | -0.004327 | -23.200 | $2.33 \times 10^{-113}$ | $5.77 \times 10^{-112}$ |
| ABBV | Elastic Net-Text | 1903 | $-6.00 \times 10^{-6}$ | -10.545 | $2.64 \times 10^{-25}$ | $9.34 \times 10^{-25}$ |
| ABBV | HAR-AAII | 1903 | $-1.44 \times 10^{-6}$ | -3.350 | 0.0008 | 0.0012 |
| ABBV | HAR-Full | 1903 | $-1.55 \times 10^{-5}$ | -10.288 | $3.39 \times 10^{-24}$ | $1.15 \times 10^{-23}$ |
| ABBV | HAR-Text | 1903 | $-2.34 \times 10^{-6}$ | -8.989 | $5.89 \times 10^{-19}$ | $1.75 \times 10^{-18}$ |
| ABBV | HAR-VIX | 1903 | $-3.35 \times 10^{-7}$ | -0.715 | 0.4745 | 0.5361 |
| ABBV | Naive | 1903 | $-1.24 \times 10^{-6}$ | -2.869 | 0.0042 | 0.0057 |
| ABBV | RF-Text | 1903 | -0.001261 | -14.043 | $1.07 \times 10^{-42}$ | $6.16 \times 10^{-42}$ |
| ABBV | Stacking | 1903 | -0.003763 | -27.027 | $2.02 \times 10^{-136}$ | $5.92 \times 10^{-135}$ |
| ABBV | XGB-Full | 1903 | -0.001408 | -15.147 | $5.11 \times 10^{-49}$ | $3.29 \times 10^{-48}$ |
| ABBV | XGB-Text | 1903 | -0.001427 | -15.206 | $2.29 \times 10^{-49}$ | $1.53 \times 10^{-48}$ |
| ABT | Elastic Net-Text | 5172 | $-1.29 \times 10^{-6}$ | -4.033 | $5.59 \times 10^{-5}$ | $8.69 \times 10^{-5}$ |
| ABT | HAR-AAII | 5172 | $-6.74 \times 10^{-7}$ | -7.847 | $5.15 \times 10^{-15}$ | $1.31 \times 10^{-14}$ |
| ABT | HAR-Full | 5172 | $-3.40 \times 10^{-6}$ | -11.319 | $2.35 \times 10^{-29}$ | $9.25 \times 10^{-29}$ |
| ABT | HAR-Text | 5172 | $-1.61 \times 10^{-6}$ | -13.014 | $4.03 \times 10^{-38}$ | $2.07 \times 10^{-37}$ |
| ABT | HAR-VIX | 5172 | $1.40 \times 10^{-7}$ | 0.965 | 0.3348 | 0.3832 |
| ABT | Naive | 5172 | $-1.21 \times 10^{-6}$ | -2.925 | 0.0035 | 0.0047 |
| ABT | RF-Text | 5172 | -0.000598 | -18.889 | $5.21 \times 10^{-77}$ | $5.57 \times 10^{-76}$ |
| ABT | Stacking | 5172 | -0.006105 | -27.876 | $1.92 \times 10^{-159}$ | $1.26 \times 10^{-157}$ |
| ABT | XGB-Full | 5172 | -0.000707 | -18.139 | $2.47 \times 10^{-71}$ | $2.33 \times 10^{-70}$ |
| ABT | XGB-Text | 5172 | -0.000716 | -18.095 | $5.18 \times 10^{-71}$ | $4.82 \times 10^{-70}$ |
| AMZN | Elastic Net-Text | 5172 | $-8.38 \times 10^{-6}$ | -10.460 | $2.37 \times 10^{-25}$ | $8.40 \times 10^{-25}$ |
| AMZN | HAR-AAII | 5172 | $-5.14 \times 10^{-5}$ | -19.740 | $1.11 \times 10^{-83}$ | $1.35 \times 10^{-82}$ |
| AMZN | HAR-Full | 5172 | -0.000240 | -15.450 | $1.12 \times 10^{-52}$ | $8.28 \times 10^{-52}$ |
| AMZN | HAR-Text | 5172 | $-7.88 \times 10^{-6}$ | -8.896 | $7.88 \times 10^{-19}$ | $2.32 \times 10^{-18}$ |
| AMZN | HAR-VIX | 5172 | $-5.08 \times 10^{-6}$ | -13.777 | $1.95 \times 10^{-42}$ | $1.11 \times 10^{-41}$ |
| AMZN | Naive | 5172 | $3.10 \times 10^{-6}$ | 6.719 | $2.03 \times 10^{-11}$ | $4.49 \times 10^{-11}$ |
| AMZN | RF-Text | 5172 | -0.045605 | -23.370 | $6.51 \times 10^{-115}$ | $1.66 \times 10^{-113}$ |
| AMZN | Stacking | 5172 | -0.027623 | -33.083 | $7.01 \times 10^{-218}$ | $1.39 \times 10^{-215}$ |
| AMZN | XGB-Full | 5172 | -0.051447 | -23.980 | $1.38 \times 10^{-120}$ | $3.64 \times 10^{-119}$ |
| AMZN | XGB-Text | 5172 | -0.044342 | -23.165 | $4.89 \times 10^{-113}$ | $1.17 \times 10^{-111}$ |

| Stock | Model | $n$ | $\Delta$ | Statistic | $p$ | $q^{global}$ |
|---|---|---|---|---|---|---|
| JNJ | Elastic Net-Text | 5172 | $-1.42 \times 10^{-5}$ | -22.433 | $1.93 \times 10^{-106}$ | $4.03 \times 10^{-105}$ |
| JNJ | HAR-AAII | 5172 | $-1.56 \times 10^{-6}$ | -14.938 | $1.98 \times 10^{-49}$ | $1.34 \times 10^{-48}$ |
| JNJ | HAR-Full | 5172 | $-2.52 \times 10^{-5}$ | -19.199 | $2.06 \times 10^{-79}$ | $2.30 \times 10^{-78}$ |
| JNJ | HAR-Text | 5172 | $-5.15 \times 10^{-6}$ | -9.510 | $2.85 \times 10^{-21}$ | $9.07 \times 10^{-21}$ |
| JNJ | HAR-VIX | 5172 | $-1.78 \times 10^{-5}$ | -19.379 | $7.98 \times 10^{-81}$ | $9.16 \times 10^{-80}$ |
| JNJ | Naive | 5172 | $2.90 \times 10^{-6}$ | 8.546 | $1.66 \times 10^{-17}$ | $4.68 \times 10^{-17}$ |
| JNJ | RF-Text | 5172 | -0.001293 | -21.699 | $5.33 \times 10^{-100}$ | $9.59 \times 10^{-99}$ |
| JNJ | Stacking | 5172 | -0.004596 | -31.750 | $2.84 \times 10^{-202}$ | $4.50 \times 10^{-200}$ |
| JNJ | XGB-Full | 5172 | -0.001297 | -21.864 | $1.97 \times 10^{-101}$ | $3.62 \times 10^{-100}$ |
| JNJ | XGB-Text | 5172 | -0.001261 | -21.322 | $9.32 \times 10^{-97}$ | $1.51 \times 10^{-95}$ |
| JPM | Elastic Net-Text | 5172 | $-1.27 \times 10^{-5}$ | -10.253 | $1.97 \times 10^{-24}$ | $6.74 \times 10^{-24}$ |
| JPM | HAR-AAII | 5172 | $-4.41 \times 10^{-7}$ | -4.559 | $5.26 \times 10^{-6}$ | $8.74 \times 10^{-6}$ |
| JPM | HAR-Full | 5172 | $-1.11 \times 10^{-5}$ | -10.587 | $6.30 \times 10^{-26}$ | $2.28 \times 10^{-25}$ |
| JPM | HAR-Text | 5172 | $-4.58 \times 10^{-7}$ | -3.592 | 0.0003 | 0.0005 |
| JPM | HAR-VIX | 5172 | $7.54 \times 10^{-7}$ | 2.173 | 0.0298 | 0.0379 |
| JPM | Naive | 5172 | $-4.94 \times 10^{-6}$ | -5.026 | $5.16 \times 10^{-7}$ | $9.05 \times 10^{-7}$ |
| JPM | RF-Text | 5172 | -0.010709 | -21.479 | $4.23 \times 10^{-98}$ | $7.28 \times 10^{-97}$ |
| JPM | Stacking | 5172 | -0.047519 | -21.276 | $2.32 \times 10^{-96}$ | $3.68 \times 10^{-95}$ |
| JPM | XGB-Full | 5172 | -0.010598 | -21.445 | $8.25 \times 10^{-98}$ | $1.39 \times 10^{-96}$ |
| JPM | XGB-Text | 5172 | -0.010575 | -21.436 | $9.95 \times 10^{-98}$ | $1.64 \times 10^{-96}$ |
| LLY | Elastic Net-Text | 5172 | $-1.46 \times 10^{-6}$ | -9.201 | $5.05 \times 10^{-20}$ | $1.54 \times 10^{-19}$ |
| LLY | HAR-AAII | 5172 | $-9.53 \times 10^{-8}$ | -2.112 | 0.0348 | 0.0440 |
| LLY | HAR-Full | 5172 | $-4.25 \times 10^{-6}$ | -8.840 | $1.29 \times 10^{-18}$ | $3.77 \times 10^{-18}$ |
| LLY | HAR-Text | 5172 | $-2.65 \times 10^{-6}$ | -7.105 | $1.37 \times 10^{-12}$ | $3.16 \times 10^{-12}$ |
| LLY | HAR-VIX | 5172 | $2.67 \times 10^{-7}$ | 2.541 | 0.0111 | 0.0146 |
| LLY | Naive | 5172 | $-3.48 \times 10^{-7}$ | -2.533 | 0.0113 | 0.0149 |
| LLY | RF-Text | 5172 | -0.000201 | -15.571 | $1.84 \times 10^{-53}$ | $1.43 \times 10^{-52}$ |
| LLY | Stacking | 5172 | -0.007938 | -27.789 | $1.60 \times 10^{-158}$ | $9.73 \times 10^{-157}$ |
| LLY | XGB-Full | 5172 | -0.000220 | -15.510 | $4.58 \times 10^{-53}$ | $3.49 \times 10^{-52}$ |
| LLY | XGB-Text | 5172 | -0.000300 | -17.468 | $1.95 \times 10^{-66}$ | $1.80 \times 10^{-65}$ |
| MA | Elastic Net-Text | 3565 | $-1.65 \times 10^{-5}$ | -10.252 | $2.51 \times 10^{-24}$ | $8.57 \times 10^{-24}$ |
| MA | HAR-AAII | 3565 | $-3.11 \times 10^{-6}$ | -11.255 | $6.72 \times 10^{-29}$ | $2.62 \times 10^{-28}$ |
| MA | HAR-Full | 3565 | $-4.36 \times 10^{-5}$ | -14.142 | $3.22 \times 10^{-44}$ | $1.89 \times 10^{-43}$ |
| MA | HAR-Text | 3565 | $-2.04 \times 10^{-5}$ | -14.423 | $7.10 \times 10^{-46}$ | $4.33 \times 10^{-45}$ |
| MA | HAR-VIX | 3565 | $-1.19 \times 10^{-5}$ | -12.158 | $2.35 \times 10^{-33}$ | $1.03 \times 10^{-32}$ |
| MA | Naive | 3565 | $4.00 \times 10^{-7}$ | 0.538 | 0.5907 | 0.6608 |
| MA | RF-Text | 3565 | -0.002437 | -21.798 | $5.45 \times 10^{-99}$ | $9.59 \times 10^{-98}$ |
| MA | Stacking | 3565 | -0.011228 | -28.759 | $9.43 \times 10^{-164}$ | $8.30 \times 10^{-162}$ |
| MA | XGB-Full | 3565 | -0.001809 | -18.981 | $1.30 \times 10^{-76}$ | $1.38 \times 10^{-75}$ |
| MA | XGB-Text | 3565 | -0.001787 | -18.789 | $3.55 \times 10^{-75}$ | $3.52 \times 10^{-74}$ |
| UNH | Elastic Net-Text | 5172 | $-7.34 \times 10^{-6}$ | -9.690 | $5.11 \times 10^{-22}$ | $1.67 \times 10^{-21}$ |
| UNH | HAR-AAII | 5172 | $-2.10 \times 10^{-7}$ | -9.010 | $2.86 \times 10^{-19}$ | $8.63 \times 10^{-19}$ |
| UNH | HAR-Full | 5172 | $-6.41 \times 10^{-6}$ | -13.561 | $3.44 \times 10^{-41}$ | $1.89 \times 10^{-40}$ |
| UNH | HAR-Text | 5172 | $-2.80 \times 10^{-6}$ | -13.576 | $2.82 \times 10^{-41}$ | $1.56 \times 10^{-40}$ |
| UNH | HAR-VIX | 5172 | $1.84 \times 10^{-7}$ | 1.477 | 0.1397 | 0.1659 |
| UNH | Naive | 5172 | $-2.24 \times 10^{-6}$ | -5.795 | $7.25 \times 10^{-9}$ | $1.39 \times 10^{-8}$ |
| UNH | RF-Text | 5172 | -0.005999 | -12.774 | $8.18 \times 10^{-37}$ | $3.93 \times 10^{-36}$ |
| UNH | Stacking | 5172 | -0.018828 | -19.206 | $1.83 \times 10^{-79}$ | $2.07 \times 10^{-78}$ |
| UNH | XGB-Full | 5172 | -0.005778 | -12.448 | $4.56 \times 10^{-35}$ | $2.10 \times 10^{-34}$ |
| UNH | XGB-Text | 5172 | -0.005763 | -12.390 | $9.15 \times 10^{-35}$ | $4.19 \times 10^{-34}$ |
| WMT | Elastic Net-Text | 5171 | $-2.63 \times 10^{-6}$ | -5.781 | $7.88 \times 10^{-9}$ | $1.51 \times 10^{-8}$ |

| Stock | Model | $n$ | $\Delta$ | Statistic | $p$ | $q^{global}$ |
|---|---|---|---|---|---|---|
| WMT | HAR-AAII | 5171 | $-3.32 \times 10^{-7}$ | -10.192 | $3.65 \times 10^{-24}$ | $1.23 \times 10^{-23}$ |
| WMT | HAR-Full | 5171 | $-3.91 \times 10^{-7}$ | -3.991 | $6.67 \times 10^{-5}$ | 0.0001 |
| WMT | HAR-Text | 5171 | $-7.88 \times 10^{-7}$ | -11.640 | $6.27 \times 10^{-31}$ | $2.61 \times 10^{-30}$ |
| WMT | HAR-VIX | 5171 | $1.21 \times 10^{-7}$ | 1.587 | 0.1126 | 0.1353 |
| WMT | Naive | 5171 | $-9.48 \times 10^{-7}$ | -3.595 | 0.0003 | 0.0005 |
| WMT | RF-Text | 5171 | -0.001545 | -20.105 | $1.26 \times 10^{-86}$ | $1.64 \times 10^{-85}$ |
| WMT | Stacking | 5171 | -0.005706 | -27.392 | $2.23 \times 10^{-154}$ | $1.26 \times 10^{-152}$ |
| WMT | XGB-Full | 5171 | -0.001521 | -19.720 | $1.60 \times 10^{-83}$ | $1.92 \times 10^{-82}$ |
| WMT | XGB-Text | 5171 | -0.001528 | -19.646 | $6.19 \times 10^{-83}$ | $7.31 \times 10^{-82}$ |
| XOM | Elastic Net-Text | 5172 | $-2.68 \times 10^{-6}$ | -6.030 | $1.75 \times 10^{-9}$ | $3.50 \times 10^{-9}$ |
| XOM | HAR-AAII | 5172 | $-2.28 \times 10^{-9}$ | -0.106 | 0.9154 | 0.9931 |
| XOM | HAR-Full | 5172 | $-1.59 \times 10^{-7}$ | -1.431 | 0.1526 | 0.1799 |
| XOM | HAR-Text | 5172 | $-3.92 \times 10^{-7}$ | -6.917 | $5.16 \times 10^{-12}$ | $1.18 \times 10^{-11}$ |
| XOM | HAR-VIX | 5172 | $3.43 \times 10^{-7}$ | 3.229 | 0.0012 | 0.0018 |
| XOM | Naive | 5172 | $-3.29 \times 10^{-6}$ | -4.416 | $1.03 \times 10^{-5}$ | $1.68 \times 10^{-5}$ |
| XOM | RF-Text | 5172 | -0.001625 | -19.883 | $7.81 \times 10^{-85}$ | $9.81 \times 10^{-84}$ |
| XOM | Stacking | 5172 | -0.013085 | -25.494 | $3.74 \times 10^{-135}$ | $1.06 \times 10^{-133}$ |
| XOM | XGB-Full | 5172 | -0.001646 | -20.245 | $9.11 \times 10^{-88}$ | $1.20 \times 10^{-86}$ |
| XOM | XGB-Text | 5172 | -0.001601 | -20.055 | $3.23 \times 10^{-86}$ | $4.12 \times 10^{-85}$ |

Table C.5: DM, annualized-volatility squared error, $h = 5$: expanding window

| Stock | Model | $n$ | $\Delta$ | Statistic | $p$ | $q^{global}$ |
|---|---|---|---|---|---|---|
| AAPL | Elastic Net-Text | 5160 | -0.000257 | -9.283 | $2.37 \times 10^{-20}$ | $7.33 \times 10^{-20}$ |
| AAPL | HAR-AAII | 5160 | $-3.26 \times 10^{-5}$ | -8.520 | $2.08 \times 10^{-17}$ | $5.85 \times 10^{-17}$ |
| AAPL | HAR-Full | 5160 | -0.000774 | -8.472 | $3.11 \times 10^{-17}$ | $8.73 \times 10^{-17}$ |
| AAPL | HAR-Text | 5160 | -0.000241 | -8.678 | $5.31 \times 10^{-18}$ | $1.53 \times 10^{-17}$ |
| AAPL | HAR-VIX | 5160 | -0.000113 | -10.345 | $7.67 \times 10^{-25}$ | $2.67 \times 10^{-24}$ |
| AAPL | Naive | 5160 | $9.47 \times 10^{-6}$ | 5.611 | $2.12 \times 10^{-8}$ | $3.92 \times 10^{-8}$ |
| AAPL | RF-Text | 5160 | -0.004567 | -13.403 | $2.73 \times 10^{-40}$ | $1.46 \times 10^{-39}$ |
| AAPL | Stacking | 5160 | -0.012232 | -22.922 | $8.18 \times 10^{-111}$ | $1.80 \times 10^{-109}$ |
| AAPL | XGB-Full | 5160 | -0.004935 | -13.541 | $4.47 \times 10^{-41}$ | $2.44 \times 10^{-40}$ |
| AAPL | XGB-Text | 5160 | -0.004883 | -13.683 | $6.86 \times 10^{-42}$ | $3.88 \times 10^{-41}$ |
| ABBV | Elastic Net-Text | 1891 | $-3.46 \times 10^{-5}$ | -7.066 | $2.24 \times 10^{-12}$ | $5.16 \times 10^{-12}$ |
| ABBV | HAR-AAII | 1891 | $-2.03 \times 10^{-6}$ | -4.561 | $5.41 \times 10^{-6}$ | $8.97 \times 10^{-6}$ |
| ABBV | HAR-Full | 1891 | $-4.21 \times 10^{-5}$ | -6.977 | $4.14 \times 10^{-12}$ | $9.51 \times 10^{-12}$ |
| ABBV | HAR-Text | 1891 | $-6.37 \times 10^{-6}$ | -6.719 | $2.42 \times 10^{-11}$ | $5.31 \times 10^{-11}$ |
| ABBV | HAR-VIX | 1891 | $-1.43 \times 10^{-6}$ | -0.623 | 0.5332 | 0.5973 |
| ABBV | Naive | 1891 | $-3.98 \times 10^{-6}$ | -1.900 | 0.0576 | 0.0712 |
| ABBV | RF-Text | 1891 | -0.001448 | -9.171 | $1.19 \times 10^{-19}$ | $3.62 \times 10^{-19}$ |
| ABBV | Stacking | 1891 | -0.003798 | -15.742 | $1.47 \times 10^{-52}$ | $1.07 \times 10^{-51}$ |
| ABBV | XGB-Full | 1891 | -0.001612 | -9.767 | $5.14 \times 10^{-22}$ | $1.68 \times 10^{-21}$ |
| ABBV | XGB-Text | 1891 | -0.001752 | -10.540 | $2.80 \times 10^{-25}$ | $9.86 \times 10^{-25}$ |
| ABT | Elastic Net-Text | 5160 | $-5.75 \times 10^{-6}$ | -3.599 | 0.0003 | 0.0005 |
| ABT | HAR-AAII | 5160 | $-2.57 \times 10^{-6}$ | -5.370 | $8.21 \times 10^{-8}$ | $1.48 \times 10^{-7}$ |
| ABT | HAR-Full | 5160 | $-7.36 \times 10^{-6}$ | -6.260 | $4.17 \times 10^{-10}$ | $8.55 \times 10^{-10}$ |
| ABT | HAR-Text | 5160 | $-1.33 \times 10^{-5}$ | -8.228 | $2.40 \times 10^{-16}$ | $6.48 \times 10^{-16}$ |
| ABT | HAR-VIX | 5160 | $3.88 \times 10^{-7}$ | 0.677 | 0.4984 | 0.5607 |
| ABT | Naive | 5160 | $-3.25 \times 10^{-6}$ | -1.952 | 0.0510 | 0.0634 |

| Stock | Model | $n$ | $\Delta$ | Statistic | $p$ | $q^{global}$ |
|---|---|---|---|---|---|---|
| ABT | RF-Text | 5160 | -0.000640 | -10.983 | $9.35 \times 10^{-28}$ | $3.53 \times 10^{-27}$ |
| ABT | Stacking | 5160 | -0.006100 | -16.064 | $1.07 \times 10^{-56}$ | $9.00 \times 10^{-56}$ |
| ABT | XGB-Full | 5160 | -0.000669 | -10.779 | $8.38 \times 10^{-27}$ | $3.09 \times 10^{-26}$ |
| ABT | XGB-Text | 5160 | -0.000734 | -10.562 | $8.17 \times 10^{-26}$ | $2.94 \times 10^{-25}$ |
| AMZN | Elastic Net-Text | 5160 | $-2.94 \times 10^{-5}$ | -8.966 | $4.22 \times 10^{-19}$ | $1.27 \times 10^{-18}$ |
| AMZN | HAR-AAII | 5160 | -0.000187 | -11.597 | $1.02 \times 10^{-30}$ | $4.17 \times 10^{-30}$ |
| AMZN | HAR-Full | 5160 | -0.000781 | -9.339 | $1.42 \times 10^{-20}$ | $4.43 \times 10^{-20}$ |
| AMZN | HAR-Text | 5160 | $-5.56 \times 10^{-5}$ | -7.360 | $2.14 \times 10^{-13}$ | $5.17 \times 10^{-13}$ |
| AMZN | HAR-VIX | 5160 | $-2.28 \times 10^{-5}$ | -10.329 | $9.11 \times 10^{-25}$ | $3.16 \times 10^{-24}$ |
| AMZN | Naive | 5160 | $2.69 \times 10^{-6}$ | 1.561 | 0.1185 | 0.1418 |
| AMZN | RF-Text | 5160 | -0.044808 | -13.579 | $2.72 \times 10^{-41}$ | $1.52 \times 10^{-40}$ |
| AMZN | Stacking | 5160 | -0.031006 | -18.810 | $2.14 \times 10^{-76}$ | $2.23 \times 10^{-75}$ |
| AMZN | XGB-Full | 5160 | -0.071262 | -15.652 | $5.52 \times 10^{-54}$ | $4.32 \times 10^{-53}$ |
| AMZN | XGB-Text | 5160 | -0.049819 | -13.789 | $1.67 \times 10^{-42}$ | $9.60 \times 10^{-42}$ |
| JNJ | Elastic Net-Text | 5160 | $-3.00 \times 10^{-5}$ | -13.271 | $1.51 \times 10^{-39}$ | $8.01 \times 10^{-39}$ |
| JNJ | HAR-AAII | 5160 | $-5.78 \times 10^{-6}$ | -8.817 | $1.58 \times 10^{-18}$ | $4.62 \times 10^{-18}$ |
| JNJ | HAR-Full | 5160 | $-5.01 \times 10^{-5}$ | -11.284 | $3.46 \times 10^{-29}$ | $1.36 \times 10^{-28}$ |
| JNJ | HAR-Text | 5160 | $-1.25 \times 10^{-5}$ | -5.625 | $1.95 \times 10^{-8}$ | $3.61 \times 10^{-8}$ |
| JNJ | HAR-VIX | 5160 | $-5.30 \times 10^{-5}$ | -11.670 | $4.43 \times 10^{-31}$ | $1.86 \times 10^{-30}$ |
| JNJ | Naive | 5160 | $1.26 \times 10^{-5}$ | 6.572 | $5.46 \times 10^{-11}$ | $1.17 \times 10^{-10}$ |
| JNJ | RF-Text | 5160 | -0.001376 | -12.826 | $4.30 \times 10^{-37}$ | $2.11 \times 10^{-36}$ |
| JNJ | Stacking | 5160 | -0.004603 | -18.360 | $5.53 \times 10^{-73}$ | $5.28 \times 10^{-72}$ |
| JNJ | XGB-Full | 5160 | -0.001343 | -12.799 | $6.03 \times 10^{-37}$ | $2.95 \times 10^{-36}$ |
| JNJ | XGB-Text | 5160 | -0.001324 | -12.607 | $6.48 \times 10^{-36}$ | $3.09 \times 10^{-35}$ |
| JPM | Elastic Net-Text | 5160 | $-2.56 \times 10^{-5}$ | -6.206 | $5.87 \times 10^{-10}$ | $1.20 \times 10^{-9}$ |
| JPM | HAR-AAII | 5160 | $-1.34 \times 10^{-6}$ | -2.719 | 0.0066 | 0.0088 |
| JPM | HAR-Full | 5160 | $-4.50 \times 10^{-5}$ | -5.740 | $9.99 \times 10^{-9}$ | $1.90 \times 10^{-8}$ |
| JPM | HAR-Text | 5160 | $-3.85 \times 10^{-6}$ | -3.774 | 0.0002 | 0.0002 |
| JPM | HAR-VIX | 5160 | $2.48 \times 10^{-6}$ | 2.240 | 0.0251 | 0.0322 |
| JPM | Naive | 5160 | $-1.59 \times 10^{-5}$ | -3.624 | 0.0003 | 0.0004 |
| JPM | RF-Text | 5160 | -0.010877 | -12.511 | $2.12 \times 10^{-35}$ | $9.81 \times 10^{-35}$ |
| JPM | Stacking | 5160 | -0.047597 | -12.344 | $1.61 \times 10^{-34}$ | $7.26 \times 10^{-34}$ |
| JPM | XGB-Full | 5160 | -0.010819 | -12.531 | $1.65 \times 10^{-35}$ | $7.68 \times 10^{-35}$ |
| JPM | XGB-Text | 5160 | -0.010878 | -12.537 | $1.54 \times 10^{-35}$ | $7.21 \times 10^{-35}$ |
| LLY | Elastic Net-Text | 5160 | $-3.26 \times 10^{-6}$ | -5.390 | $7.37 \times 10^{-8}$ | $1.34 \times 10^{-7}$ |
| LLY | HAR-AAII | 5160 | $-3.07 \times 10^{-7}$ | -1.379 | 0.1680 | 0.1974 |
| LLY | HAR-Full | 5160 | $-2.02 \times 10^{-5}$ | -6.014 | $1.93 \times 10^{-9}$ | $3.84 \times 10^{-9}$ |
| LLY | HAR-Text | 5160 | $-1.13 \times 10^{-5}$ | -4.392 | $1.15 \times 10^{-5}$ | $1.86 \times 10^{-5}$ |
| LLY | HAR-VIX | 5160 | $9.15 \times 10^{-7}$ | 1.869 | 0.0617 | 0.0761 |
| LLY | Naive | 5160 | $-1.44 \times 10^{-6}$ | -2.046 | 0.0408 | 0.0514 |
| LLY | RF-Text | 5160 | -0.000245 | -10.206 | $3.16 \times 10^{-24}$ | $1.07 \times 10^{-23}$ |
| LLY | Stacking | 5160 | -0.007991 | -16.099 | $6.17 \times 10^{-57}$ | $5.26 \times 10^{-56}$ |
| LLY | XGB-Full | 5160 | -0.000276 | -10.599 | $5.55 \times 10^{-26}$ | $2.02 \times 10^{-25}$ |
| LLY | XGB-Text | 5160 | -0.000320 | -10.781 | $8.18 \times 10^{-27}$ | $3.03 \times 10^{-26}$ |
| MA | Elastic Net-Text | 3553 | $-4.40 \times 10^{-5}$ | -6.780 | $1.41 \times 10^{-11}$ | $3.13 \times 10^{-11}$ |
| MA | HAR-AAII | 3553 | $-9.39 \times 10^{-6}$ | -7.210 | $6.80 \times 10^{-13}$ | $1.59 \times 10^{-12}$ |
| MA | HAR-Full | 3553 | -0.000110 | -8.795 | $2.19 \times 10^{-18}$ | $6.38 \times 10^{-18}$ |
| MA | HAR-Text | 3553 | $-6.59 \times 10^{-5}$ | -8.600 | $1.17 \times 10^{-17}$ | $3.35 \times 10^{-17}$ |
| MA | HAR-VIX | 3553 | $-2.48 \times 10^{-5}$ | -7.341 | $2.61 \times 10^{-13}$ | $6.26 \times 10^{-13}$ |
| MA | Naive | 3553 | $1.30 \times 10^{-6}$ | 0.353 | 0.7243 | 0.7982 |
| MA | RF-Text | 3553 | -0.002384 | -11.484 | $5.36 \times 10^{-30}$ | $2.16 \times 10^{-29}$ |

| Stock | Model | $n$ | $\Delta$ | Statistic | $p$ | $q^{global}$ |
|---|---|---|---|---|---|---|
| MA | Stacking | 3553 | -0.010981 | -16.424 | $1.74 \times 10^{-58}$ | $1.52 \times 10^{-57}$ |
| MA | XGB-Full | 3553 | -0.001944 | -11.463 | $6.73 \times 10^{-30}$ | $2.68 \times 10^{-29}$ |
| MA | XGB-Text | 3553 | -0.001954 | -11.168 | $1.74 \times 10^{-28}$ | $6.70 \times 10^{-28}$ |
| UNH | Elastic Net-Text | 5160 | $-1.52 \times 10^{-5}$ | -5.907 | $3.71 \times 10^{-9}$ | $7.31 \times 10^{-9}$ |
| UNH | HAR-AAII | 5160 | $-1.32 \times 10^{-7}$ | -1.322 | 0.1863 | 0.2182 |
| UNH | HAR-Full | 5160 | $-9.30 \times 10^{-6}$ | -7.224 | $5.77 \times 10^{-13}$ | $1.36 \times 10^{-12}$ |
| UNH | HAR-Text | 5160 | $-6.97 \times 10^{-6}$ | -8.094 | $7.13 \times 10^{-16}$ | $1.88 \times 10^{-15}$ |
| UNH | HAR-VIX | 5160 | $4.44 \times 10^{-7}$ | 0.918 | 0.3587 | 0.4088 |
| UNH | Naive | 5160 | $-7.64 \times 10^{-6}$ | -3.846 | 0.0001 | 0.0002 |
| UNH | RF-Text | 5160 | -0.006025 | -7.383 | $1.79 \times 10^{-13}$ | $4.37 \times 10^{-13}$ |
| UNH | Stacking | 5160 | -0.018772 | -11.060 | $4.03 \times 10^{-28}$ | $1.54 \times 10^{-27}$ |
| UNH | XGB-Full | 5160 | -0.005879 | -7.229 | $5.59 \times 10^{-13}$ | $1.32 \times 10^{-12}$ |
| UNH | XGB-Text | 5160 | -0.005866 | -7.227 | $5.68 \times 10^{-13}$ | $1.34 \times 10^{-12}$ |
| WMT | Elastic Net-Text | 5159 | $-4.80 \times 10^{-6}$ | -3.562 | 0.0004 | 0.0005 |
| WMT | HAR-AAII | 5159 | $-1.38 \times 10^{-6}$ | -7.323 | $2.81 \times 10^{-13}$ | $6.71 \times 10^{-13}$ |
| WMT | HAR-Full | 5159 | $-5.77 \times 10^{-6}$ | -6.593 | $4.74 \times 10^{-11}$ | $1.02 \times 10^{-10}$ |
| WMT | HAR-Text | 5159 | $-3.70 \times 10^{-6}$ | -7.303 | $3.25 \times 10^{-13}$ | $7.76 \times 10^{-13}$ |
| WMT | HAR-VIX | 5159 | $4.59 \times 10^{-7}$ | 1.256 | 0.2093 | 0.2430 |
| WMT | Naive | 5159 | $-3.11 \times 10^{-6}$ | -2.604 | 0.0092 | 0.0123 |
| WMT | RF-Text | 5159 | -0.001586 | -11.777 | $1.30 \times 10^{-31}$ | $5.52 \times 10^{-31}$ |
| WMT | Stacking | 5159 | -0.005702 | -15.836 | $3.44 \times 10^{-55}$ | $2.81 \times 10^{-54}$ |
| WMT | XGB-Full | 5159 | -0.001625 | -11.630 | $6.99 \times 10^{-31}$ | $2.88 \times 10^{-30}$ |
| WMT | XGB-Text | 5159 | -0.001577 | -11.515 | $2.59 \times 10^{-30}$ | $1.05 \times 10^{-29}$ |
| XOM | Elastic Net-Text | 5160 | $-9.78 \times 10^{-7}$ | -4.162 | $3.21 \times 10^{-5}$ | $5.06 \times 10^{-5}$ |
| XOM | HAR-AAII | 5160 | $4.28 \times 10^{-8}$ | 0.336 | 0.7371 | 0.8107 |
| XOM | HAR-Full | 5160 | $-1.31 \times 10^{-7}$ | -0.402 | 0.6876 | 0.7628 |
| XOM | HAR-Text | 5160 | $-6.74 \times 10^{-7}$ | -3.162 | 0.0016 | 0.0022 |
| XOM | HAR-VIX | 5160 | $7.90 \times 10^{-7}$ | 2.498 | 0.0125 | 0.0163 |
| XOM | Naive | 5160 | $-1.12 \times 10^{-5}$ | -2.757 | 0.0059 | 0.0079 |
| XOM | RF-Text | 5160 | -0.001645 | -11.599 | $9.95 \times 10^{-31}$ | $4.08 \times 10^{-30}$ |
| XOM | Stacking | 5160 | -0.013157 | -14.712 | $5.00 \times 10^{-48}$ | $3.12 \times 10^{-47}$ |
| XOM | XGB-Full | 5160 | -0.001769 | -11.996 | $1.01 \times 10^{-32}$ | $4.37 \times 10^{-32}$ |
| XOM | XGB-Text | 5160 | -0.001649 | -11.636 | $6.53 \times 10^{-31}$ | $2.71 \times 10^{-30}$ |

Table C.6: DM, annualized-volatility squared error, $h = 22$: expanding window

| Stock | Model | $n$ | $\Delta$ | Statistic | $p$ | $q^{global}$ |
|---|---|---|---|---|---|---|
| AAPL | Elastic Net-Text | 5109 | -0.001852 | -4.692 | $2.77 \times 10^{-6}$ | $4.67 \times 10^{-6}$ |
| AAPL | HAR-AAII | 5109 | -0.000269 | -4.346 | $1.41 \times 10^{-5}$ | $2.26 \times 10^{-5}$ |
| AAPL | HAR-Full | 5109 | -0.002271 | -5.544 | $3.10 \times 10^{-8}$ | $5.69 \times 10^{-8}$ |
| AAPL | HAR-Text | 5109 | -0.001785 | -4.672 | $3.06 \times 10^{-6}$ | $5.13 \times 10^{-6}$ |
| AAPL | HAR-VIX | 5109 | -0.000772 | -5.572 | $2.65 \times 10^{-8}$ | $4.87 \times 10^{-8}$ |
| AAPL | Naive | 5109 | 0.000350 | 4.532 | $5.98 \times 10^{-6}$ | $9.89 \times 10^{-6}$ |
| AAPL | RF-Text | 5109 | -0.005012 | -7.856 | $4.82 \times 10^{-15}$ | $1.23 \times 10^{-14}$ |
| AAPL | Stacking | 5109 | -0.012828 | -12.043 | $5.90 \times 10^{-33}$ | $2.55 \times 10^{-32}$ |
| AAPL | XGB-Full | 5109 | -0.030180 | -12.858 | $2.91 \times 10^{-37}$ | $1.44 \times 10^{-36}$ |
| AAPL | XGB-Text | 5109 | -0.005287 | -7.664 | $2.14 \times 10^{-14}$ | $5.32 \times 10^{-14}$ |
| ABBV | Elastic Net-Text | 1840 | $-9.46 \times 10^{-5}$ | -5.140 | $3.05 \times 10^{-7}$ | $5.40 \times 10^{-7}$ |
| ABBV | HAR-AAII | 1840 | $-3.65 \times 10^{-5}$ | -1.275 | 0.2026 | 0.2360 |

| Stock | Model | $n$ | $\Delta$ | Statistic | $p$ | $q^{global}$ |
|---|---|---|---|---|---|---|
| ABBV | HAR-Full | 1840 | -0.000196 | -5.005 | $6.13 \times 10^{-7}$ | $1.07 \times 10^{-6}$ |
| ABBV | HAR-Text | 1840 | $-3.10 \times 10^{-5}$ | -3.826 | 0.0001 | 0.0002 |
| ABBV | HAR-VIX | 1840 | $-2.95 \times 10^{-5}$ | -1.545 | 0.1225 | 0.1461 |
| ABBV | Naive | 1840 | $-1.83 \times 10^{-5}$ | -1.052 | 0.2930 | 0.3358 |
| ABBV | RF-Text | 1840 | -0.002309 | -5.898 | $4.37 \times 10^{-9}$ | $8.52 \times 10^{-9}$ |
| ABBV | Stacking | 1840 | -0.003815 | -8.119 | $8.49 \times 10^{-16}$ | $2.23 \times 10^{-15}$ |
| ABBV | XGB-Full | 1840 | -0.002705 | -6.913 | $6.52 \times 10^{-12}$ | $1.48 \times 10^{-11}$ |
| ABBV | XGB-Text | 1840 | -0.002541 | -6.451 | $1.42 \times 10^{-10}$ | $2.99 \times 10^{-10}$ |
| ABT | Elastic Net-Text | 5109 | $-3.22 \times 10^{-5}$ | -4.468 | $8.06 \times 10^{-6}$ | $1.33 \times 10^{-5}$ |
| ABT | HAR-AAII | 5109 | $-1.15 \times 10^{-5}$ | -2.414 | 0.0158 | 0.0206 |
| ABT | HAR-Full | 5109 | $-5.32 \times 10^{-5}$ | -2.736 | 0.0062 | 0.0084 |
| ABT | HAR-Text | 5109 | $-9.15 \times 10^{-5}$ | -4.649 | $3.41 \times 10^{-6}$ | $5.72 \times 10^{-6}$ |
| ABT | HAR-VIX | 5109 | $1.11 \times 10^{-6}$ | 0.220 | 0.8256 | 0.9019 |
| ABT | Naive | 5109 | $-8.20 \times 10^{-6}$ | -0.754 | 0.4510 | 0.5110 |
| ABT | RF-Text | 5109 | -0.001068 | -5.712 | $1.18 \times 10^{-8}$ | $2.24 \times 10^{-8}$ |
| ABT | Stacking | 5109 | -0.006077 | -8.229 | $2.37 \times 10^{-16}$ | $6.44 \times 10^{-16}$ |
| ABT | XGB-Full | 5109 | -0.000885 | -5.814 | $6.45 \times 10^{-9}$ | $1.25 \times 10^{-8}$ |
| ABT | XGB-Text | 5109 | -0.001169 | -5.743 | $9.81 \times 10^{-9}$ | $1.87 \times 10^{-8}$ |
| AMZN | Elastic Net-Text | 5109 | -0.001610 | -5.385 | $7.55 \times 10^{-8}$ | $1.37 \times 10^{-7}$ |
| AMZN | HAR-AAII | 5109 | -0.000346 | -5.124 | $3.10 \times 10^{-7}$ | $5.48 \times 10^{-7}$ |
| AMZN | HAR-Full | 5109 | -0.001130 | -5.237 | $1.70 \times 10^{-7}$ | $3.03 \times 10^{-7}$ |
| AMZN | HAR-Text | 5109 | -0.001498 | -4.882 | $1.08 \times 10^{-6}$ | $1.86 \times 10^{-6}$ |
| AMZN | HAR-VIX | 5109 | $-9.29 \times 10^{-6}$ | -0.690 | 0.4901 | 0.5522 |
| AMZN | Naive | 5109 | $-2.74 \times 10^{-5}$ | -1.805 | 0.0711 | 0.0875 |
| AMZN | RF-Text | 5109 | -0.045898 | -7.130 | $1.14 \times 10^{-12}$ | $2.64 \times 10^{-12}$ |
| AMZN | Stacking | 5109 | -0.054605 | -8.574 | $1.31 \times 10^{-17}$ | $3.71 \times 10^{-17}$ |
| AMZN | XGB-Full | 5109 | -0.067573 | -8.003 | $1.49 \times 10^{-15}$ | $3.84 \times 10^{-15}$ |
| AMZN | XGB-Text | 5109 | -0.087627 | -8.944 | $5.16 \times 10^{-19}$ | $1.54 \times 10^{-18}$ |
| JNJ | Elastic Net-Text | 5109 | $4.98 \times 10^{-5}$ | 3.358 | 0.0008 | 0.0011 |
| JNJ | HAR-AAII | 5109 | -0.000582 | -5.704 | $1.24 \times 10^{-8}$ | $2.34 \times 10^{-8}$ |
| JNJ | HAR-Full | 5109 | -0.000220 | -3.367 | 0.0008 | 0.0011 |
| JNJ | HAR-Text | 5109 | 0.000191 | 5.691 | $1.33 \times 10^{-8}$ | $2.51 \times 10^{-8}$ |
| JNJ | HAR-VIX | 5109 | -0.000511 | -6.285 | $3.54 \times 10^{-10}$ | $7.34 \times 10^{-10}$ |
| JNJ | Naive | 5109 | 0.000259 | 5.610 | $2.14 \times 10^{-8}$ | $3.94 \times 10^{-8}$ |
| JNJ | RF-Text | 5109 | -0.001412 | -7.197 | $7.05 \times 10^{-13}$ | $1.64 \times 10^{-12}$ |
| JNJ | Stacking | 5109 | -0.004591 | -9.365 | $1.11 \times 10^{-20}$ | $3.49 \times 10^{-20}$ |
| JNJ | XGB-Full | 5109 | -0.001363 | -6.639 | $3.48 \times 10^{-11}$ | $7.56 \times 10^{-11}$ |
| JNJ | XGB-Text | 5109 | -0.001469 | -6.702 | $2.28 \times 10^{-11}$ | $5.01 \times 10^{-11}$ |
| JPM | Elastic Net-Text | 5109 | -0.000332 | -8.579 | $1.26 \times 10^{-17}$ | $3.58 \times 10^{-17}$ |
| JPM | HAR-AAII | 5109 | $-3.82 \times 10^{-5}$ | -2.077 | 0.0379 | 0.0478 |
| JPM | HAR-Full | 5109 | $7.47 \times 10^{-5}$ | 2.286 | 0.0223 | 0.0287 |
| JPM | HAR-Text | 5109 | $7.79 \times 10^{-5}$ | 4.458 | $8.44 \times 10^{-6}$ | $1.39 \times 10^{-5}$ |
| JPM | HAR-VIX | 5109 | $2.80 \times 10^{-5}$ | 2.988 | 0.0028 | 0.0039 |
| JPM | Naive | 5109 | $9.21 \times 10^{-5}$ | 1.939 | 0.0526 | 0.0653 |
| JPM | RF-Text | 5109 | -0.011491 | -6.615 | $4.11 \times 10^{-11}$ | $8.91 \times 10^{-11}$ |
| JPM | Stacking | 5109 | -0.047644 | -6.373 | $2.02 \times 10^{-10}$ | $4.22 \times 10^{-10}$ |
| JPM | XGB-Full | 5109 | -0.011401 | -6.610 | $4.24 \times 10^{-11}$ | $9.17 \times 10^{-11}$ |
| JPM | XGB-Text | 5109 | -0.012235 | -6.789 | $1.26 \times 10^{-11}$ | $2.81 \times 10^{-11}$ |
| LLY | Elastic Net-Text | 5109 | $-2.27 \times 10^{-5}$ | -2.917 | 0.0035 | 0.0049 |
| LLY | HAR-AAII | 5109 | $-2.13 \times 10^{-5}$ | -3.080 | 0.0021 | 0.0029 |
| LLY | HAR-Full | 5109 | -0.000201 | -3.984 | $6.87 \times 10^{-5}$ | 0.0001 |

| Stock | Model | $n$ | $\Delta$ | Statistic | $p$ | $q^{global}$ |
|---|---|---|---|---|---|---|
| LLY | HAR-Text | 5109 | $-3.12 \times 10^{-5}$ | -3.484 | 0.0005 | 0.0007 |
| LLY | HAR-VIX | 5109 | $1.23 \times 10^{-5}$ | 2.648 | 0.0081 | 0.0108 |
| LLY | Naive | 5109 | $-8.72 \times 10^{-6}$ | -1.153 | 0.2488 | 0.2877 |
| LLY | RF-Text | 5109 | -0.000405 | -5.604 | $2.20 \times 10^{-8}$ | $4.05 \times 10^{-8}$ |
| LLY | Stacking | 5109 | -0.008055 | -8.253 | $1.94 \times 10^{-16}$ | $5.34 \times 10^{-16}$ |
| LLY | XGB-Full | 5109 | -0.000394 | -5.239 | $1.68 \times 10^{-7}$ | $3.01 \times 10^{-7}$ |
| LLY | XGB-Text | 5109 | -0.000516 | -5.719 | $1.13 \times 10^{-8}$ | $2.15 \times 10^{-8}$ |
| MA | Elastic Net-Text | 3502 | -0.000481 | -4.133 | $3.66 \times 10^{-5}$ | $5.73 \times 10^{-5}$ |
| MA | HAR-AAII | 3502 | $-2.17 \times 10^{-5}$ | -1.577 | 0.1148 | 0.1378 |
| MA | HAR-Full | 3502 | -0.000669 | -4.284 | $1.89 \times 10^{-5}$ | $3.00 \times 10^{-5}$ |
| MA | HAR-Text | 3502 | -0.000341 | -3.907 | $9.54 \times 10^{-5}$ | 0.0001 |
| MA | HAR-VIX | 3502 | -0.000188 | -4.220 | $2.50 \times 10^{-5}$ | $3.96 \times 10^{-5}$ |
| MA | Naive | 3502 | $3.27 \times 10^{-5}$ | 1.079 | 0.2804 | 0.3228 |
| MA | RF-Text | 3502 | -0.003277 | -6.269 | $4.08 \times 10^{-10}$ | $8.38 \times 10^{-10}$ |
| MA | Stacking | 3502 | -0.014388 | -7.946 | $2.58 \times 10^{-15}$ | $6.63 \times 10^{-15}$ |
| MA | XGB-Full | 3502 | -0.004008 | -4.760 | $2.01 \times 10^{-6}$ | $3.43 \times 10^{-6}$ |
| MA | XGB-Text | 3502 | -0.002972 | -6.436 | $1.39 \times 10^{-10}$ | $2.95 \times 10^{-10}$ |
| UNH | Elastic Net-Text | 5109 | $-3.11 \times 10^{-5}$ | -4.452 | $8.70 \times 10^{-6}$ | $1.42 \times 10^{-5}$ |
| UNH | HAR-AAII | 5109 | $-1.10 \times 10^{-6}$ | -0.636 | 0.5246 | 0.5893 |
| UNH | HAR-Full | 5109 | $-2.94 \times 10^{-5}$ | -4.571 | $4.97 \times 10^{-6}$ | $8.27 \times 10^{-6}$ |
| UNH | HAR-Text | 5109 | $-1.64 \times 10^{-5}$ | -3.486 | 0.0005 | 0.0007 |
| UNH | HAR-VIX | 5109 | $-9.58 \times 10^{-7}$ | -0.241 | 0.8098 | 0.8858 |
| UNH | Naive | 5109 | $-4.35 \times 10^{-5}$ | -2.252 | 0.0244 | 0.0313 |
| UNH | RF-Text | 5109 | -0.006212 | -3.810 | 0.0001 | 0.0002 |
| UNH | Stacking | 5109 | -0.018487 | -5.637 | $1.83 \times 10^{-8}$ | $3.41 \times 10^{-8}$ |
| UNH | XGB-Full | 5109 | -0.006187 | -3.843 | 0.0001 | 0.0002 |
| UNH | XGB-Text | 5109 | -0.006347 | -3.933 | $8.48 \times 10^{-5}$ | 0.0001 |
| WMT | Elastic Net-Text | 5108 | $-5.99 \times 10^{-6}$ | -1.472 | 0.1410 | 0.1672 |
| WMT | HAR-AAII | 5108 | $-6.72 \times 10^{-6}$ | -3.478 | 0.0005 | 0.0007 |
| WMT | HAR-Full | 5108 | $-7.85 \times 10^{-5}$ | -3.840 | 0.0001 | 0.0002 |
| WMT | HAR-Text | 5108 | $-2.32 \times 10^{-5}$ | -3.487 | 0.0005 | 0.0007 |
| WMT | HAR-VIX | 5108 | $9.45 \times 10^{-7}$ | 0.335 | 0.7380 | 0.8107 |
| WMT | Naive | 5108 | $-1.50 \times 10^{-5}$ | -2.214 | 0.0269 | 0.0343 |
| WMT | RF-Text | 5108 | -0.001972 | -6.416 | $1.52 \times 10^{-10}$ | $3.21 \times 10^{-10}$ |
| WMT | Stacking | 5108 | -0.005775 | -8.236 | $2.25 \times 10^{-16}$ | $6.14 \times 10^{-16}$ |
| WMT | XGB-Full | 5108 | -0.001915 | -6.282 | $3.62 \times 10^{-10}$ | $7.48 \times 10^{-10}$ |
| WMT | XGB-Text | 5108 | -0.002324 | -6.149 | $8.37 \times 10^{-10}$ | $1.70 \times 10^{-9}$ |
| XOM | Elastic Net-Text | 5109 | $-3.82 \times 10^{-5}$ | -3.119 | 0.0018 | 0.0025 |
| XOM | HAR-AAII | 5109 | $3.78 \times 10^{-6}$ | 1.620 | 0.1053 | 0.1267 |
| XOM | HAR-Full | 5109 | $-4.23 \times 10^{-5}$ | -3.347 | 0.0008 | 0.0012 |
| XOM | HAR-Text | 5109 | $-5.61 \times 10^{-6}$ | -1.139 | 0.2546 | 0.2940 |
| XOM | HAR-VIX | 5109 | $6.25 \times 10^{-6}$ | 1.631 | 0.1030 | 0.1241 |
| XOM | Naive | 5109 | $-5.94 \times 10^{-5}$ | -2.157 | 0.0311 | 0.0394 |
| XOM | RF-Text | 5109 | -0.001786 | -6.113 | $1.05 \times 10^{-9}$ | $2.11 \times 10^{-9}$ |
| XOM | Stacking | 5109 | -0.013445 | -7.574 | $4.27 \times 10^{-14}$ | $1.06 \times 10^{-13}$ |
| XOM | XGB-Full | 5109 | -0.001871 | -6.384 | $1.87 \times 10^{-10}$ | $3.92 \times 10^{-10}$ |
| XOM | XGB-Text | 5109 | -0.001719 | -6.128 | $9.57 \times 10^{-10}$ | $1.93 \times 10^{-9}$ |

Table C.7: CW, annualized-volatility squared error, $h = 1$: expanding window

| Stock | Model | $n$ | $\Delta$ | Statistic | $p$ | $q^{global}$ |
|---|---|---|---|---|---|---|
| AAPL | HAR-AAII | 5172 | $-2.79 \times 10^{-6}$ | -7.210 | 1.0000 | 1.0000 |
| AAPL | HAR-Full | 5172 | $-3.83 \times 10^{-6}$ | -1.443 | 0.9254 | 1.0000 |
| AAPL | HAR-Text | 5172 | $-2.11 \times 10^{-6}$ | -1.859 | 0.9685 | 1.0000 |
| AAPL | HAR-VIX | 5172 | $-5.79 \times 10^{-6}$ | -7.235 | 1.0000 | 1.0000 |
| ABBV | HAR-AAII | 1903 | $-2.38 \times 10^{-7}$ | -1.533 | 0.9373 | 1.0000 |
| ABBV | HAR-Full | 1903 | $-1.90 \times 10^{-6}$ | -1.906 | 0.9716 | 1.0000 |
| ABBV | HAR-Text | 1903 | $-8.78 \times 10^{-7}$ | -4.227 | 1.0000 | 1.0000 |
| ABBV | HAR-VIX | 1903 | $1.59 \times 10^{-6}$ | 2.675 | 0.0038 | 0.0052 |
| ABT | HAR-AAII | 5172 | $2.86 \times 10^{-8}$ | 0.362 | 0.3585 | 0.4088 |
| ABT | HAR-Full | 5172 | $1.97 \times 10^{-6}$ | 7.766 | $4.85 \times 10^{-15}$ | $1.24 \times 10^{-14}$ |
| ABT | HAR-Text | 5172 | $-6.61 \times 10^{-7}$ | -8.410 | 1.0000 | 1.0000 |
| ABT | HAR-VIX | 5172 | $4.23 \times 10^{-7}$ | 2.768 | 0.0028 | 0.0039 |
| AMZN | HAR-AAII | 5172 | $-2.04 \times 10^{-5}$ | -17.093 | 1.0000 | 1.0000 |
| AMZN | HAR-Full | 5172 | $-3.58 \times 10^{-5}$ | -11.360 | 1.0000 | 1.0000 |
| AMZN | HAR-Text | 5172 | $1.56 \times 10^{-5}$ | 14.804 | $6.73 \times 10^{-49}$ | $4.26 \times 10^{-48}$ |
| AMZN | HAR-VIX | 5172 | $-2.86 \times 10^{-6}$ | -8.053 | 1.0000 | 1.0000 |
| JNJ | HAR-AAII | 5172 | $-1.27 \times 10^{-6}$ | -13.838 | 1.0000 | 1.0000 |
| JNJ | HAR-Full | 5172 | $-1.22 \times 10^{-5}$ | -16.384 | 1.0000 | 1.0000 |
| JNJ | HAR-Text | 5172 | $-2.23 \times 10^{-6}$ | -6.209 | 1.0000 | 1.0000 |
| JNJ | HAR-VIX | 5172 | $-9.79 \times 10^{-6}$ | -16.377 | 1.0000 | 1.0000 |
| JPM | HAR-AAII | 5172 | $6.25 \times 10^{-9}$ | 0.072 | 0.4715 | 0.5334 |
| JPM | HAR-Full | 5172 | $-1.39 \times 10^{-6}$ | -3.226 | 0.9994 | 1.0000 |
| JPM | HAR-Text | 5172 | $2.10 \times 10^{-6}$ | 14.951 | $8.22 \times 10^{-50}$ | $5.71 \times 10^{-49}$ |
| JPM | HAR-VIX | 5172 | $1.54 \times 10^{-6}$ | 4.181 | $1.47 \times 10^{-5}$ | $2.36 \times 10^{-5}$ |
| LLY | HAR-AAII | 5172 | $1.91 \times 10^{-7}$ | 4.255 | $1.06 \times 10^{-5}$ | $1.73 \times 10^{-5}$ |
| LLY | HAR-Full | 5172 | $1.22 \times 10^{-6}$ | 5.688 | $6.77 \times 10^{-9}$ | $1.30 \times 10^{-8}$ |
| LLY | HAR-Text | 5172 | $2.25 \times 10^{-7}$ | 2.104 | 0.0177 | 0.0229 |
| LLY | HAR-VIX | 5172 | $7.99 \times 10^{-7}$ | 6.696 | $1.18 \times 10^{-11}$ | $2.65 \times 10^{-11}$ |
| MA | HAR-AAII | 3565 | $-2.32 \times 10^{-6}$ | -10.019 | 1.0000 | 1.0000 |
| MA | HAR-Full | 3565 | $-1.43 \times 10^{-5}$ | -10.276 | 1.0000 | 1.0000 |
| MA | HAR-Text | 3565 | $-9.16 \times 10^{-6}$ | -11.427 | 1.0000 | 1.0000 |
| MA | HAR-VIX | 3565 | $-5.55 \times 10^{-6}$ | -8.148 | 1.0000 | 1.0000 |
| UNH | HAR-AAII | 5172 | $-8.53 \times 10^{-8}$ | -3.822 | 0.9999 | 1.0000 |
| UNH | HAR-Full | 5172 | $-7.45 \times 10^{-7}$ | -3.854 | 0.9999 | 1.0000 |
| UNH | HAR-Text | 5172 | $-6.18 \times 10^{-7}$ | -6.308 | 1.0000 | 1.0000 |
| UNH | HAR-VIX | 5172 | $4.29 \times 10^{-7}$ | 3.124 | 0.0009 | 0.0013 |
| WMT | HAR-AAII | 5171 | $-8.52 \times 10^{-8}$ | -3.300 | 0.9995 | 1.0000 |
| WMT | HAR-Full | 5171 | $3.55 \times 10^{-7}$ | 3.639 | 0.0001 | 0.0002 |
| WMT | HAR-Text | 5171 | $-2.24 \times 10^{-7}$ | -4.555 | 1.0000 | 1.0000 |
| WMT | HAR-VIX | 5171 | $3.42 \times 10^{-7}$ | 4.225 | $1.21 \times 10^{-5}$ | $1.96 \times 10^{-5}$ |
| XOM | HAR-AAII | 5172 | $2.75 \times 10^{-8}$ | 1.283 | 0.0998 | 0.1205 |
| XOM | HAR-Full | 5172 | $4.60 \times 10^{-7}$ | 3.993 | $3.31 \times 10^{-5}$ | $5.22 \times 10^{-5}$ |
| XOM | HAR-Text | 5172 | $1.76 \times 10^{-7}$ | 3.637 | 0.0001 | 0.0002 |
| XOM | HAR-VIX | 5172 | $5.80 \times 10^{-7}$ | 5.177 | $1.17 \times 10^{-7}$ | $2.10 \times 10^{-7}$ |

Table C.8: CW, annualized-volatility squared error, $h = 5$: expanding window

| Stock | Model | $n$ | $\Delta$ | Statistic | $p$ | $q^{global}$ |
|---|---|---|---|---|---|---|
| AAPL | HAR-AAII | 5160 | $-1.60 \times 10^{-5}$ | -7.010 | 1.0000 | 1.0000 |
| AAPL | HAR-Full | 5160 | $-5.72 \times 10^{-5}$ | -4.503 | 1.0000 | 1.0000 |
| AAPL | HAR-Text | 5160 | $-3.98 \times 10^{-5}$ | -6.163 | 1.0000 | 1.0000 |
| AAPL | HAR-VIX | 5160 | $-4.02 \times 10^{-5}$ | -7.771 | 1.0000 | 1.0000 |
| ABBV | HAR-AAII | 1891 | $-1.29 \times 10^{-6}$ | -2.961 | 0.9984 | 1.0000 |
| ABBV | HAR-Full | 1891 | $-6.25 \times 10^{-6}$ | -1.639 | 0.9494 | 1.0000 |
| ABBV | HAR-Text | 1891 | $-1.81 \times 10^{-6}$ | -2.255 | 0.9879 | 1.0000 |
| ABBV | HAR-VIX | 1891 | $4.75 \times 10^{-6}$ | 1.739 | 0.0411 | 0.0517 |
| ABT | HAR-AAII | 5160 | $2.61 \times 10^{-7}$ | 0.606 | 0.2722 | 0.3138 |
| ABT | HAR-Full | 5160 | $5.49 \times 10^{-6}$ | 5.026 | $2.59 \times 10^{-7}$ | $4.61 \times 10^{-7}$ |
| ABT | HAR-Text | 5160 | $-4.95 \times 10^{-6}$ | -6.905 | 1.0000 | 1.0000 |
| ABT | HAR-VIX | 5160 | $1.39 \times 10^{-6}$ | 2.280 | 0.0113 | 0.0149 |
| AMZN | HAR-AAII | 5160 | $-5.77 \times 10^{-5}$ | -10.169 | 1.0000 | 1.0000 |
| AMZN | HAR-Full | 5160 | $-8.91 \times 10^{-5}$ | -7.303 | 1.0000 | 1.0000 |
| AMZN | HAR-Text | 5160 | $5.88 \times 10^{-5}$ | 9.883 | $3.95 \times 10^{-23}$ | $1.32 \times 10^{-22}$ |
| AMZN | HAR-VIX | 5160 | $-1.15 \times 10^{-5}$ | -6.431 | 1.0000 | 1.0000 |
| JNJ | HAR-AAII | 5160 | $-4.76 \times 10^{-6}$ | -8.157 | 1.0000 | 1.0000 |
| JNJ | HAR-Full | 5160 | $-2.86 \times 10^{-5}$ | -9.697 | 1.0000 | 1.0000 |
| JNJ | HAR-Text | 5160 | $-6.04 \times 10^{-6}$ | -3.772 | 0.9999 | 1.0000 |
| JNJ | HAR-VIX | 5160 | $-3.14 \times 10^{-5}$ | -10.229 | 1.0000 | 1.0000 |
| JPM | HAR-AAII | 5160 | $5.74 \times 10^{-7}$ | 1.320 | 0.0935 | 0.1136 |
| JPM | HAR-Full | 5160 | $-5.11 \times 10^{-6}$ | -2.907 | 0.9982 | 1.0000 |
| JPM | HAR-Text | 5160 | $1.13 \times 10^{-5}$ | 10.974 | $5.17 \times 10^{-28}$ | $1.97 \times 10^{-27}$ |
| JPM | HAR-VIX | 5160 | $5.40 \times 10^{-6}$ | 4.435 | $4.71 \times 10^{-6}$ | $7.85 \times 10^{-6}$ |
| LLY | HAR-AAII | 5160 | $8.24 \times 10^{-7}$ | 3.642 | 0.0001 | 0.0002 |
| LLY | HAR-Full | 5160 | $5.13 \times 10^{-6}$ | 4.758 | $1.00 \times 10^{-6}$ | $1.73 \times 10^{-6}$ |
| LLY | HAR-Text | 5160 | $1.10 \times 10^{-6}$ | 2.054 | 0.0200 | 0.0258 |
| LLY | HAR-VIX | 5160 | $2.83 \times 10^{-6}$ | 4.938 | $4.08 \times 10^{-7}$ | $7.16 \times 10^{-7}$ |
| MA | HAR-AAII | 3553 | $-7.13 \times 10^{-6}$ | -6.640 | 1.0000 | 1.0000 |
| MA | HAR-Full | 3553 | $-4.35 \times 10^{-5}$ | -7.313 | 1.0000 | 1.0000 |
| MA | HAR-Text | 3553 | $-3.09 \times 10^{-5}$ | -7.478 | 1.0000 | 1.0000 |
| MA | HAR-VIX | 3553 | $-1.34 \times 10^{-5}$ | -5.504 | 1.0000 | 1.0000 |
| UNH | HAR-AAII | 5160 | $2.06 \times 10^{-7}$ | 2.134 | 0.0165 | 0.0214 |
| UNH | HAR-Full | 5160 | $-3.57 \times 10^{-7}$ | -0.545 | 0.7072 | 0.7823 |
| UNH | HAR-Text | 5160 | $-1.38 \times 10^{-6}$ | -3.495 | 0.9998 | 1.0000 |
| UNH | HAR-VIX | 5160 | $1.23 \times 10^{-6}$ | 2.269 | 0.0117 | 0.0153 |
| WMT | HAR-AAII | 5159 | $-3.28 \times 10^{-7}$ | -2.535 | 0.9944 | 1.0000 |
| WMT | HAR-Full | 5159 | $-2.41 \times 10^{-7}$ | -0.439 | 0.6696 | 0.7449 |
| WMT | HAR-Text | 5159 | $-9.03 \times 10^{-7}$ | -3.110 | 0.9991 | 1.0000 |
| WMT | HAR-VIX | 5159 | $1.32 \times 10^{-6}$ | 3.404 | 0.0003 | 0.0005 |
| XOM | HAR-AAII | 5160 | $2.20 \times 10^{-7}$ | 1.722 | 0.0426 | 0.0534 |
| XOM | HAR-Full | 5160 | $1.39 \times 10^{-6}$ | 3.898 | $4.91 \times 10^{-5}$ | $7.66 \times 10^{-5}$ |
| XOM | HAR-Text | 5160 | $7.15 \times 10^{-7}$ | 2.624 | 0.0044 | 0.0059 |
| XOM | HAR-VIX | 5160 | $1.51 \times 10^{-6}$ | 4.532 | $2.99 \times 10^{-6}$ | $5.02 \times 10^{-6}$ |

Table C.9: CW, annualized-volatility squared error, $h = 22$: expanding window

| Stock | Model | $n$ | $\Delta$ | Statistic | $p$ | $q^{global}$ |
|---|---|---|---|---|---|---|
| AAPL | HAR-AAII | 5109 | -0.000205 | -4.056 | 1.0000 | 1.0000 |
| AAPL | HAR-Full | 5109 | -0.000904 | -4.655 | 1.0000 | 1.0000 |
| AAPL | HAR-Text | 5109 | -0.000616 | -3.202 | 0.9993 | 1.0000 |
| AAPL | HAR-VIX | 5109 | -0.000480 | -4.853 | 1.0000 | 1.0000 |
| ABBV | HAR-AAII | 1840 | $-3.53 \times 10^{-6}$ | -0.469 | 0.6805 | 0.7559 |
| ABBV | HAR-Full | 1840 | $-3.80 \times 10^{-5}$ | -1.519 | 0.9356 | 1.0000 |
| ABBV | HAR-Text | 1840 | $-3.48 \times 10^{-6}$ | -0.505 | 0.6932 | 0.7678 |
| ABBV | HAR-VIX | 1840 | $1.58 \times 10^{-5}$ | 0.813 | 0.2080 | 0.2419 |
| ABT | HAR-AAII | 5109 | $6.70 \times 10^{-6}$ | 1.467 | 0.0713 | 0.0875 |
| ABT | HAR-Full | 5109 | $1.90 \times 10^{-5}$ | 1.729 | 0.0419 | 0.0526 |
| ABT | HAR-Text | 5109 | $-3.65 \times 10^{-5}$ | -3.833 | 0.9999 | 1.0000 |
| ABT | HAR-VIX | 5109 | $8.27 \times 10^{-6}$ | 1.650 | 0.0495 | 0.0616 |
| AMZN | HAR-AAII | 5109 | 0.000182 | 4.716 | $1.24 \times 10^{-6}$ | $2.12 \times 10^{-6}$ |
| AMZN | HAR-Full | 5109 | 0.000210 | 2.552 | 0.0054 | 0.0072 |
| AMZN | HAR-Text | 5109 | -0.000223 | -4.914 | 1.0000 | 1.0000 |
| AMZN | HAR-VIX | 5109 | $9.38 \times 10^{-5}$ | 5.689 | $6.74 \times 10^{-9}$ | $1.30 \times 10^{-8}$ |
| JNJ | HAR-AAII | 5109 | -0.000406 | -5.762 | 1.0000 | 1.0000 |
| JNJ | HAR-Full | 5109 | $-8.38 \times 10^{-5}$ | -1.764 | 0.9611 | 1.0000 |
| JNJ | HAR-Text | 5109 | 0.000287 | 6.158 | $3.98 \times 10^{-10}$ | $8.20 \times 10^{-10}$ |
| JNJ | HAR-VIX | 5109 | -0.000343 | -5.746 | 1.0000 | 1.0000 |
| JPM | HAR-AAII | 5109 | $-1.17 \times 10^{-5}$ | -0.708 | 0.7605 | 0.8330 |
| JPM | HAR-Full | 5109 | 0.000266 | 4.656 | $1.66 \times 10^{-6}$ | $2.83 \times 10^{-6}$ |
| JPM | HAR-Text | 5109 | 0.000115 | 5.534 | $1.65 \times 10^{-8}$ | $3.09 \times 10^{-8}$ |
| JPM | HAR-VIX | 5109 | $4.85 \times 10^{-5}$ | 4.565 | $2.55 \times 10^{-6}$ | $4.31 \times 10^{-6}$ |
| LLY | HAR-AAII | 5109 | $2.47 \times 10^{-6}$ | 0.555 | 0.2894 | 0.3322 |
| LLY | HAR-Full | 5109 | $5.60 \times 10^{-5}$ | 3.262 | 0.0006 | 0.0008 |
| LLY | HAR-Text | 5109 | $2.78 \times 10^{-5}$ | 3.387 | 0.0004 | 0.0005 |
| LLY | HAR-VIX | 5109 | $3.00 \times 10^{-5}$ | 5.214 | $9.63 \times 10^{-8}$ | $1.73 \times 10^{-7}$ |
| MA | HAR-AAII | 3502 | $-1.29 \times 10^{-5}$ | -1.006 | 0.8427 | 0.9193 |
| MA | HAR-Full | 3502 | -0.000276 | -3.677 | 0.9999 | 1.0000 |
| MA | HAR-Text | 3502 | -0.000154 | -3.197 | 0.9993 | 1.0000 |
| MA | HAR-VIX | 3502 | -0.000106 | -3.538 | 0.9998 | 1.0000 |
| UNH | HAR-AAII | 5109 | $3.85 \times 10^{-6}$ | 2.335 | 0.0098 | 0.0130 |
| UNH | HAR-Full | 5109 | $5.24 \times 10^{-6}$ | 1.341 | 0.0901 | 0.1096 |
| UNH | HAR-Text | 5109 | $7.30 \times 10^{-6}$ | 2.276 | 0.0114 | 0.0150 |
| UNH | HAR-VIX | 5109 | $4.63 \times 10^{-6}$ | 1.142 | 0.1268 | 0.1510 |
| WMT | HAR-AAII | 5108 | $5.03 \times 10^{-6}$ | 3.000 | 0.0014 | 0.0019 |
| WMT | HAR-Full | 5108 | $1.56 \times 10^{-5}$ | 2.607 | 0.0046 | 0.0062 |
| WMT | HAR-Text | 5108 | $6.44 \times 10^{-6}$ | 1.914 | 0.0278 | 0.0355 |
| WMT | HAR-VIX | 5108 | $8.04 \times 10^{-6}$ | 2.627 | 0.0043 | 0.0059 |
| XOM | HAR-AAII | 5109 | $1.04 \times 10^{-5}$ | 3.641 | 0.0001 | 0.0002 |
| XOM | HAR-Full | 5109 | $1.32 \times 10^{-5}$ | 1.577 | 0.0574 | 0.0711 |
| XOM | HAR-Text | 5109 | $4.89 \times 10^{-6}$ | 1.098 | 0.1361 | 0.1618 |
| XOM | HAR-VIX | 5109 | $1.17 \times 10^{-5}$ | 2.940 | 0.0016 | 0.0023 |

# Appendix D

## D.1 Window and inference sensitivity summaries

The rolling arm uses the same initial hyperparameters, models, outcomes, and test dates as the primary expanding arm, changing the training window to the latest 252 eligible observations. The joint 1,584-test adjustment includes both arms. Longer-HAC and BY-adjusted results should be consulted before treating small short-bandwidth probabilities as robust evidence. Table D.1 compares estimation windows, and Table D.8 contrasts the two HAC rules for HAR-Text. Tables D.9 and D.10 report compact fold and regime summaries. The complete fold-level QLIKE differences for $h = 1$, $h = 5$, and $h = 22$ are reported in Tables D.2– D.7; the corresponding cell-level ex-post regime diagnostics are reported in Tables D.11 and D.12.

Table D.1: Matched comparisons by estimation window

| Window | Model | QLIKE better | DM better | DM worse | DM joint |
|---|---|---|---|---|---|
| Expanding | HAR-AAII | 2 | 0 | 25 | 0 |
| Expanding | HAR-Full | 2 | 1 | 30 | 1 |
| Expanding | HAR-Text | 2 | 2 | 30 | 2 |
| Expanding | HAR-VIX | 18 | 8 | 11 | 8 |
| Rolling (252) | HAR-AAII | 3 | 2 | 25 | 2 |
| Rolling (252) | HAR-Full | 1 | 1 | 31 | 1 |
| Rolling (252) | HAR-Text | 2 | 0 | 29 | 0 |
| Rolling (252) | HAR-VIX | 8 | 7 | 22 | 7 |

*Note:* All counts are out of 33. DM better/worse uses the window-specific 792-test BH adjustment; DM joint uses the combined 1,584-test adjustment. All use QLIKE and bandwidth $h$.

Table D.2: HAR-Text versus HAR: fold-level QLIKE outcomes, $h = 1$, Expanding window

| Stock | 1 | 2 | 3 | 4 | 5 |
|---|---|---|---|---|---|
| AAPL | -0.003242 | -0.000029 | -0.000017 | +0.000015 | +0.000007 |
| ABBV | -0.000137 | -0.000043 | -0.000024 | +0.000001 | -0.000033 |
| ABT | -0.000369 | -0.000013 | -0.000001 | 0.000000 | +0.000003 |
| AMZN | -0.000328 | +0.000001 | -0.000018 | +0.000001 | -0.000005 |
| JNJ | -0.000911 | -0.000044 | -0.000015 | -0.000013 | -0.000018 |
| JPM | -0.000216 | -0.000003 | -0.000002 | -0.000006 | -0.000007 |
| LLY | -0.000576 | -0.000040 | -0.000004 | +0.000006 | -0.000002 |
| MA | -0.001548 | -0.000027 | -0.000001 | +0.000005 | +0.000006 |
| UNH | -0.000472 | 0.000000 | -0.000003 | -0.000002 | -0.000008 |
| WMT | -0.000250 | +0.000002 | 0.000000 | +0.000004 | +0.000001 |
| XOM | -0.000080 | 0.000000 | -0.000002 | -0.000003 | +0.000006 |

*Note:* $\Delta QLIKE = QLIKE_{HAR} - QLIKE_{HAR\text{-}Text}$. Positive values favor HAR-Text; negative values favor HAR. This window reports 55 stock-fold comparisons (11 stocks × 5 folds); both windows together report 110 at this horizon.

Table D.3: HAR-Text versus HAR: fold-level QLIKE outcomes, $h = 1$, Rolling (252) window

| Stock | 1 | 2 | 3 | 4 | 5 |
|---|---|---|---|---|---|
| AAPL | -0.009670 | -0.000081 | -0.003093 | -0.000492 | -0.000205 |
| ABBV | -0.000234 | -0.000752 | -0.000096 | -0.000230 | -0.000144 |
| ABT | -0.000794 | -0.000147 | -0.000384 | -0.002498 | +0.000030 |
| AMZN | -0.000072 | -0.000343 | -0.000126 | -0.000229 | -0.000169 |
| JNJ | -0.000488 | +0.000047 | -0.000179 | -0.000522 | -0.000700 |
| JPM | +0.000268 | -0.000822 | -0.000217 | -0.000155 | -0.000295 |
| LLY | -0.000494 | -0.000242 | -0.000240 | -0.001148 | -0.000808 |
| MA | -0.003490 | +0.000015 | -0.000804 | -0.006111 | -0.000172 |
| UNH | -0.000661 | -0.002846 | -0.000155 | -0.000755 | -0.000030 |
| WMT | -0.000076 | +0.000180 | -0.000193 | -0.000069 | -0.000198 |
| XOM | -0.000434 | -0.000358 | -0.000466 | -0.000640 | -0.000053 |

*Note:* $\Delta QLIKE = QLIKE_{HAR} - QLIKE_{HAR\text{-}Text}$. Positive values favor HAR-Text; negative values favor HAR. This window reports 55 stock-fold comparisons (11 stocks × 5 folds); both windows together report 110 at this horizon.

Table D.4: HAR-Text versus HAR: fold-level QLIKE outcomes, $h = 5$, Expanding window

| Stock | 1 | 2 | 3 | 4 | 5 |
|---|---|---|---|---|---|
| AAPL | -0.012310 | -0.000102 | -0.000055 | +0.000086 | +0.000050 |
| ABBV | -0.000270 | -0.000182 | -0.000102 | -0.000004 | -0.000147 |
| ABT | -0.002974 | -0.000053 | -0.000004 | -0.000010 | +0.000018 |
| AMZN | -0.002573 | -0.000005 | -0.000059 | 0.000000 | -0.000007 |
| JNJ | -0.002121 | -0.000185 | -0.000060 | -0.000044 | -0.000071 |
| JPM | -0.001989 | -0.000006 | -0.000012 | -0.000026 | -0.000013 |
| LLY | -0.002334 | -0.000125 | -0.000018 | +0.000028 | -0.000004 |
| MA | -0.004603 | -0.000046 | +0.000013 | +0.000029 | +0.000036 |
| UNH | -0.001163 | 0.000000 | -0.000014 | -0.000002 | -0.000027 |
| WMT | -0.001297 | +0.000009 | 0.000000 | +0.000015 | +0.000024 |
| XOM | -0.000121 | +0.000004 | -0.000008 | -0.000012 | +0.000029 |

*Note:* $\Delta QLIKE = QLIKE_{HAR} - QLIKE_{HAR\text{-}Text}$. Positive values favor HAR-Text; negative values favor HAR. This window reports 55 stock-fold comparisons (11 stocks × 5 folds); both windows together report 110 at this horizon.

Table D.5: HAR-Text versus HAR: fold-level QLIKE outcomes, $h = 5$, Rolling (252) window

| Stock | 1 | 2 | 3 | 4 | 5 |
|---|---|---|---|---|---|
| AAPL | -0.023772 | -0.000494 | -0.018554 | -0.000667 | -0.000244 |

| Stock | 1 | 2 | 3 | 4 | 5 |
|---|---|---|---|---|---|
| ABBV | -0.000587 | -0.002863 | -0.000498 | -0.001317 | -0.000436 |
| ABT | -0.006952 | -0.000503 | -0.002566 | -0.007434 | -0.000180 |
| AMZN | -0.002849 | -0.000867 | -0.000463 | -0.001130 | -0.000854 |
| JNJ | -0.002263 | +0.000260 | -0.001523 | -0.001774 | -0.000523 |
| JPM | -0.001514 | -0.006549 | -0.000846 | -0.000492 | -0.000044 |
| LLY | -0.001386 | -0.000337 | -0.000811 | -0.002349 | +0.006263 |
| MA | -0.009997 | -0.000049 | -0.001686 | -0.013024 | -0.000246 |
| UNH | -0.001445 | -0.007517 | -0.000564 | -0.003007 | -0.000217 |
| WMT | -0.000509 | +0.001132 | -0.000955 | -0.000792 | -0.000140 |
| XOM | -0.000731 | -0.001665 | -0.003428 | -0.004891 | -0.000073 |

*Note:* $\Delta QLIKE = QLIKE_{HAR} - QLIKE_{HAR\text{-}Text}$. Positive values favor HAR-Text; negative values favor HAR. This window reports 55 stock-fold comparisons (11 stocks × 5 folds); both windows together report 110 at this horizon.

Table D.6: HAR-Text versus HAR: fold-level QLIKE outcomes, $h = 22$, Expanding window

| Stock | 1 | 2 | 3 | 4 | 5 |
|---|---|---|---|---|---|
| AAPL | -0.064993 | -0.000363 | -0.000649 | +0.001162 | +0.001010 |
| ABBV | -0.000418 | -0.000844 | -0.001094 | +0.000225 | -0.001422 |
| ABT | -0.017136 | -0.000302 | -0.000428 | -0.000157 | +0.000588 |
| AMZN | -0.056369 | -0.000270 | -0.000970 | -0.000112 | +0.000107 |
| JNJ | +0.044643 | +0.000286 | -0.000966 | -0.000551 | -0.000112 |
| JPM | +0.016018 | +0.000049 | +0.000054 | -0.000277 | +0.001040 |
| LLY | -0.006443 | -0.000085 | -0.000065 | +0.000366 | +0.000084 |
| MA | -0.017451 | -0.000167 | +0.000619 | +0.000478 | +0.000758 |
| UNH | -0.002240 | -0.000027 | -0.000067 | +0.000014 | +0.000024 |
| WMT | -0.009831 | +0.000091 | -0.000156 | +0.000135 | +0.000678 |
| XOM | -0.002510 | +0.000161 | -0.000158 | -0.000303 | +0.000846 |

*Note:* $\Delta QLIKE = QLIKE_{HAR} - QLIKE_{HAR\text{-}Text}$. Positive values favor HAR-Text; negative values favor HAR. This window reports 55 stock-fold comparisons (11 stocks × 5 folds); both windows together report 110 at this horizon.

Table D.7: HAR-Text versus HAR: fold-level QLIKE outcomes, $h = 22$, Rolling (252) window

| Stock | 1 | 2 | 3 | 4 | 5 |
|---|---|---|---|---|---|
| AAPL | -0.069078 | +0.006653 | -0.048426 | -0.000970 | -0.004384 |
| ABBV | -0.003991 | -0.023143 | -0.013785 | -0.010667 | -0.023999 |
| ABT | -0.041115 | -0.002230 | -0.027096 | -0.020590 | -0.031383 |

| Stock | 1 | 2 | 3 | 4 | 5 |
|---|---|---|---|---|---|
| AMZN | -0.058631 | +0.000976 | -0.003520 | -0.019787 | -0.035988 |
| JNJ | +0.072930 | +0.005293 | -0.017695 | -0.034294 | -0.000903 |
| JPM | +0.048113 | -0.075440 | -0.003218 | -0.001496 | -0.003506 |
| LLY | -0.005019 | +0.002242 | -0.005825 | -0.010824 | -0.008717 |
| MA | -0.052296 | -0.027048 | -0.002958 | -0.035826 | -0.007567 |
| UNH | -0.001859 | -0.038090 | -0.006713 | -0.009967 | -0.016032 |
| WMT | -0.011181 | +0.022916 | -0.012571 | -0.001182 | -0.002565 |
| XOM | -0.004501 | -0.022319 | -0.027241 | -0.054010 | -0.006035 |

*Note:* $\Delta QLIKE = QLIKE_{HAR} - QLIKE_{HAR\text{-}Text}$. Positive values favor HAR-Text; negative values favor HAR. This window reports 55 stock-fold comparisons (11 stocks × 5 folds); both windows together report 110 at this horizon.

Table D.8: HAR-Text inferential sensitivity

| Window | HAC rule | BH better | BH worse | BY better | BY worse |
|---|---|---|---|---|---|
| Expanding | $h$ | 2 | 30 | 2 | 29 |
| Expanding | $252 + h - 1$ | 0 | 3 | 0 | 0 |
| Rolling (252) | $h$ | 0 | 29 | 0 | 29 |
| Rolling (252) | $252 + h - 1$ | 0 | 24 | 0 | 6 |

*Note:* Each count is out of 33. The $h$ rule is retained for protocol comparability. The $252 + h - 1$ rule is the dependence-robust rule used for thesis-level inference. The BH and BY columns use the window-specific 792-test family.

Table D.9: HAR-Text versus HAR: fold-level QLIKE wins

| Window | $h = 1$ | $h = 5$ | $h = 22$ |
|---|---|---|---|
| Expanding | 15/55 | 14/55 | 23/55 |
| Rolling (252) | 5/55 | 3/55 | 7/55 |

*Note:* Each denominator is 11 stocks × 5 outer folds. A win means lower fold-level QLIKE for HAR-Text. The complete 330-row table for both windows is retained as `tables/fold_metrics.csv`.

Table D.10: HAR-Text versus HAR: ex-post regime diagnostic

| Window | High-state wins | Low-state wins | High larger | Median improvement: high / low |
|---|---|---|---|---|
| Expanding | 10/33 | 2/33 | 22/33 | -8.73% / -56.28% |
| Rolling (252) | 5/33 | 2/33 | 21/33 | -100.49% / -194.89% |

*Note:* A win means lower regime-specific QLIKE for HAR-Text. "High larger" counts cells in which the relative QLIKE improvement is numerically larger in the high state. Regimes are estimated ex post and are descriptive heterogeneity diagnostics, not causal or confirmatory results. The complete cell-by-regime table is retained as `tables/regime_metrics.csv`.

Table D.11: HAR-Text versus HAR: cell-level ex-post regime diagnostics (expanding)

| Stock | $h$ | High ΔQLIKE | Low ΔQLIKE | High (%) | Low (%) | High larger |
|---|---|---|---|---|---|---|
| AAPL | 1 | -0.001632 | -0.000201 | -796.55 | -50.12 | No |

| Stock | $h$ | High ΔQLIKE | Low ΔQLIKE | High (%) | Low (%) | High larger |
|---|---|---|---|---|---|---|
| AAPL | 5 | -0.006118 | -0.000788 | -1070.60 | -90.24 | No |
| AAPL | 22 | -0.032262 | -0.003967 | -416.46 | -62.63 | No |
| ABBV | 1 | -0.000061 | -0.000033 | -16.55 | -14.64 | No |
| ABBV | 5 | -0.000138 | -0.000144 | -16.23 | -29.06 | Yes |
| ABBV | 22 | -0.000489 | -0.000967 | -10.05 | -42.10 | Yes |
| ABT | 1 | -0.000007 | -0.000114 | -3.01 | -67.74 | Yes |
| ABT | 5 | -0.000033 | -0.000915 | -4.84 | -225.11 | Yes |
| ABT | 22 | +0.000253 | -0.005471 | 4.63 | -247.69 | Yes |
| AMZN | 1 | -0.000108 | -0.000063 | -61.29 | -17.83 | No |
| AMZN | 5 | -0.000471 | -0.000539 | -131.89 | -85.90 | No |
| AMZN | 22 | -0.006018 | -0.012455 | -249.12 | -545.35 | Yes |
| JNJ | 1 | -0.000147 | -0.000222 | -33.39 | -36.92 | Yes |
| JNJ | 5 | -0.000281 | -0.000584 | -20.62 | -32.34 | Yes |
| JNJ | 22 | +0.003995 | +0.010495 | 35.07 | 55.22 | No |
| JPM | 1 | +0.000005 | -0.000100 | 1.03 | -50.99 | Yes |
| JPM | 5 | +0.000047 | -0.000879 | 4.38 | -159.52 | Yes |
| JPM | 22 | +0.000569 | +0.006236 | 7.25 | 34.86 | No |
| LLY | 1 | -0.000020 | -0.000269 | -8.73 | -93.79 | Yes |
| LLY | 5 | -0.000089 | -0.001054 | -18.20 | -208.71 | Yes |
| LLY | 22 | -0.000726 | -0.001922 | -28.23 | -85.23 | Yes |
| MA | 1 | -0.000280 | -0.000319 | -178.58 | -85.53 | No |
| MA | 5 | -0.000970 | -0.000902 | -179.00 | -100.58 | No |
| MA | 22 | -0.006501 | -0.002550 | -94.08 | -53.57 | No |
| UNH | 1 | -0.000002 | -0.000120 | -1.15 | -56.28 | Yes |
| UNH | 5 | -0.000009 | -0.000298 | -1.77 | -65.72 | Yes |
| UNH | 22 | +0.000003 | -0.000573 | 0.06 | -29.19 | Yes |
| WMT | 1 | +0.000002 | -0.000070 | 0.69 | -40.78 | Yes |
| WMT | 5 | +0.000025 | -0.000366 | 3.55 | -105.85 | Yes |
| WMT | 22 | +0.000610 | -0.002901 | 12.62 | -201.10 | Yes |
| XOM | 1 | -0.000003 | -0.000023 | -0.89 | -17.09 | Yes |
| XOM | 5 | -0.000008 | -0.000029 | -1.08 | -6.84 | Yes |
| XOM | 22 | +0.000718 | -0.000818 | 7.61 | -32.92 | Yes |

*Note:* $\Delta QLIKE = QLIKE_{HAR} - QLIKE_{HAR\text{-}Text}$. Positive values favor HAR-Text. Improvement is the relative QLIKE improvement of HAR-Text over HAR. States are ex-post diagnostics.

Table D.12: HAR-Text versus HAR: cell-level ex-post regime diagnostics (rolling252)

| Stock | $h$ | High ΔQLIKE | Low ΔQLIKE | High (%) | Low (%) | High larger |
|---|---|---|---|---|---|---|
| AAPL | 1 | -0.005184 | -0.001564 | -1433.59 | -310.97 | No |
| AAPL | 5 | -0.013572 | -0.006528 | -1716.34 | -672.68 | No |
| AAPL | 22 | -0.031414 | -0.019545 | -292.30 | -284.75 | No |
| ABBV | 1 | -0.000239 | -0.000346 | -64.31 | -126.86 | Yes |
| ABBV | 5 | -0.000885 | -0.001410 | -104.26 | -212.09 | Yes |
| ABBV | 22 | -0.010305 | -0.020685 | -194.20 | -624.36 | Yes |
| ABT | 1 | -0.000769 | -0.000753 | -197.24 | -359.51 | Yes |
| ABT | 5 | -0.002640 | -0.004010 | -195.71 | -658.32 | Yes |
| ABT | 22 | -0.026886 | -0.023201 | -207.94 | -529.35 | Yes |
| AMZN | 1 | -0.000298 | -0.000167 | -71.55 | -34.95 | No |
| AMZN | 5 | -0.000834 | -0.001306 | -69.68 | -130.51 | Yes |
| AMZN | 22 | -0.006328 | -0.026304 | -59.91 | -508.42 | Yes |
| JNJ | 1 | -0.000487 | -0.000320 | -58.88 | -72.41 | Yes |
| JNJ | 5 | -0.000710 | -0.001349 | -24.42 | -103.87 | Yes |
| JNJ | 22 | +0.000056 | +0.007029 | 0.20 | 26.28 | No |
| JPM | 1 | -0.000449 | -0.000033 | -50.70 | -7.83 | No |
| JPM | 5 | -0.002597 | -0.001160 | -85.93 | -84.18 | No |
| JPM | 22 | -0.030499 | +0.016730 | -141.10 | 46.41 | No |

| Stock | $h$ | High ΔQLIKE | Low ΔQLIKE | High (%) | Low (%) | High larger |
|---|---|---|---|---|---|---|
| LLY | 1 | -0.000662 | -0.000479 | -10.40 | -135.37 | Yes |
| LLY | 5 | +0.001484 | -0.001417 | 11.56 | -153.41 | Yes |
| LLY | 22 | -0.006160 | -0.004894 | -23.52 | -131.55 | Yes |
| MA | 1 | -0.001565 | -0.002220 | -444.30 | -511.24 | Yes |
| MA | 5 | -0.004783 | -0.005042 | -423.92 | -443.85 | Yes |
| MA | 22 | -0.040439 | -0.022385 | -390.11 | -315.47 | No |
| UNH | 1 | -0.002635 | -0.000466 | -646.19 | -194.89 | No |
| UNH | 5 | -0.007179 | -0.001423 | -724.21 | -269.08 | No |
| UNH | 22 | -0.042128 | -0.007702 | -693.06 | -302.72 | No |
| WMT | 1 | +0.000151 | -0.000166 | 26.10 | -83.84 | Yes |
| WMT | 5 | +0.000847 | -0.000720 | 43.14 | -166.45 | Yes |
| WMT | 22 | +0.014340 | -0.007736 | 55.05 | -325.94 | Yes |
| XOM | 1 | -0.000403 | -0.000383 | -88.94 | -179.34 | Yes |
| XOM | 5 | -0.002328 | -0.002068 | -148.77 | -227.64 | Yes |
| XOM | 22 | -0.019924 | -0.023935 | -100.49 | -173.64 | Yes |

*Note:* $\Delta QLIKE = QLIKE_{HAR} - QLIKE_{HAR\text{-}Text}$. Positive values favor HAR-Text. Improvement is the relative QLIKE improvement of HAR-Text over HAR. States are ex-post diagnostics.

### D.1.1 Numerical positivity-floor diagnostic

The primary safeguard is retained unchanged. One raw forecast is negative: HAR-Full for MA at $h = 5$, rolling window, on 30 July 2015. Its raw annualized-volatility forecast is approximately $-0.01110$, while the outcome is 0.19726. Replacing that forecast by $10^{-6}$ produces a single-observation QLIKE of approximately $3.8911 \times 10^{10}$. This explains the extreme QLIKE magnitude for that cell and the rolling horizon average; it is not deleted from either the losses or tests.

Table D.13 is a post-hoc numerical diagnostic based on the same raw forecasts. It changes no fitted model and defines no new inferential family. The original primary floor is not replaced by a more favorable value. Raw MSE for this cell is approximately $5.261 \times 10^{-4}$, compared with $5.509 \times 10^{-5}$ for HAR. The unconstrained level forecast is therefore problematic beyond the particular numerical magnitude induced by the QLIKE floor.

Table D.13: Post-hoc floor sensitivity: MA, HAR-Full, $h = 5$, rolling window

| Annualized-volatility floor | Replaced forecasts | Cell mean QLIKE |
|---|---|---|
| $1.00 \times 10^{-6}$ | 1 | $1.10 \times 10^{7}$ |
| $1.00 \times 10^{-5}$ | 1 | $1.10 \times 10^{5}$ |
| 0.000100 | 1 | $1.095740 \times 10^{3}$ |
| 0.001000 | 1 | $1.153019 \times 10^{1}$ |
| 0.010000 | 2 | $2.569860 \times 10^{-1}$ |

*Note:* The cell contains 3,553 OOS forecasts. The primary floor is $10^{-6}$. At $10^{-2}$, one additional small positive forecast is also replaced. No diagnostic setting is used to revise the main results.